\documentclass[journal]{IEEEtran}
\usepackage{lineno}
\usepackage{float}
\usepackage{cite}
\usepackage{hyperref}
\usepackage{amsmath,amssymb,amsfonts}
\usepackage{amsthm}
\DeclareMathOperator*{\argmax}{argmax}
\usepackage{graphicx}
\usepackage{textcomp}
\usepackage{xcolor}
\usepackage{graphicx}
\usepackage{subfigure}
\usepackage{amsmath}
\usepackage{amsfonts,amssymb}
\usepackage{mathrsfs}
\usepackage{mathtools}
\usepackage{algorithm}
\usepackage{algorithmicx}
\usepackage{algpseudocode}
\usepackage{bm}
\usepackage{multirow}
\usepackage{array}
\usepackage{url}
\usepackage{fancyhdr}
\usepackage{mdwmath}
\usepackage{mdwtab}
\usepackage{caption}
\usepackage{setspace}
\usepackage{bm}
\usepackage{algpseudocode}
\usepackage{mathtools}
\usepackage{dsfont}
\usepackage{bbm}
\usepackage{cases}
\usepackage{booktabs}
\newtheorem{remark}{Remark}
\theoremstyle{definition}
\newtheorem{theorem}{Theorem}

\newtheorem{lemma}{Lemma}

\newtheorem{corollary}{Corollary}

\makeatletter
\newcommand{\biggg}{\bBigg@{3}}
\newcommand{\Biggg}{\bBigg@{3.5}}
\makeatother
    \expandafter\def\expandafter\normalsize\expandafter{%
    \normalsize%
    \setlength\abovedisplayskip{4pt}%
    \setlength\belowdisplayskip{4pt}%
    \setlength\abovedisplayshortskip{2pt}%
    \setlength\belowdisplayshortskip{2pt}%
}
\begin{document}

\title{Downlink and Uplink ISAC in Continuous-Aperture Array (CAPA) Systems}

\author{Boqun Zhao, \IEEEmembership{Graduate Student Member, IEEE}, Chongjun Ouyang, \IEEEmembership{Member, IEEE},\\Xingqi Zhang, \IEEEmembership{Senior Member, IEEE}, Hyundong Shin, \IEEEmembership{Fellow, IEEE}, and Yuanwei Liu, \IEEEmembership{Fellow, IEEE}\vspace{-15pt}
	\thanks{B. Zhao and X. Zhang are with the Department of Electrical and Computer Engineering, University of Alberta, Edmonton AB, T6G 2R3, Canada (email: \{boqun1, xingqi.zhang\}@ualberta.ca).}
	\thanks{C. Ouyang is with the School of Electronic Engineering and Computer Science, Queen Mary University of London, London, E1 4NS, U.K. (e-mail: c.ouyang@qmul.ac.uk).}
    \thanks{H. Shin is with the Department of Electronics and Information Convergence Engineering, Kyung Hee University, Yongin-si, Gyeonggido 17104, Republic of Korea (e-mail: hshin@khu.ac.kr).}
	\thanks{Y. Liu is with the Department of Electrical and Electronic Engineering, The University of Hong Kong, Hong Kong (e-mail: yuanwei@hku.hk).}
}

\maketitle

\begin{abstract}
A continuous-aperture array (CAPA)-based integrated sensing and communications (ISAC) framework is proposed for both downlink and uplink scenarios. Within this framework, continuous operator-based signal models are employed to describe the sensing and communication processes. The performance of communication and sensing is analyzed using two information-theoretic metrics: the communication rate (CR) and the sensing rate (SR). 1) For downlink ISAC, three continuous beamforming designs are proposed: \romannumeral1) the communications-centric (C-C) design that maximizes the CR, \romannumeral2) the sensing-centric (S-C) design that maximizes the SR, and \romannumeral3) the Pareto-optimal design that characterizes the Pareto boundary of the CR-SR region. A low-complexity signal subspace-based approach is proposed to derive the closed-form optimal beamformers for the considered designs. On this basis, closed-form expressions are derived for the achievable CRs and SRs, and the downlink rate region achieved by CAPAs is characterized. 2) For uplink ISAC, the C-C and S-C successive interference cancellation-based methods are proposed to manage inter-functionality interference. Using the subspace approach closed-form expressions for the optimal detectors as well as the achievable CRs and SRs are derived. The uplink SR-CR region is characterized based on the time-sharing technique. Numerical results demonstrate that, for both downlink and uplink, CAPA-based ISAC achieves higher CRs and SRs as well as larger CR-SR regions compared to conventional spatially discrete array-based ISAC.
\end{abstract} 

\section{Introduction}
Sixth-generation (6G) networks are expected to play a pivotal role in shaping a connected, smart, and intelligent wireless world \cite{saad2019vision}. Achieving this vision requires a paradigm shift to support not only high-quality wireless connectivity but also high-accuracy sensing capabilities \cite{liu2022integrated}. Against this background, integrated sensing and communications (ISAC) has been identified as one of the six key usage scenarios for IMT-2030 (6G), which has sparked growing interest in this area \cite{recommendation2023framework}. Compared to frequency-division sensing and communications (FDSAC) techniques, where sensing and communications require isolated frequency bands as well as separate hardware infrastructures, ISAC enables dual-functional sensing and communications (DFSAC) via a single time-frequency-power-hardware resource. Therefore, ISAC is envisioned to be more spectral-, energy-, and hardware-efficient than FDSAC \cite{liu2022integrated}.

Among the key research topics in ISAC, a significant focus is the integration of multiple-antenna technologies. Specifically, technologies such as multiple-input multiple-output (MIMO) provide additional spatial degrees of freedom (DoFs) and enhanced array gains for both communications and sensing. These improvements lead to higher network throughput and better sensing resolution; see \cite{liu2022integrated,hua_1} and related references for more details. In general, the available spatial DoFs and array gains scale linearly with the number of antennas. Motivated by this, various evolved \emph{large-scale MIMO variants} and novel array architectures have been proposed and applied to ISAC. Examples include massive MIMO \cite{liao2024power}, holographic MIMO \cite{zhang2023holographic}, extremely large aperture arrays (ELAAs) \cite{hua_2}, reconfigurable intelligent surfaces (RISs) \cite{chepuri2023integrated}, and fluid/movable antennas \cite{lyu2025movable,zhou2024fluid}. These works provide extensive results demonstrating the significant performance gains in ISAC systems achieved by leveraging these novel multiple-antenna technologies.

Despite their diverse architectures, these novel multiple-antenna systems share a common evolutionary trend: \emph{larger} aperture sizes, \emph{denser} antenna configurations, \emph{higher} operating frequencies, and \emph{more flexible} structures. This trend leads to the development of an (approximately) continuous electromagnetic (EM) aperture, also referred to as a \emph{continuous-aperture array (CAPA)} \cite{liu2024capa}. A CAPA operates as a single large-aperture antenna with a continuous current distribution, comprising a (virtually) infinite number of radiating elements coupled with electronic circuits and fed by a limited number of radio-frequency (RF) chains \cite{liu2024capa}. Compared with conventional spatially discrete arrays (SPDAs), a CAPA fully utilizes the entire surface of the aperture and enables free control of the current distribution. This capability yields significantly enhanced spatial DoFs and array gains. Initial efforts to apply CAPAs to both communications and sensing have demonstrated that CAPA outperforms traditional SPDAs in both areas, as discussed in Section \ref{Section: Prior Works}.

Given this context, it is natural to explore CAPA technology to further enhance ISAC performance. However, due to its continuous nature, a CAPA-based system should be modeled using EM theory and based on a \emph{continuous Hilbert–Schmidt operator-based integral model} \cite{poon2005degrees,liu2024capa,migliore2008electromagnetics,migliore2018horse,capa_single_0,pizzo2022spatial}, which differs significantly from the conventional matrix-based model for SPDAs. This shift renders existing SPDA-based ISAC frameworks inapplicable to CAPA-based ISAC systems. Therefore, a new conceptual CAPA-based ISAC framework is needed to address the continuous EM nature inherent in such systems.

{ \subsection{Hardware Implementation of CAPAs}
Achieving a continuous-aperture array (CAPA) has been a longstanding goal in antenna design, with early contributions from Harold Wheeler \cite{wheeler1965simple} and David Staiman \cite{staiman1968new} in the 1960s, who laid the groundwork for this design ideal. Recently, advancements in EM and optical materials and array fabrication have enabled the development of CAPA prototypes, utilizing designs such as metasurface-based leaky-wave antennas \cite{araghi2021holographic,smith2017analysis}, optically driven tightly coupled arrays \cite{prather2017optically}, and interdigital transducer (IDT)-based grating antennas \cite{yuan2024interdigital}. Some prototypes have even reached commercialization and, according to initial test reports, demonstrate significant potential for enhancing coverage and throughput in practical wireless network environments \cite{staff2019holographic,sazegar2022full}. 

Although these implementations differ, they share a common objective: to realize CAPA as an architecture for analog EM beamforming, in contrast to traditional digital beamforming approaches \cite{liu2024capa}. Fundamentally, CAPA is characterized not only by a continuous or semi-continuous EM structure but more critically by its ability to support full analog control of both amplitude and phase across the aperture \cite{BJORNSON20193,liu2024capa}. This enables CAPA systems to require only as many RF chains as there are spatially multiplexed data streams. In contrast, conventional SPDAs lack reconfigurability at the antenna level---their radiation patterns are fixed once the antenna type is selected. As a result, beamforming in SPDAs must rely entirely on analog/digital processing before the RF front end. 

However, achieving full analog amplitude and phase control over a continuous aperture at the radio front end introduces significant hardware complexity. This remains one of the principal challenges in CAPA implementation. A detailed discussion of the associated design considerations and enabling technologies is provided in \cite{liu2024capa}.

Additionally, despite its promise for full analog EM beamforming, CAPA faces hardware impairments similar to those of conventional arrays---including component non-linearity, I/Q imbalance, quantization error, and phase noise. Additional concerns, such as EM coupling, feed network complexity, and fabrication constraints, must also be addressed \cite{zhao2024continuous}. These challenges suggest that even with current advancements, the realization of a fully functional CAPA or densely packed semi-CAPA remains a significant undertaking, requiring collaborative innovation from both the communication and antenna communities.}

\vspace{-5pt}
\subsection{Prior Works}\label{Section: Prior Works}
While the hardware implementation of an ideal CAPA is still at an early stage, growing attention has been directed toward its theoretical analysis and designs of CAPA-based communications and sensing. The performance of CAPA-based communications has been extensively studied in terms of spatial DoFs \cite{pizzo2022nyquist}, array gains \cite{ouyang2024impact}, and channel capacity \cite{zhao2024continuous}, with most of these studies focusing on line-of-sight (LoS) channels. To address this limitation, researchers have developed a novel multipath spatial fading model for CAPA \cite{pizzo2022spatial}. Using this model, the work in \cite{per5} analyzed the spatial correlation and the diversity-multiplexing trade-off in CAPA-based communications. These studies have collectively validated CAPA's effectiveness in enhancing spatial DoFs and array gains, thereby providing a theoretical foundation for its application.

Another key research area is continuous beamforming design for CAPA-based communications. Since optimizing a continuous source current function involves challenging fractional programming, various approaches have been proposed. For example, a discretization-based approach (using Fourier series expansion) has been proposed to approximate continuous signals, which simplified the optimization problem \cite{opt1,opt2,opt3}. Additionally, a calculus of variations method was introduced to design low-complexity beamforming \cite{opt5} and explore optimal multi-user beamforming structures \cite{opt4}.

In addition to its role in communications, CAPA has also been applied to enhance sensing performance. The studies in \cite{chen2023cramer} and \cite{chen2024near} derived Cram{\'e}r-Rao bounds (CRBs) and Ziv-Zakai bounds (ZZBs) for CAPA-based positioning. Building on these advancements, \cite{jiang2024cram} addressed multi-target sensing and proposed low-complexity beamforming methods to improve CRBs. Initial efforts in CAPA-based ISAC have also been reported. For example, a Fourier series expansion-based beamforming approach was proposed to balance the communication signal-to-interference-plus-noise ratio (SINR) and sensing SINR in a downlink CAPA-based ISAC system \cite{capa_isac}.
\vspace{-3pt}
\subsection{Motivation and Contributions}
In light of the above discussion, CAPA-based ISAC is still in its early stages of exploration, with its fundamental performance limits yet to be fully investigated. From an information-theoretic perspective, the performance limits of sensing and communications can be described by the sensing rate (SR) and communication rate (CR), respectively \cite{ouyang2023integrated}. SR evaluates the system's ability to extract environmental information through sensing processes, while CR quantifies the system's capacity to transmit communication data. Understanding these limits is essential for optimizing CAPA-based ISAC design and achieving a balance between communication and sensing performance.

Motivated by the above research gaps, this article presents a comprehensive performance analysis of a CAPA-based ISAC system. { We propose a signal subspace-based approach with significantly lower complexity than the widely used Fourier-based approach \cite{opt1,opt2,opt3,capa_isac}, and employ it to study the SRs, CRs, and rate regions for both downlink and uplink scenarios.} The main contributions are summarized as follows.
\begin{itemize}
    \item We propose an analytically tractable framework for CAPA-based ISAC. For both downlink and uplink scenarios, we employ continuous operator-based signal and channel models to describe the CAPA-based sensing and communication processes. Based on the proposed framework, we define information-theoretic metrics---the CR and SR---to evaluate the fundamental limits of CAPA-based ISAC.
    \item For downlink ISAC, we propose three continuous DFSAC beamforming designs: \romannumeral1) a \emph{communications-centric (C-C)} design that maximizes the CR, \romannumeral2) a \emph{sensing-centric (S-C)} design that maximizes the SR, and \romannumeral3) a \emph{Pareto-optimal} design that characterizes the Pareto boundary of the rate region. For both C-C and S-C designs, we derive closed-form expressions for SRs and CRs, and examine the asymptotic performance for the infinitely larger CAPAs to gain more physical or geometric insights. For the Pareto-optimal design, we propose a signal subspace-based approach to derive closed-form solutions for the optimal beamformer and the achievable rate region.
    \item For uplink ISAC, we propose two successive interference cancellation (SIC)-based methods to address inter-functionality interference (IFI): \romannumeral1) a \emph{C-C SIC} that prioritizes communication signal decoding and \romannumeral2) a \emph{S-C SIC} that prioritizes sensing signal decoding. For both designs, we employ the subspace approach to derive closed-form expressions for CR and SR. Besides, we apply a time-sharing technique to characterize the uplink rate region.
    \item We present simulation results to demonstrate that: \romannumeral1) the proposed subspace approach for solving the optimal Pareto beamforming offers significantly lower computational complexity compared to the existing Fourier-based method; \romannumeral2) The achievable SR-CR rate regions of SPDA-based ISAC and CAPA-based FDSAC systems are entirely encompassed within those of the CAPA-based ISAC system in both downlink and uplink scenarios.
\end{itemize}

The remainder of this paper is organized as follows. Section \ref{section_system} introduces the downlink and uplink CAPA-based ISAC framework. Sections \ref{section_downlink} and \ref{section_uplink} analyze the performance of sensing and communications in donwlink and uplink scenarios, respectively. Section \ref{section_numerical} presents numerical results to demonstrate the superior performance of CAPA-based ISAC. Finally, Section \ref{section_conclusion} concludes the paper.
\subsubsection*{Notations}
Throughout this paper, scalars, vectors, and matrices are denoted by non-bold, bold lower-case, and bold upper-case letters, respectively. For the matrix $\mathbf{A}$, ${\mathbf{A}}^{\mathsf{T}}$, ${\mathbf{A}}^{*}$ and ${\mathbf{A}}^{\mathsf{H}}$ denote its transpose, conjugate, and transpose conjugate, respectively. For the square matrix $\mathbf{B}$, $\det(\mathbf{B})$ denotes its determinant. The notations $\lvert a\rvert$ and $\lVert \mathbf{a} \rVert$ denote the magnitude and norm of scalar $a$ and vector $\mathbf{a}$, respectively. The identity matrix with dimensions $N\times N$ is represented by $\mathbf{I}_N$. The set $\mathbbmss{R}$ and $\mathbbmss{C}$ stand for the real and complex spaces, respectively, and notation $\mathbbmss{E}\{\cdot\}$ represents mathematical expectation. The mutual information between random variables $X$ and $Y$ conditioned on $Z$ is shown by $I\left(X;Y|Z\right)$. The floor function is represented by $\lfloor \cdot \rfloor $. Finally, ${\mathcal{CN}}({\bm\mu},\mathbf{X})$ is used to denote the circularly-symmetric complex Gaussian distribution with mean $\bm\mu$ and covariance matrix $\mathbf{X}$.

\begin{figure}[!t]
 \centering
\setlength{\abovecaptionskip}{2pt}
\includegraphics[height=0.24\textwidth]{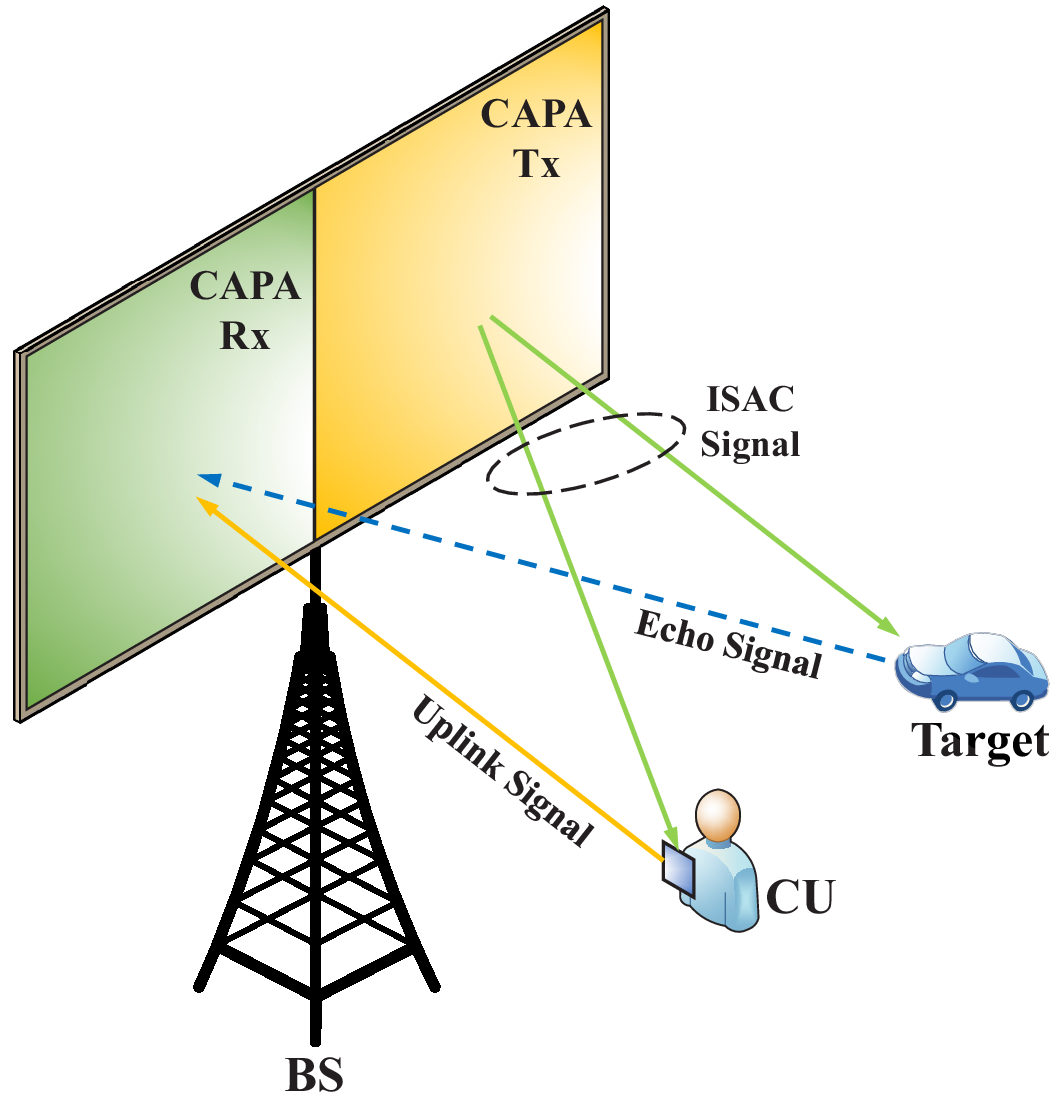}
\caption{Illustration of CAPA-based downlink/uplink ISAC.}
\label{model_a}
\vspace{-10pt}
\end{figure}

\section{System Model}\label{section_system}
\subsection{System Description}
We consider a donwlink/uplink CAPA-based ISAC system, where a DFSAC BS serves a single CU while simultaneously sensing a single point-like target, as illustrated in {\figurename} {\ref{model_a}}. The BS is equipped with two CAPAs of the same physical size: a transmit CAPA with aperture ${\mathcal{A}}_{\rm{t}}\subseteq{\mathbbmss{R}}^{3\times1}$ and a receive CAPA with aperture ${\mathcal{A}}_{\rm{r}}\subseteq{\mathbbmss{R}}^{3\times1}$. The CU has a single discrete antenna element with aperture ${\mathcal{A}}\subseteq{\mathbbmss{R}}^{3\times1}$ for both transmitting and receiving signals. Without loss of generality, we assume that the aperture sizes satisfy  $\lvert{\mathcal{A}}_{\rm{t}}\rvert=\lvert{\mathcal{A}}_{\rm{r}}\rvert\gg \lvert{\mathcal{A}}\rvert$. Furthermore, all antennas and arrays in the system are assumed to be \emph{uni-polarized}, meaning they generate one-dimensional (1D) source currents and receive 1D electric fields.

\begin{figure}[!t]
 \centering
\setlength{\abovecaptionskip}{1pt}
\includegraphics[height=0.26\textwidth]{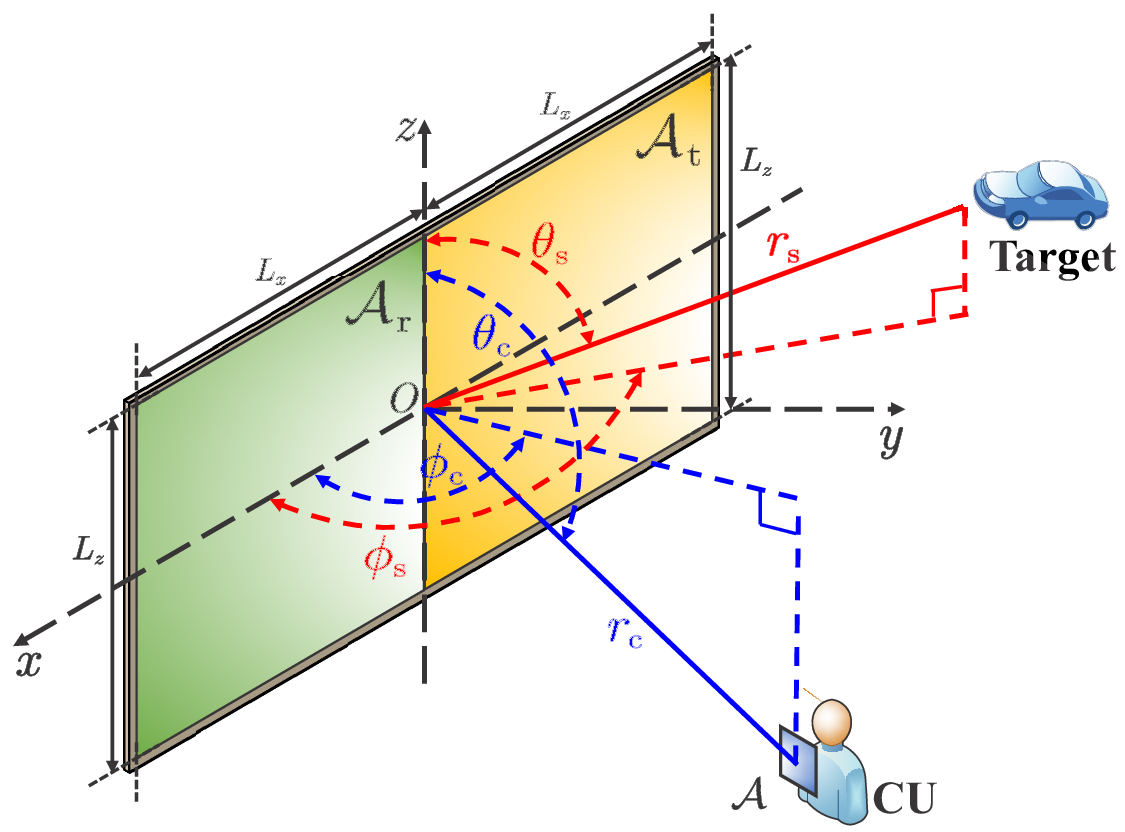}
\caption{Illustration of CAPA-based channels.}
\label{model_b}
\vspace{-7pt}
\end{figure}

As shown in {\figurename} \ref{model_b}, the transmit and receive CAPAs are positioned edge-to-edge on the $xz$ plane, symmetrically around the origin. The edges of the CAPAs are parallel to the coordinate axes, with physical dimensions $L_x$ and $L_z$ along the $x$- and $z$-axes, respectively. Therefore, the apertures of the CAPAs are given as follows:
\begin{subequations}
	\begin{align}
		&\mathcal{A} _{\mathrm{t}}=\{[x,0,z]^{\mathsf{T}}|x\in [0,L_x],z\in [-{L_z}/{2},{L_z}/{2}]\},\\
		&\mathcal{A} _{\mathrm{r}}=\{[x,0,z]^{\mathsf{T}}|x\in [-L_x,0],z\in [-{L_z}/{2},{L_z}/{2}]\},
	\end{align}
\end{subequations}
where the sizes of both apertures are $\lvert{\mathcal{A}}_{\rm{t}}\rvert=\lvert{\mathcal{A}}_{\rm{r}}\rvert=L_xL_z$. We denote the distance from the origin to the center of the CU's aperture (denoted by the subscript ``$\rm{c}$'') and the sensing target (denoted by the subscript ``$\rm{s}$'') by $r_k$. The corresponding elevation and azimuth angles are represented by $\theta_k \in \left[ 0,\pi \right] $ and $\phi_k \in \left[ 0,\pi \right] $, respectively, where $k\in\{\rm{c},\rm{s}\}$. As a result, the positions of the centers of the CU's aperture and the target are given by ${\mathbf{p}}_{k}=[r_k\Phi_k, r_k\Psi_k, r_k\Theta_k]^{\mathsf{T}}$ for $k\in\{\rm{c},\rm{s}\}$, where 
\begin{align}
\Phi_k\triangleq\cos{\phi_k}\sin{\theta_k}, \Psi_k\triangleq\sin{\phi_k}\sin{\theta_k}, \Theta_k\triangleq\cos{\theta_k}.
\end{align}

\subsection{Downlink ISAC}
We commence with the downlink ISAC model. Let ${\mathbf{x}}({\mathbf{t}})=[x_1({\mathbf{t}}),\ldots,x_L({\mathbf{t}})]\in{\mathbbmss{C}}^{1\times L}$ denote the DFSAC signal generated at the transmit CAPA for ${\mathbf{t}}\in{\mathcal{A}}_{\rm{t}}$, where $L$ is the length of the communication frame or sensing pulses. In the context of communications, $x_{\ell}({\mathbf{t}})$ represents the $\ell$th data symbol signal for $\ell=1,\ldots,L$. For sensing, $x_{\ell}({\mathbf{t}})$ corresponds to the sensing snapshot transmitted at the $\ell$th time interval. 

In the considered CAPA-based ISAC system, the continuous DFSAC signal is designed as follows:
\begin{align}
{\mathbf{x}}({\mathbf{t}})=\sqrt{P}w({\mathbf{t}}){{\mathbf{s}}^{\mathsf{T}}},
\end{align}
where $P$ represents the power budget, ${\mathbf{s}}\in{\mathbbmss{C}}^{L\times1}$ is the unit-power data stream intended for the CU, and $w({\mathbf{t}})$ is the normalized source current (i.e., transmit beamformer). The beamformer satisfies the condition 
\begin{align}
\int_{{\mathcal{A}}_{\rm{t}}}\lvert{w}({\mathbf{t}})\rvert^2{\rm{d}}{\mathbf{t}}=1. 
\end{align}
In this context, $w(\cdot)$ is a scalar field on ${\mathbbmss{R}}^{3\times 1}$, which maps each point ${\mathbf{t}}\in{\mathbbmss{R}}^{3\times 1}$ on the transmit array to a scalar value ${{w}}({\mathbf{t}})$. Additionally, $\mathbf{s}=[s_1,\ldots,s_L]^{\mathsf{T}}$ satisfies $\frac{1}{L}\lVert{\mathbf{s}}\rVert^2=1$, where $s_\ell$ denotes the coded data symbol transmitted at the $\ell$th time interval. As a result, the transmitted signal at the $\ell$th time interval is given by 
\begin{align}
{{x}}_\ell({\mathbf{t}})=\sqrt{P}s_\ell{{w}}({\mathbf{t}}),\quad \ell=1,\ldots,L.
\end{align}

\subsubsection{Communication Model}
At the $\ell$th time interval, the received signal at point $\mathbf{u}\in{\mathcal{A}}$ is expressed as follows \cite{poon2005degrees}:
\begin{align}\label{do_User_k_electric_radiation_field}
y_{{\rm{c}},\ell}(\mathbf{u})=\int_{{\mathcal{A}}_{\rm{t}}}h_{\rm{c}}(\mathbf{u},{\mathbf{t}})x_\ell({\mathbf{t}}){\rm{d}}{\mathbf{t}}
+n_{{\rm{c}},\ell}(\mathbf{u}),
\end{align}
for $\ell=1,\ldots,L$, where $y_{{\rm{c}},\ell}(\cdot)$ is the receive vector field, $h_{\rm{c}}(\cdot,\cdot)$ is the spatial response, and $n_{{\rm{c}},\ell}(\cdot)$ accounts for thermal noise. The function $h_{\rm{c}}(\cdot,\cdot)$ is a complex integral kernel where its domain is the set of transmit fields and its range is the set of receive fields. The term $h_{\rm{c}}(\mathbf{u},{\mathbf{t}})$ gives the channel response between the transmit
position ${\mathbf{t}}\in{\mathcal{A}}_{\rm{t}}$ and the receive position $\mathbf{u}\in{\mathcal{A}}$. The noise field is modeled as a zero-mean complex Gaussian random process, which satisfies 
\begin{align}
{\mathbbmss{E}}\{n_{{\rm{c}},\ell}(\mathbf{u})n_{{\rm{c}},\ell}^*(\mathbf{u}')\}={\sigma}_{\rm{c}}^2\delta(\mathbf{u}-{\mathbf{u}}'), 
\end{align}
where $\delta(\cdot)$ is the Dirac delta function and ${\sigma}_{\rm{c}}^2$ is the noise intensity. The signals $\mathbf{s}$ and the noise fields $\{n_{{\rm{c}},\ell}(\mathbf{u})\}_{\ell=1}^{L}$ are assumed to be uncorrelated. 

For LoS scenarios\footnote{{ While this work primarily focuses on LoS channels to establish an analytically tractable framework for CAPA-based ISAC and to facilitate theoretical investigations into fundamental performance limits, the proposed CAPA-based methodology is general and applicable to arbitrary channel models. In our recent work \cite{fading}, we have developed a CAPA-based multipath fading channel model, some results of which are directly applicable to ISAC scenarios. These developments provide a solid foundation for future extensions. However, a comprehensive analysis of more complex multipath fading environments is left for future work.}}, the channel response $h_{\rm{c}}(\mathbf{u},{\mathbf{t}})$ is expressed as follows \cite{ouyang2024impact}:
\begin{align}\label{Communication_Channel_LoS_Model}
h_{\rm{c}}(\mathbf{u},{\mathbf{t}})=\sqrt{\frac{\lvert{\mathbf{e}}_y^{\mathsf{T}}(\mathbf{u}-\mathbf{t})\rvert}{\lVert\mathbf{u}-\mathbf{t}\rVert}}
g(\mathbf{u},\mathbf{t}),
\end{align}
where $\sqrt{\frac{\lvert{\mathbf{e}}_{\rm{t}}^{\mathsf{T}}(\mathbf{u}-\mathbf{t})\rvert}{\lVert\mathbf{u}-\mathbf{t}\rVert}}$ accounts for the effective aperture loss caused by misalignment between the signal propagation direction and the array's normal direction. Moreover, ${\mathbf{e}}_y=[0,1,0]^\mathsf{T}$ denotes the normal vector to the transmit CAPA, and $g(\cdot,\cdot)$ is the scalar Green's function that models the effects of free-space EM propagation. By \cite{ouyang2024impact}, we have
\begin{align}
g(\mathbf{u},\mathbf{t})=-{\rm{j}}\eta k_0\frac{{\rm{e}}^{-{\rm{j}}k_0\lVert\mathbf{u}-\mathbf{t}\rVert}}{\sqrt{4\pi}\lVert\mathbf{u}-\mathbf{t}\rVert}
\left(1+\frac{{\rm{j}}k_0^{-1}}{\lVert\mathbf{u}-\mathbf{t}\rVert}
-\frac{k_0^{-2}}{\lVert\mathbf{u}-\mathbf{t}\rVert^2}\right),
\end{align}
where $\eta$ is the wave impedance and $k_0=\frac{2\pi}{\lambda}$ is the wavenumber, with $\lambda$ representing the wavelength. 

The scalar Green's function consists of three components. The first term corresponds to the radiating near- and far-field regions, while the other two represent the reactive near-field region. As shown in \cite{ouyang2024impact}, the reactive region has a negligible impact on the channel gain, even for an infinitely large aperture. Therefore, the reactive terms in the function can be omitted for subsequent analysis, which simplifies $g(\mathbf{u},\mathbf{t})$ as follows:
\begin{align}
g(\mathbf{u},\mathbf{t})=-{\rm{j}}\eta k_0\frac{{\rm{e}}^{-{\rm{j}}k_0\lVert\mathbf{u}-\mathbf{t}\rVert}}{\sqrt{4\pi}\lVert\mathbf{u}-\mathbf{t}\rVert}.
\end{align}
\subsubsection{Sensing Model}
At the $\ell$th time interval, the received echo signal at the BS at point $\mathbf{r}\in{\mathcal{A}_{\rm{r}}}$ for target sensing can be written as follows:
\begin{align}\label{sensing_Model_basic_one}
y_{{\rm{s}},\ell}(\mathbf{r})=\int_{{\mathcal{A}}_{\rm{t}}}h_{\rm{s}}(\mathbf{r},{\mathbf{t}})x_\ell({\mathbf{t}}){\rm{d}}{\mathbf{t}}
+n_{{\rm{s}},\ell}(\mathbf{r}),
\end{align}
where $h_{\rm{s}}(\mathbf{r},{\mathbf{t}})$ represents the target response, and $n_{{\rm{s}},\ell}(\mathbf{r})$ is zero-mean additive noise satisfying 
\begin{align}
{\mathbbmss{E}}\{n_{{\rm{s}},\ell}(\mathbf{r})n_{{\rm{s}},\ell}^*(\mathbf{r}')\}={\sigma}_{\rm{s}}^2\delta(\mathbf{r}-{\mathbf{r}}'). 
\end{align}
The target response can be decomposed as follows \cite{poon2005degrees,pizzo2022spatial}:
\begin{align}\label{Target_Response_General}
h_{\rm{s}}(\mathbf{r},{\mathbf{t}})=\int a(\mathbf{r},\mathbf{k})h(\mathbf{k})a(\mathbf{k},\mathbf{t}){\rm{d}}{\mathbf{k}},
\end{align}
where $a(\cdot,\cdot)$ and $h(\cdot)$ are complex integral kernels. The transmit array response $a(\mathbf{k},\mathbf{t})$ maps the excitation current distribution at ${\mathbf{t}}\in{\mathcal{A}}_{\rm{t}}$ to the radiated field pattern at point ${\mathbf{k}}\in{\mathbbmss{R}}^{3\times1}$. Similarly, the receive array response $a(\mathbf{r},\mathbf{k})$ maps the incident field pattern at ${\mathbf{k}}$ to the induced current distribution at ${\mathbf{r}}\in{\mathcal{A}}_{\rm{r}}$. The scattering response $h(\mathbf{k})$ describes the radar cross section (RCS) or reflectivity of the target at point ${\mathbf{k}}$. According to \cite{boqun_jstsp,ouyang2024impact} and \eqref{Communication_Channel_LoS_Model}, the array responses can be written as follows:
	\begin{align}
		a(\mathbf{x},\mathbf{y})&=\sqrt{\frac{\lvert{\mathbf{e}}_y^{\mathsf{T}}(\mathbf{x}-\mathbf{y})\rvert}{\lVert\mathbf{x}-\mathbf{y}\rVert}}g(\mathbf{x},\mathbf{y}).
	\end{align} 

Based on the ray-tracing model in \cite{zwick2002stochastic}, the scattering response is modeled as follows:
\begin{align}
h(\mathbf{k})=\sum\nolimits_{n=1}^{N}\beta_n\delta({\mathbf{k}}-{\mathbf{k}}_n),
\end{align}
where $N$ denotes the number of resolvable point-like targets, ${\mathbf{k}}_n\in{\mathbbmss{R}}^{3\times1}$ is the location of the $n$th target, and $\beta_n$ is the associated RCS. In this article, we consider a single point-like target (i.e., $N=1$) located at ${\mathbf{p}}_{\rm{s}}\in{\mathbbmss{R}}^{3\times1}$ with an RCS of $\beta_{\rm{s}}$. Therefore, the scattering response simplifies to
\begin{align}
h(\mathbf{k})=\beta_{\rm{s}}\delta({\mathbf{k}}-{\mathbf{p}}_{\rm{s}}),
\end{align}
which, together with \eqref{Target_Response_General}, yields
\begin{align}
h_{\rm{s}}(\mathbf{r},{\mathbf{t}})=a(\mathbf{r},{\mathbf{p}}_{\rm{s}})\beta_{\rm{s}}a({\mathbf{p}}_{\rm{s}},\mathbf{t}).
\end{align}
We define $a_{\rm{t}}(\mathbf{t})\triangleq a({\mathbf{p}}_{\rm{s}},\mathbf{t})$ and $a_{\rm{r}}(\mathbf{r})\triangleq a(\mathbf{r},{\mathbf{p}}_{\rm{s}})$. Consequently, the received echo signal can be expressed as follows:
\begin{align}
y_{{\rm{s}},\ell}(\mathbf{r})=a_{\rm{r}}(\mathbf{r})\beta_{\rm{s}}\int_{{\mathcal{A}}_{\rm{t}}}a_{\rm{t}}(\mathbf{t})
x_\ell({\mathbf{t}}){\rm{d}}{\mathbf{t}}
+n_{{\rm{s}},\ell}(\mathbf{r}).
\end{align}

{ In this work, assuming that the target's location is known or has been estimated via a separate process \cite{aoa_1,aoa_2,boqun_jstsp,sensing_MI}, we focus on estimating the RCS $\beta_{\rm{s}}$. The RCS encodes critical information about the target’s physical and EM characteristics, which plays a key role in target recognition, surface material classification, and other high-resolution sensing applications \cite{rcs,ouyang2023integrated}.} Under this scenario, the sensing task aims to extract the environmental information contained in $\beta_{\rm{s}}$ from the received echo signals $\{y_{{\rm{s}},\ell}(\mathbf{r})\}_{\ell=1}^{L}$, given knowledge of ${\mathbf{x}}({\mathbf{t}})$. According to the Swerling-{\uppercase\expandafter{\romannumeral1}} model \cite{richards2005fundamentals}, the RCS $\beta_{\rm{s}}$ is assumed to remain relatively constant across pulses and has an \emph{a prior} Rayleigh-distributed amplitude. Under this model, $\beta_{\rm{s}}$ follows a complex Gaussian distribution, i.e., $\beta_{\rm{s}}\sim{\mathcal{CN}}(0,\alpha_{\rm{s}})$, where $\alpha_{\rm{s}}>0$ is the arithmetic mean of all RCS values of the reflecting object. This parameter represents the average reflection strength \cite{richards2005fundamentals}.

\subsection{Uplink ISAC}
The uplink ISAC process consists of two stages. In the first stage, the BS broadcasts a predesigned sensing signal
\begin{equation}
x_{{\rm{s}},\ell}(\mathbf{t})=\sqrt{P_{\rm{s}}}w({\mathbf{t}})s_{{\rm{s}},\ell}
\end{equation} 
to the nearby environment at each time interval $\ell=1,\ldots,L$, where ${\mathbf{s}}_{\rm{s}}=[s_{{\rm{s}},1},\ldots,s_{{\rm{s}},L}]^{\mathsf{T}}\in{\mathbbmss{C}}^{L\times 1}$ represents the sensing pulse and satisfies $\frac{1}{L}\lVert{\mathbf{s}}_{\rm{s}}\rVert^2=1$, and $P_{\rm{s}}$ is the power allocated for sensing. In the second stage, the BS receives both the sensing echo signal reflected by the target and the uplink communication signal transmitted by the CU. { The uplink communication signal, sent from point $\mathbf{u}\in{\mathcal{A}}$ within the CU's aperture at the $\ell$th time interval, can be written as follows:
\begin{equation}
	x_{{\rm{c}},\ell}({\mathbf{u}})=\sqrt{\frac{P_{\rm{c}}}{\lvert{\mathcal{A}}\rvert}}s_{{\rm{c}},\ell}, 
\end{equation} 
where ${\mathbf{s}}_{\rm{c}}=[s_{{\rm{c}},1},\ldots,s_{{\rm{c}},L}]^{\mathsf{T}}\in{\mathbbmss{C}}^{L\times1}$ subject to ${\mathbbmss{E}}\{{\mathbf{s}}_{\rm{c}}{\mathbf{s}}_{\rm{c}}^{\mathsf{H}}\}={\mathbf{I}}_L$ represents the normalized data message sent by the CU, $P_{\rm{c}}$ denotes the power allocated for communications, and the term $\frac{1}{\lvert{\mathcal{A}}\rvert}$ is used to normalize the effect of the antenna's aperture such that $\int_{\mathcal{A}}{\mathbbmss{E}}\{\lvert{x_{{\rm{c}},\ell}({\mathbf{u}})}\rvert^2\}{}=P_{\rm{c}}$.}

The sensing and communication signals are assumed to be perfectly synchronized at the BS through the use of properly designed synchronization sequences \cite{liu2022integrated}. As a result, the BS observes the following superposed signal at point $\mathbf{r}\in{\mathcal{A}_{\rm{r}}}$ during the $\ell$th time interval:
\begin{equation}\label{up_signal}
	\begin{split}
		y_{\ell}({\mathbf{r}})=&a_{\rm{r}}(\mathbf{r})\beta_{\rm{s}}
		\int_{{\mathcal{A}}_{\rm{t}}}a_{\rm{t}}(\mathbf{t})x_{{\rm{s}},\ell}({\mathbf{t}}){\rm{d}}{\mathbf{t}}\\
		&+\int_{{\mathcal{A}}}h_{\rm{c}}(\mathbf{r},{\mathbf{u}})x_{{\rm{c}},\ell}({\mathbf{u}}){\rm{d}}{\mathbf{u}}+n_{\ell}(\mathbf{r}),
	\end{split}
\end{equation}
where the noise field $n_{\ell}(\mathbf{r})$ is modeled as a zero-mean complex Gaussian random process satisfying ${\mathbbmss{E}}\{{{n}}_{\ell}(\mathbf{r}){{n}}_{\ell}^*(\mathbf{r}')\}={\sigma}^2\delta(\mathbf{r}-{\mathbf{r}}')$, and ${\sigma}^2$ represents the noise intensity. 
\vspace{-5pt}
\begin{remark}
We note that all the channel responses and signal models discussed above, including \eqref{do_User_k_electric_radiation_field}, \eqref{Communication_Channel_LoS_Model}, \eqref{sensing_Model_basic_one}, \eqref{Target_Response_General}, and \eqref{up_signal} are based on continuous operators and integrals rather than matrices. This makes the design and analysis of CAPA-based ISAC significantly more challenging than conventional SPDA-based ISAC. This difficulty will be addressed in the following sections.
\end{remark}
\vspace{-5pt}
\section{Performance of Downlink ISAC}\label{section_downlink}
This section analyzes the performance of communications and sensing in CAPA-based downlink ISAC system.
\subsection{Performance Evaluation Metrics}
\subsubsection{Performance Metric for Communications}\label{Section: Performance of Downlink ISAC: Performance Evaluation Metrics: Performance Metric for Communications}
Given that $\lvert{\mathcal{A}}_{\rm{t}}\rvert\gg \lvert{\mathcal{A}}\rvert$ and $\lvert{\mathcal{A}}\rvert$ is negligible compared to the propagation distance, signal variations within ${\mathcal{A}}$ can be ignored. This allows us to simplify \eqref{do_User_k_electric_radiation_field} as follows:
\begin{align}
{{y}}_{{\rm{c}},\ell}(\mathbf{u})\approx \sqrt{P}s_\ell\int_{{\mathcal{A}}_{\rm{t}}}h_{\rm{d}}({\mathbf{t}})
w({\mathbf{t}}){\rm{d}}{\mathbf{t}}
+n_{{\rm{c}},\ell}(\mathbf{u}).
\end{align}
where $h_{\rm{d}}({\mathbf{t}})\triangleq h_{\rm{c}}(\mathbf{p}_{\rm{c}},{\mathbf{t}})$. The CU can then apply a detector ${{v}}_{\rm{c}}({\mathbf{u}})$ and use maximum likelihood (ML) decoding to recover the data contained in $s_\ell$, which yields
\begin{align}
\int_{\mathcal{A}}{{v}}_{\rm{c}}^{*}({\mathbf{u}}){{y}}_{{\rm{c}},\ell}(\mathbf{u}){\rm{d}}{\mathbf{u}}
=&\sqrt{P}s_\ell\int_{\mathcal{A}}{{v}}_{\rm{c}}^{*}({\mathbf{u}})\int_{{\mathcal{A}}_{\rm{t}}}\!h_{\rm{d}}({\mathbf{t}})
{{w}}({\mathbf{t}}){\rm{d}}{\mathbf{t}}{\rm{d}}{\mathbf{u}}\nonumber\\
&+\int_{\mathcal{A}}{{v}}_{\rm{c}}^{*}({\mathbf{u}}){{n}}_{{\rm{c}},\ell}(\mathbf{u}){\rm{d}}{\mathbf{u}}.
\end{align}
As proved in \cite{zhao2024continuous}, $\int_{\mathcal{A}}{{v}}_{\rm{c}}^{*}({\mathbf{u}}){{n}}_{{\rm{c}},\ell}(\mathbf{u}){\rm{d}}{\mathbf{u}}$ is a complex Gaussian random variable with $\int_{\mathcal{A}}{{v}}_{\rm{c}}^{*}({\mathbf{u}}){{n}}_{{\rm{c}},\ell}(\mathbf{u}){\rm{d}}{\mathbf{u}}
\sim{\mathcal{CN}}(0,{\sigma}_{\rm{c}}^2\int_{\mathcal{A}}\lvert{{v}}_{\rm{c}}({\mathbf{u}})\rvert^2{\rm{d}}{\mathbf{u}})$. The resulting signal-to-noise ratio (SNR) for decoding $s_{\ell}$ is given by:
\begin{align}\label{Downlink_Communication_SNR_pre}
\gamma_{\rm{c}}({{w}}({\mathbf{t}}))=\frac{P\lvert\int_{\mathcal{A}}{{v}}_{\rm{c}}^{*}({\mathbf{u}})\int_{{\mathcal{A}}_{\rm{t}}}h_{\rm{d}}({\mathbf{t}})
{{w}}({\mathbf{t}}){\rm{d}}{\mathbf{t}}{\rm{d}}{\mathbf{u}}\rvert^2}{{\sigma}_{\rm{c}}^2\int_{\mathcal{A}}\lvert{{v}}_{\rm{c}}({\mathbf{u}})\rvert^2{\rm{d}}{\mathbf{u}}}.
\end{align}

To maximize the received SNR, maximal-ratio combining (MRC) can be used by setting ${{v}}_{\rm{c}}({\mathbf{u}})=\int_{{\mathcal{A}}_{\rm{t}}}h_{\rm{d}}({\mathbf{t}})
{{w}}({\mathbf{t}}){\rm{d}}{\mathbf{t}}$ for ${\mathbf{u}}\in{\mathcal{A}}$. Substituting this into \eqref{Downlink_Communication_SNR_pre}, we obtain
\begin{subequations}\label{Downlink_Communication_SNR}
\begin{align}
\gamma _{\mathrm{c}}(w(\mathbf{t}))&=\frac{P \lvert \int_{\mathcal{A}}{\lvert \int_{\mathcal{A} _{\mathrm{t}}}{}h_{\rm{d}}({\mathbf{t}})w(\mathbf{t})\mathrm{d}\mathbf{t} \rvert^2}\mathrm{d}\mathbf{u}  \rvert^2}{\sigma _{\mathrm{c}}^{2}\int_{\mathcal{A}}{\lvert \int_{\mathcal{A} _{\mathrm{t}}}{}h_{\rm{d}}({\mathbf{t}})w(\mathbf{t})\mathrm{d}\mathbf{t} \rvert^2}\mathrm{d}\mathbf{u}}\\
&=\frac{P}{\sigma _{\mathrm{c}}^{2}}\int_{\mathcal{A}}{ \left| \int_{\mathcal{A} _{\mathrm{t}}}{}h_{\rm{d}}({\mathbf{t}})w(\mathbf{t})\mathrm{d}\mathbf{t}  \right|^2}\mathrm{d}\mathbf{u}\\
&=\overline{\gamma}_{\mathrm{c}} \left| \int_{\mathcal{A} _{\mathrm{t}}}{h_{\mathrm{d}}}(\mathbf{t})w(\mathbf{t})\mathrm{d}\mathbf{t}  \right|^2,
\end{align}
\end{subequations}
where $\overline{\gamma}_{\mathrm{c}}\triangleq\frac{P\left| \mathcal{A} \right|}{\sigma _{\mathrm{c}}^{2}}$, and the last equality follows from the fact that $\int_{\mathcal{A}}{\rm{d}}{\mathbf{u}}=\lvert{\mathcal{A}}\rvert$. The communication performance of the downlink ISAC system can be evaluated using the CR, which is expressed as follows:
\begin{align}\label{CR_define}
{\mathcal{R}}_{\rm{d},{\rm{c}}}({{w}}({\mathbf{t}}))&=\log_2(1+\gamma_{\rm{c}}({{w}}({\mathbf{t}}))).
\end{align} 
\subsubsection{Performance Metric for Sensing}
The sensing task aims to extract environmental information contained in $\beta_{\rm{s}}$ from the reflected echo signals $\{y_{{\rm{s}},\ell}(\mathbf{r})\}_{\ell=1}^{L}$, using the known signal ${\mathbf{x}}({\mathbf{t}})$. To this end, the receiver employs a detector ${{v}}_{\rm{s}}({\mathbf{r}})$ for $\mathbf{r}\in\mathcal{A}_{\rm{r}}$, which yields
\begin{equation}
\begin{split}
\int_{{\mathcal{A}}_{\rm{r}}}{}{v}_{\rm{s}}^{*}({\mathbf{r}}){{y}}_{{\rm{s}},\ell}(\mathbf{r}){\rm{d}}{\mathbf{r}}=\sqrt{P}s_{\ell}\beta_{\rm{s}}\int_{{\mathcal{A}}_{\rm{r}}}{{v}}_{\rm{s}}^{*}({\mathbf{r}})a_{\rm{r}}(\mathbf{r}){\rm{d}}{\mathbf{r}}&\\
\times\int_{{\mathcal{A}}_{\rm{t}}}a_{\rm{t}}(\mathbf{t})w({\mathbf{t}}){\rm{d}}{\mathbf{t}}
+\int_{{\mathcal{A}}_{\rm{r}}}{{v}}_{\rm{s}}^{*}({\mathbf{r}}){{n}}_{{\rm{s}},\ell}(\mathbf{r}){\rm{d}}{\mathbf{r}}&.
\end{split}
\end{equation}
For clarity, we denote $\hat{y}_{{\rm{s}},\ell}=\int_{{\mathcal{A}}_{\rm{r}}}{{v}}_{\rm{s}}^{*}({\mathbf{r}}){{y}}_{{\rm{s}},\ell}(\mathbf{r}){\rm{d}}{\mathbf{r}}$, $\hat{{v}}_{\rm{s}}=\int_{{\mathcal{A}}_{\rm{r}}}{{v}}_{\rm{s}}^{*}({\mathbf{r}})a_{\rm{r}}(\mathbf{r}){\rm{d}}{\mathbf{r}}$, $\hat{{w}}_{\rm{s}}=\int_{{\mathcal{A}}_{\rm{t}}}a_{\rm{t}}(\mathbf{t})w({\mathbf{t}}){\rm{d}}{\mathbf{t}}$, and $\hat{n}_{{\rm{s}},\ell}=\int_{{\mathcal{A}}_{\rm{r}}}{{v}}_{\rm{s}}^{*}({\mathbf{r}}){{n}}_{{\rm{s}},\ell}(\mathbf{r}){\rm{d}}{\mathbf{r}}
\sim{\mathcal{CN}}(0,{\sigma}_{\rm{s}}^2\int_{\mathcal{A}_{\rm{r}}}\lvert{{v}}_{\rm{s}}({\mathbf{r}})\rvert^2{\rm{d}}{\mathbf{r}})$. It follows that
\begin{align}
\hat{y}_{{\rm{s}},\ell}=\sqrt{P}\hat{{v}}_{\rm{s}}\beta_{\rm{s}}\hat{{w}}_{\rm{s}}s_{\ell}+\hat{n}_{{\rm{s}},\ell},~\ell=1,\ldots,L,
\end{align}
which can be rewritten as follows:
\begin{align}\label{Sensing_Model_Transformed}
\hat{\mathbf{y}}_{{\rm{s}}}=\sqrt{P}\hat{{v}}_{\rm{s}}\hat{{w}}_{\rm{s}}{\mathbf{s}}\beta_{\rm{s}}+\hat{\mathbf{n}}_{{\rm{s}}},
\end{align}
where ${\hat{\mathbf{y}}_{{\rm{s}}}}=[\hat{y}_{{\rm{s}},1},\ldots,\hat{y}_{{\rm{s}},L}]^{\mathsf{T}}\in{\mathbbmss{C}}^{L\times1}$ and ${\hat{\mathbf{n}}_{{\rm{s}}}}=[\hat{n}_{{\rm{s}},1},\ldots,\hat{n}_{{\rm{s}},L}]^{\mathsf{T}}\sim{\mathcal{CN}}({\mathbf{0}},{\sigma}_{\rm{s}}^2\int_{{\mathcal{A}}_{\rm{r}}}\lvert{{v}}_{\rm{s}}({\mathbf{r}})\rvert^2{\rm{d}}{\mathbf{r}}{\mathbf{I}}_L)$. 

The next task is to recover $\beta_{\rm{s}}$ from $\hat{\mathbf{y}}_{{\rm{s}}}$ given ${\mathbf{s}}$. The information-theoretic limits of this sensing task are measured by the sensing mutual information (MI), which quantifies the MI between $\hat{\mathbf{y}}_{{\rm{s}}}$ and $\beta_{\rm{s}}$, conditioned on the ISAC signal ${\mathbf{s}}$ \cite{ouyang2023integrated,sensing_MI,bcrb_2}. To evaluate sensing performance, we use the SR as the performance metric, which is defined as the sensing MI per unit time \cite{ouyang2023integrated}. Since each DFSAC symbol lasts one unit of time, the SR is expressed as follows:
\begin{align}\label{SR_define}
{\mathcal{R}}_{\rm{d},\rm{s}}({{w}}({\mathbf{t}}))=\frac{1}{L}I(\hat{\mathbf{y}}_{{\rm{s}}};\beta_{\rm{s}}|{\mathbf{s}}),
\end{align}
where $I(\hat{\mathbf{y}}_{{\rm{s}}};\beta_{\rm{s}}|{\mathbf{s}})$ denotes the sensing MI. Interpreting \eqref{Sensing_Model_Transformed} as a single-input multiple-output (SIMO) channel, the virtual channel has the effective channel gain $\hat{{v}}_{\rm{s}}\hat{{w}}_{\rm{s}}{\mathbf{s}}$, input $\beta_{\rm{s}}$, noise $\hat{\mathbf{n}}_{{\rm{s}}}$, and output $\hat{\mathbf{y}}_{{\rm{s}}}$. The sensing MI, therefore, corresponds to the channel capacity of this virtual SIMO channel, which can be calculated as follows:
\begin{subequations}\label{MI}
\begin{align}
I(\hat{\mathbf{y}}_{{\rm{s}}};\beta_{\rm{s}}|{\mathbf{s}})
&=\log_2\det\left({\mathbf{I}}_L+\frac{P\alpha_{\rm{s}}\lvert\hat{{v}}_{\rm{s}}\rvert^2\lvert\hat{{w}}_{\rm{s}}\rvert^2}{{\sigma}_{\rm{s}}^2
\int_{{\mathcal{A}}_{\rm{r}}}\lvert{{v}}_{\rm{s}}({\mathbf{r}})\rvert^2{\rm{d}}{\mathbf{r}}}
{\mathbf{s}}{\mathbf{s}}^{\mathsf{H}}\right)\\
&=\log_2\left(1+\frac{P\alpha_{\rm{s}}\lvert\hat{{v}}_{\rm{s}}\rvert^2\lvert\hat{{w}}_{\rm{s}}\rvert^2}{{\sigma}_{\rm{s}}^2
\int_{{\mathcal{A}}_{\rm{r}}}\lvert{{v}}_{\rm{s}}({\mathbf{r}})\rvert^2{\rm{d}}{\mathbf{r}}}
\lVert{\mathbf{s}}\rVert^2\right),
\end{align}
\end{subequations}
where the last equality is based on the Sylvester's determinant identity. Substituting \eqref{MI} into \eqref{SR_define} and applying the fact that $\frac{1}{L}\lVert{\mathbf{s}}\rVert^2=1$, we can write the SR as follows:
\begin{align}\label{SR_define_2}
{\mathcal{R}}_{\rm{d},\rm{s}}({{w}}({\mathbf{t}}))=\frac{1}{L}\log_2\left(1+\frac{P\alpha_{\rm{s}}\lvert\hat{{v}}_{\rm{s}}\rvert^2\lvert\hat{{w}}_{\rm{s}}\rvert^2}{{\sigma}_{\rm{s}}^2
\int_{\mathcal{A}_{\rm{r}}}\lvert{{v}}_{\rm{s}}({\mathbf{r}})\rvert^2{\rm{d}}{\mathbf{r}}}L\right).
\end{align}

Note that the SR depends on the detector ${{v}}_{\rm{s}}({\mathbf{r}})$. The SR-optimal detector can be determined by solving the following optimization problem:
\begin{align}\label{Downlink_SR_Pre_Opt1}
\max_{{{v}}_{\rm{s}}({\mathbf{r}})}\frac{\lvert\hat{{v}}_{\rm{s}}\rvert^2}{\int_{\mathcal{A}_{\rm{r}}}\lvert{{v}}_{\rm{s}}({\mathbf{r}})\rvert^2{\rm{d}}{\mathbf{r}}}\Leftrightarrow
\max_{{{v}}_{\rm{s}}({\mathbf{r}})}\frac{\left\lvert\int_{{\mathcal{A}}_{\rm{r}}}{{v}}_{\rm{s}}^{*}({\mathbf{r}})a_{\rm{r}}(\mathbf{r}){\rm{d}}{\mathbf{r}}\right\rvert^2}{\int_{\mathcal{A}_{\rm{r}}}\lvert{{v}}_{\rm{s}}({\mathbf{r}})\rvert^2{\rm{d}}{\mathbf{r}}}.
\end{align}
The optimal solution to this problem is proportional to $a_{\rm{r}}(\mathbf{r})$, i.e., ${{v}}_{\rm{s}}({\mathbf{r}})\propto a_{\rm{r}}(\mathbf{r})$, which corresponds to the MRC detector. Therefore, we have
\begin{align}\label{Downlink_SR_Pre_Opt2}
\max_{{{v}}_{\rm{s}}({\mathbf{r}})}\frac{\lvert\hat{{v}}_{\rm{s}}\rvert^2}{\int_{\mathcal{A}_{\rm{r}}}\lvert{{v}}_{\rm{s}}({\mathbf{r}})\rvert^2{\rm{d}}{\mathbf{r}}}
=\frac{\lvert\int_{\mathcal{A}_{\rm{r}}}\lvert a_{\rm{r}}(\mathbf{r})\rvert^2{\rm{d}}{\mathbf{r}}\rvert^2}{\int_{\mathcal{A}_{\rm{r}}}\lvert a_{\rm{r}}(\mathbf{r})\rvert^2{\rm{d}}{\mathbf{r}}}=\int_{\mathcal{A}_{\rm{r}}}\lvert a_{\rm{r}}(\mathbf{r})\rvert^2{\rm{d}}{\mathbf{r}}. 
\end{align}
By inserting \eqref{Downlink_SR_Pre_Opt2} into \eqref{SR_define_2} and defining $\overline{\gamma }_{\mathrm{s}}\triangleq\frac{P\alpha_{\mathrm{s}}}{\sigma _{\mathrm{s}}^{2}}$, we have
\begin{align}\label{SR_define_2_step1}
{\mathcal{R}}_{\rm{d},\rm{s}}({{w}}({\mathbf{t}}))=\frac{1}{L}\log_2\left(1+L{\overline{\gamma }_{\mathrm{s}}\lvert\hat{{w}}_{\rm{s}}\rvert^2\int_{\mathcal{A}_{\rm{r}}}\lvert a_{\rm{r}}(\mathbf{r})\rvert^2{\rm{d}}{\mathbf{r}}}\right).
\end{align}

{ Notably, the following theorem reveals the connection between the SR and both the mean-squared error (MSE) and the Bayesian Cram\'{e}r-Rao bound (BCRB) under the considered sensing model.
\vspace{-5pt}
\begin{theorem}\label{the_mse}
Under the considered sensing model, maximizing the achievable SR is equivalent to minimizing the MSE and BCRB in estimating the RCS.
\end{theorem}
\vspace{-5pt}
\begin{IEEEproof}
Please refer to Appendix \ref{proof_mse} for more details.
\end{IEEEproof}}
The results in \eqref{CR_define} and \eqref{SR_define_2_step1} indicate that the performance of both communications and sensing in the downlink ISAC system is closely tied to the source current distribution $w(\mathbf{t})$. However, it is challenging to determine an optimal $w(\mathbf{t})$ that can simultaneously maximize both the CR and the SR. To address this challenge, we propose three continuous beamforming designs to explore the fundamental performance limits of CAPA-based downlink ISAC systems: \romannumeral1) \emph{the C-C design} that maximizes ${\mathcal{R}}_{\rm{d},{\rm{c}}}({{w}}({\mathbf{t}}))$; \romannumeral2) \emph{the S-C design} that maximizes ${\mathcal{R}}_{\rm{d},\rm{s}}({{w}}({\mathbf{t}}))$; \romannumeral3) \emph{the Pareto-optimal design} that characterizes the Pareto boundary of the CR-SR region.
\subsection{Communications-Centric Design}
\subsubsection{Beamforming Design}
In the C-C design, the continuous beamformer is configured to maximize the CR, which is expressed as follows:
\begin{subequations}\label{CC_Beamforming_Design}
\begin{align}
w_{\rm{c}}({\mathbf{t}})&=\argmax\nolimits_{\int_{{\mathcal{A}}_{\rm{t}}}\lvert{w}({\mathbf{t}})\rvert^2{\rm{d}}{\mathbf{t}}=1}{\mathcal{R}}_{\rm{d},{\rm{c}}}({{w}}({\mathbf{t}}))\\
&=\argmax\nolimits_{\int_{{\mathcal{A}}_{\rm{t}}}\lvert{w}({\mathbf{t}})\rvert^2{\rm{d}}{\mathbf{t}}=1}\hat{\gamma}_{\rm{c}}({{w}}({\mathbf{t}})),
\end{align} 
\end{subequations}
where $\hat{\gamma}_{\rm{c}}({{w}}({\mathbf{t}}))\triangleq\lvert\int_{{\mathcal{A}}_{\rm{t}}}h_{\rm{d}}({\mathbf{t}}){{w}}({\mathbf{t}}){\rm{d}}{\mathbf{t}}\rvert^2$. The optimal beamformer that maximizes $\hat{\gamma}_{\rm{c}}({{w}}({\mathbf{t}}))$ is given by
\begin{align}\label{Communications_Centric_Beamforming}
w_{\rm{c}}({\mathbf{t}})={h_{\rm{d}}^{*}({\mathbf{t}})}
{\left({\int_{{\mathcal{A}}_{\rm{t}}}\lvert h_{\rm{d}}({\mathbf{t}})\rvert^2{\rm{d}}{\mathbf{t}}}\right)^{-\frac{1}{2}}}.
\end{align}
This C-C beamformer corresponds to the maximum-ratio transmission (MRT) beamformer, which aligns with the downlink communication channel $h_{\rm{d}}({\mathbf{t}})$.
\subsubsection{Performance of Communications}
By substituting $w({\mathbf{t}})=w_{\rm{c}}({\mathbf{t}})$ into \eqref{CR_define}, the downlink CR under the C-C design can be expressed as follows:
\begin{align}
{\mathcal{R} _{\mathrm{d},\mathrm{c}}^{\mathrm{c}}}=\log _2\left( 1+\overline{\gamma}_{\mathrm{c}}\int_{\mathcal{A} _{\mathrm{t}}}{\left| h_{\rm{d}}({\mathbf{t}}) \right|}^2\mathrm{d}\mathbf{t} \right) .
\end{align}
Theorem \ref{the_do_cc_cr} provides a closed-form expression for ${\mathcal{R} _{\mathrm{d},\mathrm{c}}^{\mathrm{c}}}$.
\vspace{-5pt}
\begin{theorem}\label{the_do_cc_cr}
In the C-C design, the downlink CR is given by
\begin{align}\label{do_cc_cr}
	{\mathcal{R} _{\mathrm{d},{\mathrm{c}}}^{\mathrm{c}}}=\log _2\left( 1+\overline{\gamma}_{\mathrm{c}}g_{\mathrm{d}} \right),
\end{align}
where
\begin{align}\label{g_d}
	g_{\mathrm{d}}=\int_{\mathcal{A} _{\mathrm{t}}}{\left| h_{\rm{d}}({\mathbf{t}}) \right|}^2\mathrm{d}\mathbf{t}=\frac{\eta^2k_0^2}{4\pi }\sum_{x\in \mathcal{X} _\mathrm{d}}\sum_{z\in \mathcal{Z} _\mathrm{d}}\zeta_{\rm{c}}(x,z),
\end{align}
with $\zeta_{\rm{c}}(x,z)\triangleq{\arctan}\Big( \frac{xz}{\Psi_\mathrm{c} \sqrt{\Psi _{\mathrm{c}}^{2}+x^2+z^2}} \Big)$, ${\mathcal{X}}_\mathrm{d}=\{\Phi _{\mathrm{c}},\frac{L_x}{r_{\mathrm{c}}}-\Phi _{\mathrm{c}}\}$, and ${\mathcal{Z}}_\mathrm{d}=\{\frac{L_z}{2r_\mathrm{c}}\pm \Theta_\mathrm{c}\}$.
\end{theorem}
\vspace{-5pt}
\begin{IEEEproof}
Please refer to Appendix \ref{proof_do_cc_cr} for more details.
\end{IEEEproof}

{ To further elucidate the properties of CAPAs in ISAC and gain more physical or geometric insights, we examine the asymptotic performance by considering the infinitely larger CAPAs, i.e., $L_x,L_z\rightarrow \infty$. The following corollary provides the asymptotic CR under the C-C design.
\vspace{-5pt}
\begin{corollary}\label{cor_asym}
As  $L_x,L_z\rightarrow \infty$, the downlink CR achieved by the  C-C design satisfies
\begin{align}
\underset{L_x,L_z\rightarrow \infty}{\lim}\mathcal{R} _{\mathrm{d},\mathrm{c}}^{\mathrm{c}}=\log _2\left( 1+\frac{\overline{\gamma }_{\mathrm{c}}\eta ^2k_{0}^{2}}{2} \right) .	
\end{align}
\end{corollary}
\vspace{-5pt}
\begin{IEEEproof}
This can be obtained based on $\underset{L_x,L_z\rightarrow \infty}{\lim}g_{\mathrm{d}}=\frac{\eta ^2k_{0}^{2}}{4\pi}\times 4\frac{\pi}{2}=\frac{\eta ^2k_{0}^{2}}{2}$.
\end{IEEEproof}
\vspace{-5pt}
\begin{remark}\label{rem_asy}
	The results of Corollary \ref{cor_asym} suggest that the CR will converge to a finite value
	as the aperture size grows, which is intuitively reasonable from the perspective of energy conservation.
\end{remark}}
\vspace{-5pt}
%Next, to further explore the communication performance of CAPA-based ISAC, we investigate the asymptotic CR by considering the CAPAs with infinitely large apertures, i.e., $L_x,L_z\rightarrow\infty$.
%\begin{corollary}\label{cor_do_cc_cr}
%When $L_x,L_z\rightarrow\infty$, the downlink CR under the C-C design satisfies
%\begin{align}
%\lim_{L_x,L_z\rightarrow \infty} {\mathcal{R} _{\mathrm{d},{\mathrm{c}}}^{\mathrm{c}}}=\log _2\left( 1+\overline{\gamma}_{\mathrm{c}}\frac{\pi-\phi_{\mathrm{c}}}{2\pi } \right). 	
%\end{align}
%\end{corollary}
%\vspace{-5pt}
%\begin{IEEEproof}
%In \eqref{g_d}, when $x=\Phi_{\mathrm{c}}$, we have $\frac{xz/\Psi _k}{\sqrt{\Psi _{\mathrm{c}}^{2}+x^2+z^2}}=\frac{z\cot \phi_{\mathrm{c}}}{\sqrt{\sin ^2\theta_{\mathrm{c}} +z^2}}$, and $\lim_{z\rightarrow \infty} \arctan \Big( \frac{z\cot \phi_{\mathrm{c}}}{\sqrt{\sin ^2\theta_{\mathrm{c}} +z^2}} \Big) =\arctan(\cot \phi_{\mathrm{c}})=\frac{\pi}{2}-\phi_{\mathrm{c}} $. When $x=\frac{L_x}{r_{\mathrm{c}}}-\Phi _{\mathrm{c}}$, we have $\lim_{L_x,L_z\rightarrow \infty} \arctan \Big( \frac{xz/\Psi _{\mathrm{c}}}{\sqrt{\Psi _{\mathrm{c}}^{2}+x^2+z^2}} \Big) =\frac{\pi}{2}$. Taken together, we can obtain the results of Corollary~\ref{cor_do_cc_cr}.
%\end{IEEEproof}  

\subsubsection{Performance of Sensing}
By inserting $w(\mathbf{t})=w_\mathrm{c}(\mathbf{t})$ into \eqref{SR_define_2_step1}, the downlink SR can be written as follows:
\begin{equation}
		{\mathcal{R} _{\mathrm{d},{\mathrm{s}}}^{\mathrm{c}}}=\frac{1}{L}\log _2\!\left( \!1+\!L\overline{\gamma }_{\mathrm{s}}\frac{\lvert\int_{\mathcal{A} _{\mathrm{t}}}\!{a_{\mathrm{t}}(\mathbf{t})h_{\mathrm{d}}^{*}(\mathbf{t})}\mathrm{d}\mathbf{t}\rvert^2\!\int_{\mathcal{A} _{\mathrm{r}}}\!{\left| a_{\mathrm{r}}(\mathbf{r}) \right|^2}\mathrm{d}\mathbf{r}}{\int_{\mathcal{A} _{\mathrm{t}}}{\left| h_{\rm{d}}({\mathbf{t}}) \right|}^2\mathrm{d}\mathbf{t}} \right).
\end{equation}
A closed-form expression for ${\mathcal{R} _{\mathrm{d},{\mathrm{s}}}^{\mathrm{c}}}$ is given as follows.
\vspace{-5pt}
\begin{theorem}\label{the_do_cc_sr}
In the C-C design, the downlink SR is given by
\begin{align}
{\mathcal{R} _{\mathrm{d},{\mathrm{s}}}^{\mathrm{c}}}=\frac{1}{L}\log _2\left( 1+L\overline{\gamma }_{\mathrm{s}}g_{\mathrm{d}}^{-1}g_{\mathrm{r}}\lvert\rho _{\mathrm{d}}\rvert^2 \right),
\end{align}
where $g_{\mathrm{r}}$ is expressed as follows:
\begin{align}
g_{\mathrm{r}}=\!\int_{\mathcal{A} _{\mathrm{r}}}\!{\left| a_{\mathrm{r}}(\mathbf{r}) \right|^2}\mathrm{d}\mathbf{r}=\frac{\eta^2k_0^2}{4\pi }\sum\nolimits_{x\in \mathcal{X} _\mathrm{r}}\sum\nolimits_{z\in \mathcal{Z} _\mathrm{r}}\zeta_{\rm{s}}(x,z),\label{g_r}
\end{align}
with $\zeta_{\rm{s}}(x,z)\triangleq{\arctan}\Big( \frac{xz}{\Psi _\mathrm{s}\sqrt{\Psi _{\mathrm{s}}^{2}+x^2+z^2}} \Big)$, $\mathcal{X} _{\mathrm{r}}=\{-\Phi _{\mathrm{s}},\frac{L_x}{r_{\mathrm{s}}}+\Phi _{\mathrm{s}}\}$, and ${\mathcal{Z}}_\mathrm{r}=\{\frac{L_z}{2r_\mathrm{s}}\pm \Theta_\mathrm{s}\}$. Additionally, the term $\rho _{\mathrm{d}}=\int_{\mathcal{A} _{\mathrm{t}}}{a_{\mathrm{t}}(\mathbf{t})h_{\mathrm{d}}^{*}(\mathbf{t})}\mathrm{d}\mathbf{t}$ is calculated as follows:
\begin{equation}\label{rho_d}
	\begin{split}
	\rho _{\mathrm{d}} =&\frac{\pi ^2L_xL_z}{4N^2}\sum\nolimits_{n=1}^N{\sum\nolimits_{n^{\prime}=1}^N{\sqrt{\left( 1-\xi _{n}^{2} \right) (1-\xi _{n^{\prime}}^{2})}}}\\
	&\times \tilde{h}_{\mathrm{c}}^{*}\Big(\frac{\xi_n+1}{2}L_x,\frac{\xi _{n^\prime}}{2}L_z\Big) \tilde{h}_{\mathrm{s}}\Big(\frac{\xi _n+1}{2}L_x,\frac{\xi _{n^\prime}}{2}L_z\Big),
	\end{split}
\end{equation}
where $N$ is a complexity-vs-accuracy tradeoff parameter, 
\begin{align}
\tilde{h}_k(x,z)\triangleq \frac{\mathrm{j}\eta k_0\sqrt{\frac{r_k\Psi _k}{4\pi}}\mathrm{e}^{-\mathrm{j}k_0(x^2+z^2-2r_k\left( \Phi _kx+\Theta _kz \right) +r_{k}^{2})^{\frac{1}{2}}}}{(x^2\!+\!z^2\!-\!2r_k\!\left( \Phi _kx\!+\!\Theta _kz \right) \!+\!r_{k}^{2})^{\frac{3}{4}}}\nonumber
\end{align}
for $k\in\{\rm{c},\rm{s}\}$, and $\xi _n=\cos \left( \frac{ 2n-1  }{2N}\pi \right) $.
\end{theorem}
\vspace{-3pt}
\begin{IEEEproof}
The term $\int_{\mathcal{A} _{\mathrm{r}}}{\left| a_{\mathrm{r}}(\mathbf{r}) \right|^2}\mathrm{d}\mathbf{r}=g_{\mathrm{r}}$ can be computed by following steps similar to those in Appendix \ref{proof_do_cc_cr}. Additionally, the term $\rho _{\mathrm{d}}$ can be expressed as follows:
\begin{equation}
\rho _{\mathrm{d}}=\int_{-{L_z}/{2}}^{{L_z}/{2}}{\int_0^{L_x}{\tilde{h}_{\mathrm{c}}^{*}(x,z)\tilde{h}_{\mathrm{s}}(x,z)\mathrm{d}x\mathrm{d}z}}.    
\end{equation}
This integral can be computed using the \emph{Chebyshev-Gauss} quadrature rule \cite{math}, i.e., $\int_{-1}^1{\frac{f\left( x \right)}{\sqrt{1-x^2}}\mathrm{d}}x\approx \frac{\pi}{n}\sum_{j=1}^n{f\left( x_j \right)}$ with $x_j=\cos\big(\frac{\left( 2j-1 \right) \pi}{2n}\big) $, which yields the results of \eqref{rho_d}.
\end{IEEEproof}

{ We further consider the asymptotic SR for infinitely large arrays. While $\underset{L_x,L_z\rightarrow \infty}{\lim}g_{\mathrm{r}}=\frac{\eta ^2k_{0}^{2}}{2}$ can be easily obtained, determining $\underset{L_x,L_z\rightarrow \infty}{\lim}\rho _{\mathrm{d}}$ is more challenging. Although one might intuitively expect it to vanish, recent work in \cite{correlation} employs the stationary phase method to derive an expression for $\underset{L_x,L_z\rightarrow \infty}{\lim}\rho _{\mathrm{d}}$, demonstrating that it converges to a finite value close but not equal to zero. Due to the complexity of this expression, it is omitted here; however, numerical simulations illustrating the asymptotic SR behavior are presented in Section \ref{section_numerical}. These results suggest that as the aperture size increases, the SR achieved by the C-C design asymptotically converges to a finite value larger than zero.
	}

\subsection{Sensing-Centric Design}
\subsubsection{Beamforming Design}
In the S-C design, the continuous beamformer is designed to maximize the SR, which is expressed as follows:
\begin{align}\label{SC_Beamforming_Design}
w_{\rm{s}}({\mathbf{t}})&=\argmax_{\int_{{\mathcal{A}}_{\rm{t}}}\lvert{w}({\mathbf{t}})\rvert^2{\rm{d}}{\mathbf{t}}=1}{\mathcal{R}}_{\rm{d},{\rm{s}}}({{w}}({\mathbf{t}}))
=\argmax_{\int_{{\mathcal{A}}_{\rm{t}}}\lvert{w}({\mathbf{t}})\rvert^2{\rm{d}}{\mathbf{t}}=1}{\lvert\hat{{w}}_{\rm{s}}\rvert^2}\notag\\
&=\argmax\nolimits_{\int_{{\mathcal{A}}_{\rm{t}}}\lvert{w}({\mathbf{t}})\rvert^2{\rm{d}}{\mathbf{t}}=1}\hat{\gamma}_{\rm{s}}({{w}}({\mathbf{t}})),
\end{align} 
where $\hat{\gamma}_{\rm{s}}({{w}}({\mathbf{t}}))\triangleq\lvert\int_{{\mathcal{A}}_{\rm{t}}}a_{\rm{t}}(\mathbf{t})w({\mathbf{t}}){\rm{d}}{\mathbf{t}}\rvert^2$. The optimal beamformer that maximizes $\hat{\gamma}_{\rm{c}}({{w}}({\mathbf{t}}))$ is given by
\begin{align}\label{Sensing_Centric_Beamforming}
w_{\rm{s}}({\mathbf{t}})=a_{\rm{t}}^{*}(\mathbf{t})
{\left({\int_{{\mathcal{A}}_{\rm{t}}}\lvert a_{\rm{t}}(\mathbf{t})\rvert^2{\rm{d}}{\mathbf{t}}}\right)^{-\frac{1}{2}}}.
\end{align}
This S-C beamformer corresponds to the MRT beamformer aligned with the sensing channel $a_{\rm{t}}(\mathbf{t})$.
\subsubsection{Performance of Sensing}
Inserting $w({\mathbf{t}})=w_{\rm{s}}({\mathbf{t}})$ into \eqref{SR_define_2_step1} gives the SR as follows:
\begin{align}
{\mathcal{R}}_{\rm{d},\rm{s}}^{\rm{s}}\!=\frac{1}{L}\log_2\!\left(1\!+L\overline{\gamma}_{\mathrm{s}}
\int_{{\mathcal{A}}_{\rm{t}}}\!\lvert a_{\rm{t}}(\mathbf{t})\rvert^2{\rm{d}}{\mathbf{t}}\int_{\mathcal{A}_{\rm{r}}}\!\lvert a_{\rm{r}}(\mathbf{r})\rvert^2{\rm{d}}{\mathbf{r}}\right).
\end{align}
A closed-form expression of ${\mathcal{R}}_{\rm{d},\rm{s}}^{\rm{s}}$ is given as follows.
\vspace{-5pt}
\begin{theorem}
The SR achieved by the donwlink S-C design can be expressed as follows:
\begin{align}
{\mathcal{R} _{\mathrm{d},{\mathrm{s}}}^{\mathrm{s}}}=\frac{1}{L}\log _2\left( 1+L\overline{\gamma }_{\mathrm{s}}g_{\mathrm{t}}g_{\mathrm{r}} \right),
\end{align}
where
\begin{align}
	g_{\mathrm{t}}=\int_{\mathcal{A} _{\mathrm{t}}}\!{\left| a_{\mathrm{t}}(\mathbf{t}) \right|^2}\mathrm{d}\mathbf{t}=\frac{\eta^2k_0^2}{4\pi }\sum\nolimits_{x\in \mathcal{X} _\mathrm{t}}\sum\nolimits_{z\in \mathcal{Z} _\mathrm{t}}\zeta_{\rm{s}}(x,z)\label{g_t},
\end{align}
with ${\mathcal{X}}_\mathrm{t}=\{\Phi _{\mathrm{s}},\frac{L_x}{r_{\mathrm{s}}}-\Phi _{\mathrm{s}}\}$ and ${\mathcal{Z}}_\mathrm{t}=\{\frac{L_z}{2r_\mathrm{s}}\pm \Theta_\mathrm{s}\}$.
\end{theorem}
\vspace{-3pt}
\begin{IEEEproof}
Similar to the proof of Theorem~\ref{the_do_cc_cr}.
\end{IEEEproof}

{ Note that the asymptotic analysis of the S-C design is similar to that of the C-C case, which is omitted here for brevity, while the associate simulation results are presented in Section \ref{section_numerical}.}

\subsubsection{Performance of Communications}
When the S-C beamformer $w_{\mathrm{s}}(\mathbf{t})$ is employed, the downlink CR is given by
\begin{align}
	{\mathcal{R} _{\mathrm{d},{\mathrm{c}}}^{\mathrm{s}}}=\log _2\left( 1+\overline{\gamma }_{\mathrm{c}}\frac{\lvert\int_{\mathcal{A} _{\mathrm{t}}}{a_{\mathrm{t}}(\mathbf{t})h_{\mathrm{d}}^{*}(\mathbf{t})}\mathrm{d}\mathbf{t}\rvert^2}{\int_{\mathcal{A} _{\mathrm{t}}}{\lvert a_{\rm{t}}({\mathbf{t}}) \rvert}^2\mathrm{d}\mathbf{t}} \right).
\end{align}	
Based on this, the following theorem is found.
\vspace{-5pt}
\begin{theorem}
The donwlink CR achieved by the S-C design can be calculated as follows:
\begin{align}
{\mathcal{R} _{\mathrm{d},{\mathrm{c}}}^{\mathrm{s}}}=\log _2\left( 1+\overline{\gamma }_{\mathrm{c}}g_{\mathrm{t}}^{-1}\lvert\rho _{\mathrm{d}}\rvert^2 \right),
\end{align}
\end{theorem}
\vspace{-5pt}
\begin{IEEEproof}
Similar to the proof of Theorem~\ref{the_do_cc_sr}.
\end{IEEEproof}

\subsection{Pareto-Optimal Design}
In practical applications, the beamformer $w({\mathbf{t}})$ can be designed to meet various quality-of-service (QoS) requirements, which creates a trade-off between communication and sensing performance. To evaluate this trade-off, we investigate the \emph{Pareto boundary} of the achievable SR-CR region in the downlink CAPA-based ISAC system. The Pareto boundary comprises SR-CR pairs where any improvement in one rate can only be achieved by sacrificing performance in the other \cite{zhang2010cooperative}. This boundary characterizes \emph{the optimal trade-off} between communication and sensing capabilities, and provides valuable insights for performance optimization. 

It is worth noting that the donwlink CR is a monotone increasing function with respect to $\hat{\gamma}_{\rm{c}}({{w}}({\mathbf{t}}))=\lvert\int_{{\mathcal{A}}_{\rm{t}}}h_{\rm{d}}({\mathbf{t}})
{{w}}({\mathbf{t}}){\rm{d}}{\mathbf{t}}\rvert^2$, while the SR is a monotone increasing function with respect to $\hat{\gamma}_{\rm{s}}({{w}}({\mathbf{t}}))=\lvert\int_{{\mathcal{A}}_{\rm{t}}}a_{\rm{t}}(\mathbf{t})w({\mathbf{t}}){\rm{d}}{\mathbf{t}}\rvert^2$. Therefore, the achievable SR-CR region can be determined from the region defined by $\hat{\gamma}_{\rm{c}}({{w}}({\mathbf{t}}))$ and $\hat{\gamma}_{\rm{s}}({{w}}({\mathbf{t}}))$. In particular, any pair located on the Pareto boundary of this region can be identified by solving the following problem:
\begin{subequations}\label{pareto}
\begin{align}
\max_{w({\mathbf{t}}),\gamma}~~&\gamma\\
{\rm{s.t.}}~~&\hat{\gamma}_{\rm{c}}({{w}}({\mathbf{t}}))\geq\epsilon\gamma,~\hat{\gamma}_{\rm{s}}({{w}}({\mathbf{t}}))\geq(1-\epsilon)\gamma,\\
&\int_{{\mathcal{A}}_{\rm{t}}}\lvert{w}({\mathbf{t}})\rvert^2{\rm{d}}{\mathbf{t}}=1,
\end{align}
\end{subequations}
where $\epsilon\in[0,1]$ is a trade-off parameter. The entire Pareto boundary can be traced by solving this problem while varying $\epsilon$ varying from $0$ to $1$. This optimization problem belongs to the class of non-convex integral-based \emph{functional programming}\footnote{A functional is a specific type of function that takes another function as its input and returns a scalar as its output, i.e., a ``function of a function''.}. Solving such a problem is challenging. To address this, we introduce the following lemma.
\vspace{-3pt}
\begin{lemma}[Subspace Approach]\label{Lemma_Subspace}
Given $\epsilon$, the Pareto-optimal beamformer must lie in the signal subspace spanned by $\{h_{\rm{d}}^{*}({\mathbf{t}}),a_{\rm{t}}^{*}(\mathbf{t})\}$, i.e., $w({\mathbf{t}})=ah_{\rm{d}}^{*}({\mathbf{t}})+ba_{\rm{t}}^{*}(\mathbf{t})$ with $a,b\in\mathbbmss{C}$.
\end{lemma}
\vspace{-3pt}
\begin{IEEEproof}
Please refer to Appendix \ref{proof_Lemma_Subspace} for more details.
\end{IEEEproof}
Based on Lemma \ref{Lemma_Subspace}, we define $\{\phi_1(\mathbf{t}),\phi_2(\mathbf{t})\}$ as an orthonormal basis for the signal subspace spanned by $\{h_{\rm{d}}^{*}({\mathbf{t}}),a_{\rm{t}}^{*}(\mathbf{t})\}$. This basis can be obtained by applying the \emph{Gram-Schmidt process} to $\{h_{\rm{d}}^{*}({\mathbf{t}}),a_{\rm{t}}^{*}(\mathbf{t})\}$. It follows that
\begin{align}
\int_{{\mathcal{A}}_{\rm{t}}}\phi_{i}^{*}(\mathbf{t})\phi_{j}(\mathbf{t}){\rm{d}}{\mathbf{t}}=\delta_{i,j},\quad\forall i,j\in\{1,2\},
\end{align}
where $\delta_{i,j}$ denotes the Kronecker delta. By expressing the beamformer and channels in terms of this basis, we have
\begin{subequations}\label{Subspace_Approach_Transform_Basic_Variables}
\begin{align}
w({\mathbf{t}})&=w_1\phi_1(\mathbf{t})+w_2\phi_2(\mathbf{t}),\\
h_{\rm{d}}^{*}({\mathbf{t}})&=h_{\mathrm{d},1}\phi_1(\mathbf{t})+h_{\mathrm{d},2}\phi_2(\mathbf{t}),\\
a_{\rm{t}}^{*}(\mathbf{t})&=a_{\mathrm{t},1}\phi_1(\mathbf{t})+a_{\mathrm{t},2}\phi_2(\mathbf{t}).
\end{align}
\end{subequations}
These expressions lead to
\begin{subequations}
\begin{align}
	&\int_{{\mathcal{A}}_{\rm{t}}}h_{\rm{d}}({\mathbf{t}})
	{{w}}({\mathbf{t}}){\rm{d}}{\mathbf{t}}=h_{\mathrm{d},1}w_1+h_{\mathrm{d},2}w_2={\mathbf{h}}_{\mathrm{d}}^{\mathsf{T}}{\mathbf{w}},\\
	&\int_{{\mathcal{A}}_{\rm{t}}}a_{\rm{t}}(\mathbf{t})w({\mathbf{t}}){\rm{d}}{\mathbf{t}}=a_{\mathrm{t},1}w_1+a_{\mathrm{t},2}w_2
	={\mathbf{a}}_{\mathrm{t}}^{\mathsf{T}}{\mathbf{w}},\\
	&\int_{{\mathcal{A}}_{\rm{t}}}\lvert{w}({\mathbf{t}})\rvert^2{\rm{d}}{\mathbf{t}}=\lVert{\mathbf{w}}\rVert^2=1,
\end{align}
\end{subequations}
where we define the vector representations: ${\mathbf{w}}=[w_1,w_2]^{\mathsf{T}}$, ${\mathbf{h}}_{\mathrm{d}}=[h_{\mathrm{d},1},h_{\mathrm{d},2}]^{\mathsf{T}}$, and ${\mathbf{a}}_{\mathrm{t}}=[a_{\mathrm{t},1},a_{\mathrm{t},2}]^{\mathsf{T}}$. On this basis, problem \eqref{pareto} can be equivalently rewritten as follows:
\begin{align}\label{pareto_problme_2}
\max_{{\mathbf{w}},\gamma}\gamma~{\rm{s.t.}}~\lvert{\mathbf{h}}_{\mathrm{d}}^{\mathsf{T}}{\mathbf{w}}\rvert^2\!\geq\!\epsilon\gamma,
\lvert{\mathbf{a}}_{\mathrm{t}}^{\mathsf{T}}{\mathbf{w}}\rvert^2\!\geq\!(1-\epsilon)\gamma,
\lVert{\mathbf{w}}\rVert^2=1.
\end{align}
Additionally, we have $\left\| \mathbf{h}_{\mathrm{d}} \right\| ^2=\int_{\mathcal{A} _{\mathrm{t}}}{\!}\left| h_{\mathrm{d}}(\mathbf{t}) \right|^2\mathrm{d}\mathbf{t}=g_{\mathrm{d}}$, $\left\| \mathbf{a}_{\mathrm{t}} \right\| ^2=\int_{\mathcal{A} _{\mathrm{t}}}{\!}\left| a_{\mathrm{t}}(\mathbf{t}) \right|^2\mathrm{d}\mathbf{t}=g_{\mathrm{t}}$, and $\mathbf{h}_{\mathrm{d}}^{\mathsf{H}}\mathbf{a}_{\mathrm{t}}=\int_{\mathcal{A} _{\mathrm{t}}}{a_{\mathrm{t}}(\mathbf{t})h_{\mathrm{d}}^{*}(\mathbf{t})}\mathrm{d}\mathbf{t}=\rho _{\mathrm{d}}$.
\vspace{-3pt}
\begin{remark}
By comparing \eqref{pareto} with \eqref{pareto_problme_2}, it is evident that the subspace approach transforms the intractable functional programming problem into a classical optimization problem involving vectors. This demonstrates the effectiveness of the subspace approach in optimizing CAPA-based systems \cite{liu2024capa}.
\end{remark}
\vspace{-5pt}
Although problem \eqref{pareto_problme_2} is more mathematically tractable than problem \eqref{pareto}, it remains non-convex. To solve this, we derive a closed-form solution using the Karush–Kuhn–Tucker (KKT) conditions.
\vspace{-5pt}
\begin{lemma}\label{lem_w_star}
For a given $\epsilon$, the Pareto-optimal solution for problem \eqref{pareto_problme_2}, denoted by $\mathbf{w}_{\epsilon}$, can be expressed as follows:
\begin{align}\label{w_star}
\mathbf{w}_{\epsilon}\!=\!\begin{cases}
	\!\frac{\mathbf{a}_{\mathrm{t}}^*}{\sqrt{g_{\mathrm{t}}}}&		\epsilon\in\! \left[ 0,\frac{\left| \rho _{\mathrm{d}} \right|^2}{\left| \rho _{\mathrm{d}} \right|^2+g_{\mathrm{t}}^{2}} \right]\\
	\!\frac{\upsilon _1}{\tau}\mathbf{h}_{\mathrm{d}}^*+\frac{\upsilon _2\mathrm{e}^{-\mathrm{j}\angle \rho _{\mathrm{d}}}}{\tau}\mathbf{a}_{\mathrm{t}}^*&		\epsilon \in\! \left( \frac{\left| \rho _{\mathrm{d}} \right|^2}{\left| \rho _{\mathrm{d}} \right|^2+g_{\mathrm{t}}^{2}},\frac{g_{\mathrm{d}}^{2}}{g_{\mathrm{d}}^{2}+\left| \rho _{\mathrm{d}} \right|^2} \right)\\
	\!\frac{\mathbf{h}_{\mathrm{d}}^*}{\sqrt{g_{\mathrm{d}}}}&		\epsilon \in\! \left[ \frac{g_{\mathrm{d}}^{2}}{g_{\mathrm{d}}^{2}+\left| \rho _{\mathrm{d}} \right|^2},1 \right]\\
\end{cases}.    
\end{align}
The terms in \eqref{w_star} are given as follows:
\begin{subequations}\label{Appendix_Used_Results}
\begin{align}
&\upsilon_1=\frac{\sqrt{\epsilon}g_{\mathrm{t}}-\sqrt{\left( 1-\epsilon \right)}\left| \rho _{\mathrm{d}} \right|}{\left( 1-\epsilon \right) g_{\mathrm{d}}+\epsilon g_{\mathrm{t}}-2\sqrt{\epsilon \left( 1-\epsilon \right)}\left| \rho _{\mathrm{d}} \right|},\\
&\upsilon _2=\frac{\sqrt{\left( 1-\epsilon \right)}g_{\mathrm{d}}-\sqrt{\epsilon}\left| \rho _{\mathrm{d}} \right|}{\left( 1-\epsilon \right) g_{\mathrm{d}}+\epsilon g_{\mathrm{t}}-2\sqrt{\epsilon \left( 1-\epsilon \right)}\left| \rho _{\mathrm{d}} \right|},
\end{align}
\end{subequations}
and $\tau =\sqrt{\upsilon _{1}^{2}g_{\mathrm{d}}+\upsilon _{2}^{2}g_{\mathrm{t}}+2\upsilon _1\upsilon _2\left| \rho _{\mathrm{d}} \right|}$ is for normalization.
\end{lemma}
\vspace{-3pt}
\begin{IEEEproof}
Please refer to Appendix \ref{proof_lem_w_star} for more details.
\end{IEEEproof}
By substituting $\mathbf{w}=\mathbf{w}_{\epsilon}$ into \eqref{Subspace_Approach_Transform_Basic_Variables} and performing some basic mathematical manipulations, we derive the Pareto-optimal continuous beamformer for problem \eqref{pareto}.
\vspace{-3pt}
\begin{theorem}
For a given $\epsilon$, the Pareto-optimal continuous beamformer, denoted by $w_{\epsilon}( \mathbf{t} )$, is expressed as follows:
\begin{equation}\label{Pareto_Optimal_Continuous_Beamformer}
\resizebox{1\hsize}{!}{$w_{\epsilon}( \mathbf{t} ) =\begin{cases}
	w_{\rm{s}}({\mathbf{t}})&		\epsilon \in \left[ 0,\frac{\left| \rho _{\mathrm{d}} \right|^2}{\left| \rho _{\mathrm{d}} \right|^2+g_{\mathrm{t}}^{2}} \right]\\
	\frac{\upsilon _1h_{\mathrm{d}}^{*}( \mathbf{t} )}{\tau} +\frac{\upsilon _2\mathrm{e}^{-\mathrm{j}\angle \rho _{\mathrm{d}}}a_{\mathrm{t}}^{*}( \mathbf{t} )}{\tau}&		\epsilon \in \left( \frac{\left| \rho _{\mathrm{d}} \right|^2}{\left| \rho _{\mathrm{d}} \right|^2+g_{\mathrm{t}}^{2}},\frac{g_{\mathrm{d}}^{2}}{g_{\mathrm{d}}^{2}+\left| \rho _{\mathrm{d}} \right|^2} \right)\\
	w_{\rm{c}}({\mathbf{t}})&		\epsilon \in \left[ \frac{g_{\mathrm{d}}^{2}}{g_{\mathrm{d}}^{2}+\left| \rho _{\mathrm{d}} \right|^2},1 \right]\\
\end{cases}.$}
\end{equation}
\end{theorem}
\vspace{-5pt}
\vspace{-5pt}
\begin{remark}
From \eqref{Pareto_Optimal_Continuous_Beamformer}, it can be observed that the Pareto-optimal beamformer reduces to the C-C beamformer when $\epsilon \in \left[ \frac{g_{\mathrm{d}}^{2}}{g_{\mathrm{d}}^{2}+\left| \rho _{\mathrm{d}} \right|^2},1 \right]$ and the S-C beamformer when $\epsilon \in \left[ 0,\frac{\left| \rho _{\mathrm{d}} \right|^2}{\left| \rho _{\mathrm{d}} \right|^2+g_{\mathrm{t}}^{2}} \right]$. Furthermore, the Pareto-optimal beamformer is a weighted sum of $\{h_{\rm{d}}^{*}({\mathbf{t}}),a_{\rm{t}}^{*}(\mathbf{t})\}$, with the weighting coefficients determined by the channel gains $\{g_{\mathrm{t}},g_{\mathrm{d}}\}$ and the channel correlation $\rho _{\mathrm{d}}$.
\end{remark}
\vspace{-5pt}
With $w_{\epsilon}( \mathbf{t} )$ determined, the expressions for the SR-CR pairs $(\mathcal{R}_{\rm{d},\rm{s}}^\epsilon,\mathcal{R}_{\rm{d},\rm{c}}^\epsilon)$ on the Pareto boundary can also be derived. However, these expressions are omitted here due to space limitations. The achievable SR-CR region for the downlink CAPA-based ISAC system is characterized as follows:
\begin{equation}\label{do_capa_region}
\mathcal{C}_{\mathrm{d}}=\left\{\left({\mathcal{R}}_{\mathrm{s}},{\mathcal{R}}_{\mathrm{c}}\right)\left|\begin{matrix}{\mathcal{R}}_{\mathrm{s}}\in[0,\mathcal{R}_{\mathrm{d},\mathrm{s}}^{\epsilon}],
{\mathcal{R}}_{\mathrm{c}}\in[0,\mathcal{R}_{\mathrm{d},\mathrm{c}}^{\epsilon}],\\\epsilon\in\left[0,1\right]\end{matrix}\right.\right\}.
\end{equation}

{ Note that problem \eqref{pareto} can alternatively be solved using the Fourier-based approach \cite{opt1,opt2,opt3,capa_isac}, where the spatial-domain CAPA channel is transformed into its wavenumber-domain representation, converting the original continuous problem into a discretized one. However, such discretization introduces inherent approximation loss and scales poorly with CAPA aperture size and carrier frequency, resulting in high computational complexity. Specifically, obtaining the Fourier coefficients for both sensing and communication channels requires calculating $2\times \left( 2\lceil \frac{L_x}{\lambda} \rceil +1 \right)\times \left( 2\lceil \frac{L_z}{\lambda} \rceil +1 \right) $ integrals. In contrast, our subspace-based formulation directly optimizes the continuous source current pattern, which requires calculating only one integral for the channel correlation $\rho_{\rm{d}}$, significantly reducing complexity. Moreover, it provides a closed-form solution that better captures the system’s continuous nature and offers clearer insight into performance limits.
}

\section{Performance of Uplink ISAC}\label{section_uplink}
In this section, we analyze the performance of CAPA-based uplink ISAC systems.
\subsection{SIC-Based Receive Processing}
As mentioned in Section \ref{Section: Performance of Downlink ISAC: Performance Evaluation Metrics: Performance Metric for Communications}, signal variations within the user's antenna aperture ${\mathcal{A}}$ can be ignored. This allows the simplification of \eqref{up_signal} as follows:
\begin{equation}
	\begin{split}
		y_{\ell}({\mathbf{r}})&\approx\sqrt{P_{\rm{s}}}s_{{\rm{s}},\ell}a_{\rm{r}}(\mathbf{r})\beta_{\rm{s}}
		\int_{{\mathcal{A}}_{\rm{t}}}a_{\rm{t}}(\mathbf{t})w({\mathbf{t}}){\rm{d}}{\mathbf{t}}\\
		&+\sqrt{P_{\rm{c}}}s_{{\rm{c}},\ell}\sqrt{\lvert{\mathcal{A}}\rvert} h_{\rm{u}}(\mathbf{r})+n_{\ell}(\mathbf{r}).
	\end{split}
\end{equation}

Upon observing $\{y_{\ell}({\mathbf{r}})\}_{\ell=1}^{L}$, the BS needs to: \romannumeral1) decode the data information in the communication signal, and \romannumeral2) extract the environmental information from the sensing signal. To handle the IFI between the communication and sensing tasks, the BS can use SIC. Two different SIC orders can be employed \cite{ouyang2023revealing}. \romannumeral1) \emph{C-C SIC}: the BS first decodes the sensing signal by treating the communication signal as interference. Once the sensing signal is decoded, it is subtracted from the superposed signal. The remaining signal is then used for communication signal detection. \romannumeral2) \emph{S-C SIC}: the BS first decodes the communication signal by treating the sensing signal as interference. After detecting and subtracting the communication signal, the remainder is used to recover the sensing information. The \emph{C-C SIC} order is more effective in optimizing \emph{communication performance}, while the \emph{S-C SIC} order prioritizes \emph{sensing performance}.
\subsection{Communications-Centric SIC}
We begin with the C-C SIC, where the target response is estimated and removed first, which allows the communication signal to be detected without interference. 
\subsubsection{Performance of Sensing} 
To estimate the target response, the receiver employs a detector $v_{\rm{s}}({\mathbf{r}})$, which yields
\begin{align}
&\int_{{\mathcal{A}}_{\rm{r}}}v_{\rm{s}}^{*}({\mathbf{r}})y_{\ell}({\mathbf{r}}){\rm{d}}{\mathbf{r}}\nonumber\\
&\approx\sqrt{P_{\rm{s}}}s_{{\rm{s}},\ell}
\int_{{\mathcal{A}}_{\rm{r}}}v_{\rm{s}}^{*}({\mathbf{r}})a_{\rm{r}}(\mathbf{r}){\rm{d}}{\mathbf{r}}\beta_{\rm{s}}
\int_{{\mathcal{A}}_{\rm{t}}}a_{\rm{t}}(\mathbf{t})w({\mathbf{t}}){\rm{d}}{\mathbf{t}}\\
&+\sqrt{P_{\rm{c}}}s_{{\rm{c}},\ell}\sqrt{\lvert{\mathcal{A}}\rvert} \int_{{\mathcal{A}}_{\rm{r}}}v_{\rm{s}}^{*}({\mathbf{r}})h_{\rm{u}}(\mathbf{r}){\rm{d}}{\mathbf{r}}
+\int_{{\mathcal{A}}_{\rm{r}}}v_{\rm{s}}^{*}({\mathbf{r}})n_{\ell}(\mathbf{r}){\rm{d}}{\mathbf{r}}.\nonumber
\end{align}
{ We assume that the following aggregate interference-plus-noise term: 
\begin{align}
z_{{\rm{c}},\ell}=\sqrt{P_{\rm{c}}}s_{{\rm{c}},\ell}\sqrt{\lvert{\mathcal{A}}\rvert} \int_{{\mathcal{A}}_{\rm{r}}}v_{\rm{s}}^{*}({\mathbf{r}})h_{\rm{u}}(\mathbf{r}){\rm{d}}{\mathbf{r}}
+\int_{{\mathcal{A}}_{\rm{r}}}v_{\rm{s}}^{*}({\mathbf{r}})n_{\ell}(\mathbf{r}){\rm{d}}{\mathbf{r}} \nonumber
\end{align}
is Gaussian noise\footnote{{ We assume $z_{c,l}$ as Gaussian noise for two reasons. First, the transmitted data symbol $s_{{\rm{c}},\ell}$ is often considered as complex Gaussian random variables with zero mean and unit variance, i.e., $s_{{\rm{c}},\ell}\sim \mathcal{CN}(0,1)$ \cite{data_distribution_1,data_distribution_2}. This assumption is widely adopted as it reflects the capacity-achieving input distribution under power constraints \cite{goldsmith}. Also, modeling the transmit signal as Gaussian is a valid approximation when many modulated symbols are superimposed (e.g., in large-scale antenna arrays) due to the central limit theorem \cite{clt}. The second reason is that the Gaussian distribution maximizes the entropy among all distributions with the same power, and thus minimizes the mutual information between the input and output of the channel \cite{goldsmith,gaussian_noise}. Therefore, modeling the aggregated interference-plus-noise as Gaussian represents the worst-case noise, which provides a conservative and analytically tractable lower bound of the system performance.} }}, with $z_{{\rm{c}},\ell}\sim{\mathcal{CN}}(0,\sigma_{{\rm{c}},\ell}^2)$ and 
\begin{align}
\sigma_{{\rm{c}},\ell}^2={P_{\rm{c}}}{\lvert{\mathcal{A}}\rvert} \left\lvert\int_{{\mathcal{A}}_{\rm{r}}}\!v_{\rm{s}}^{*}({\mathbf{r}})h_{\rm{u}}(\mathbf{r}){\rm{d}}{\mathbf{r}}\right\rvert^2
+\sigma^2\!\int_{{\mathcal{A}}_{\rm{r}}}\!\lvert v_{\rm{s}}({\mathbf{r}})\rvert^2{\rm{d}}{\mathbf{r}}.
\end{align}
As a result, by setting $w({\mathbf{t}})=w_{\rm{s}}({\mathbf{t}})$ (as per \eqref{Sensing_Centric_Beamforming}), the SR can be expressed as follows:
\begin{equation}\label{up_cc_sr}
	\begin{split}
	{\mathcal{R}}^{\rm{c}}_{\rm{u},\rm{s}}(v_{\rm{s}}({\mathbf{r}}))=\frac{1}{L}{\log_2\left(1+\frac{\lvert\int_{{\mathcal{A}}_{\rm{r}}}v_{\rm{s}}^{*}({\mathbf{r}})a_{\rm{r}}(\mathbf{r}){\rm{d}}{\mathbf{r}}\rvert^2}
{\sigma_{{\rm{c}},\ell}^{2}\frac{1}{P_{\rm{s}} \alpha_{\rm{s}}L\int_{{\mathcal{A}}_{\rm{t}}}\lvert a_{\rm{t}}(\mathbf{t})\rvert^2{\rm{d}}{\mathbf{t}}}}\right)},
	\end{split}
\end{equation}
which is monotone increasing with respect to $\frac{1}{\sigma_{{\rm{c}},\ell}^{2}}{\lvert\int_{{\mathcal{A}}_{\rm{r}}}v_{\rm{s}}^{*}({\mathbf{r}})a_{\rm{r}}(\mathbf{r}){\rm{d}}{\mathbf{r}}\rvert^2}$. Therefore, the problem of maximizing the SR can be formulated as follows:
\begin{align}\label{problem_sr}
\max_{v_{\rm{s}}({\mathbf{r}})}~\frac{\lvert\int_{{\mathcal{A}}_{\rm{r}}}v_{\rm{s}}^{*}({\mathbf{r}})a_{\rm{r}}(\mathbf{r}){\rm{d}}{\mathbf{r}}\rvert^2}{\tilde{\gamma}_{\mathrm{c}} \lvert\int_{{\mathcal{A}}_{\rm{r}}}v_{\rm{s}}^{*}({\mathbf{r}})h_{\rm{u}}(\mathbf{r}){\rm{d}}{\mathbf{r}}\rvert^2
+\int_{{\mathcal{A}}_{\rm{r}}}\lvert v_{\rm{s}}({\mathbf{r}})\rvert^2{\rm{d}}{\mathbf{r}}},
\end{align}
where $\tilde{\gamma}_{\mathrm{c}}\triangleq\frac{P_{\mathrm{c}}\left| \mathcal{A} \right|}{\sigma ^2}$. By using the fact that $\int_{{\mathcal{A}}_{\rm{r}}}\lvert v_{\rm{s}}({\mathbf{r}})\rvert^2{\rm{d}}{\mathbf{r}}=\int_{{\mathcal{A}}_{\rm{r}}}\int_{{\mathcal{A}}_{\rm{r}}}v_{\rm{s}}^{*}({\mathbf{r}})\delta(\mathbf{r}-\mathbf{r}')v_{\rm{s}}({\mathbf{r}}'){\rm{d}}{\mathbf{r}}{\rm{d}}{\mathbf{r}}'$, problem \eqref{problem_sr} can be equivalently rewritten as follows:
\begin{align}\label{Rayleigh_quotient_Basic}
\max_{v_{\rm{s}}({\mathbf{r}})}
\frac{\int_{{\mathcal{A}}_{\rm{r}}}\int_{{\mathcal{A}}_{\rm{r}}}v_{\rm{s}}^{*}({\mathbf{r}})a_{\rm{r}}(\mathbf{r})a_{\rm{r}}^{*}(\mathbf{r}')
v_{\rm{s}}({\mathbf{r}}'){\rm{d}}{\mathbf{r}}{\rm{d}}{\mathbf{r}}'}{\int_{{\mathcal{A}}_{\rm{r}}}\int_{{\mathcal{A}}_{\rm{r}}}v_{\rm{s}}^{*}({\mathbf{r}})
\left(\tilde{\gamma}_{\mathrm{c}}
h_{\rm{u}}(\mathbf{r})h_{\rm{u}}^{*}(\mathbf{r}')+\delta(\mathbf{r}-\mathbf{r}')\right)
v_{\rm{s}}({\mathbf{r}}'){\rm{d}}{\mathbf{r}}{\rm{d}}{\mathbf{r}}'}.
\end{align}

Problem \eqref{Rayleigh_quotient_Basic} is the maximization of an integral-form generalized Rayleigh quotient \cite{zhang2017matrix}, but it still belongs to the category of functional programming. To solve this problem, we can apply the subspace approach introduced in Lemma \ref{Lemma_Subspace}. 

It is trivial to show that the optimal $v_{\rm{s}}({\mathbf{r}})$ must lie in the subspace spanned by $\{a_{\rm{r}}(\mathbf{r}),h_{\rm{u}}(\mathbf{r})\}$. We define $\{\psi_1(\mathbf{r}),\psi_2(\mathbf{r})\}$ as an orthonormal basis for this subspace. By expressing the relevant functions in terms of this basis, we have
\begin{subequations}\label{transform}
\begin{align}
v_{\rm{s}}({\mathbf{r}})&=v_1\psi_1(\mathbf{r})+v_2\psi_2(\mathbf{r}),\\ 
h_{\rm{u}}(\mathbf{r})&=h_{\rm{u},1}\psi_1(\mathbf{r})+h_{\rm{u},2}\psi_2(\mathbf{r}),\\
a_{\rm{r}}(\mathbf{r})&=a_{\rm{r},1}\psi_1(\mathbf{r})+a_{\rm{r},2}\psi_2(\mathbf{r}).
\end{align}
\end{subequations}
With these definitions, problem \eqref{Rayleigh_quotient_Basic} can be equivalently transformed into the following optimization problem:
\begin{align}\label{Rayleigh_quotient_Transform}
\max_{\mathbf{v}}~\frac{{\mathbf{v}}^{\mathsf{H}}{\mathbf{a}}_{\rm{r}}{\mathbf{a}}_{\rm{r}}^{\mathsf{H}}{\mathbf{v}}}{{\mathbf{v}}^{\mathsf{H}}
(\tilde{\gamma}_{\mathrm{c}}{\mathbf{h}}_{\rm{u}}{\mathbf{h}}_{\rm{u}}^{\mathsf{H}}
+{\mathbf{I}_2}){\mathbf{v}}},
\end{align}
where ${\mathbf{v}}=[v_1,v_2]^{\mathsf{T}}$, ${\mathbf{h}}_{\rm{u}}=[h_{\rm{u},1},h_{\rm{u},2}]^{\mathsf{T}}$, and ${\mathbf{a}}_{\rm{r}}=[a_{\rm{r},1},a_{\rm{r},2}]^{\mathsf{T}}$. This is now a classical matrix-form maximization of a generalized Rayleigh quotient, whose optimal solution is given by 
\begin{align}
\mathbf{v}_\star\propto(\tilde{\gamma}_{\mathrm{c}}{\mathbf{h}}_{\rm{u}}{\mathbf{h}}_{\rm{u}}^{\mathsf{H}}
+{\mathbf{I}}_2)^{-1}{\mathbf{a}}_{\rm{r}}.
\end{align}
By applying the Woodbury matrix identity to this solution and based on \eqref{transform}, we obtain the optimal detector $v_{\rm{s}}^\star({\mathbf{r}})$ as follows.
\vspace{-5pt}
\begin{lemma}\label{lem_objective}
In the C-C SIC, the optimal detector for the sensing signal is given by
\begin{equation}\label{objective}
v_{\mathrm{s}}^{\star}(\mathbf{r})\propto a_{\mathrm{r}}(\mathbf{r})-\frac{\tilde{\gamma}_{\mathrm{c}}\int_{\mathcal{A} _{\mathrm{r}}}{a_{\mathrm{r}}(\mathbf{r})h_{\mathrm{u}}^{*}(\mathbf{r})}\mathrm{d}\mathbf{r}}{1+\tilde{\gamma}_{\mathrm{c}}\int_{\mathcal{A} _{\mathrm{r}}}{\left| h_{\mathrm{u}}(\mathbf{r}) \right|^2}\mathrm{d}\mathbf{r}}h_{\mathrm{u}}(\mathbf{r}).
\end{equation}
\end{lemma}
%\vspace{-5pt}
%\begin{IEEEproof}
%Please refer to Appendix \ref{proof_lem_objective} for more details.
%\end{IEEEproof}
By inserting the results of Lemma \ref{lem_objective} into \eqref{up_cc_sr} and defining $\tilde{\gamma}_{\mathrm{s}}\triangleq\frac{P_{\mathrm{s}}\alpha _{\mathrm{s}}}{\sigma ^2}$, the SR achieved by the C-C SIC is given by
\begin{equation}\label{up_cc_sr_2}
    \begin{split}
     {\mathcal{R}}^{\rm{c}}_{\rm{u},\rm{s}}&=\frac{1}{L}\log _2\left( 1+L\tilde{\gamma}_{\mathrm{s}}\int_{\mathcal{A} _{\mathrm{t}}}{\lvert a_{\mathrm{t}}(\mathbf{t}) \rvert^2}\mathrm{d}\mathbf{t}\right.\\
&\times\left.\left( \int_{\mathcal{A} _{\mathrm{r}}}\lvert a_{\mathrm{r}}(\mathbf{r}) \rvert^2\mathrm{d}\mathbf{r}-\frac{\tilde{\gamma}_{\mathrm{c}}\lvert \int_{\mathcal{A} _{\mathrm{r}}}{a_{\mathrm{r}}(\mathbf{r})h_{\mathrm{u}}^{*}(\mathbf{r})}\mathrm{d}\mathbf{r} \rvert^2}{1+\tilde{\gamma}_{\mathrm{c}}\int_{\mathcal{A} _{\mathrm{r}}}\lvert h_{\mathrm{u}}(\mathbf{r}) \rvert^2\mathrm{d}\mathbf{r}} \right) \right).
    \end{split}
\end{equation}
A closed-form expression for ${\mathcal{R}}^{\rm{c}}_{\rm{u},\rm{s}}$ is given as follows.
\vspace{-3pt}
\begin{theorem}
The SR under the uplink C-C design can be calculated as follows:
\begin{align}
\mathcal{R} _{\mathrm{u},\mathrm{s}}^{\mathrm{c}}=\frac{1}{L}\log _2\left( 1+L\tilde{\gamma}_{\mathrm{s}}g_{\mathrm{t}}\left( g_{\mathrm{r}}-\frac{\tilde{\gamma}_{\mathrm{c}}\lvert \rho _{\mathrm{u}} \rvert^2}{1+\tilde{\gamma}_{\mathrm{c}}g_{\mathrm{u}}} \right) \right),     
\end{align}
where
\begin{align}
g_{\mathrm{u}}=\int_{\mathcal{A} _{\mathrm{r}}}\lvert h_{\mathrm{u}}(\mathbf{r}) \rvert^2\mathrm{d}\mathbf{r}=\frac{\eta^2k_0^2}{4\pi }\sum_{x\in \mathcal{X} _\mathrm{u}}\sum_{z\in \mathcal{Z} _\mathrm{u}}\zeta_{\rm{c}}(x,z)\label{g_u},
\end{align}
with $\mathcal{X} _{\mathrm{u}}=\{-\Phi _{\mathrm{c}},\frac{L_x}{r_{\mathrm{c}}}+\Phi _{\mathrm{c}}\}$ and ${\mathcal{Z}}_\mathrm{u}=\{\frac{L_z}{2r_\mathrm{c}}\pm \Theta_\mathrm{c}\}$. Additionally, $\rho _{\mathrm{u}}=\int_{\mathcal{A} _{\mathrm{r}}}{a_{\mathrm{r}}(\mathbf{r})h_{\mathrm{u}}^{*}(\mathbf{r})}\mathrm{d}\mathbf{r}$ is given by
\begin{equation}
	\begin{split}
	\rho _{\mathrm{u}} &=\frac{\pi ^2L_xL_z}{4N^2}\sum\nolimits_{n=1}^N{\sum\nolimits_{n^{\prime}=1}^N{\sqrt{\left( 1-\xi _{n}^{2} \right) (1-\xi _{n^{\prime}}^{2})}}}\\
	&\times \tilde{h}_{\mathrm{c}}^{*}\Big(\frac{\xi_n-1}{2}L_x,\frac{\xi _{n^\prime}}{2}L_z\Big) \tilde{h}_{\mathrm{s}}\Big(\frac{\xi _n-1}{2}L_x,\frac{\xi _{n^\prime}}{2}L_z\Big).
	\end{split}
\end{equation}
\end{theorem}
\vspace{-5pt}
\begin{IEEEproof}
Similar to the proof of Theorem~\ref{the_do_cc_sr}.
\end{IEEEproof}
\subsubsection{Performance of Communications}
After removing the sensing signal, the remaining communication signal $s_{{\rm{c}},\ell}$ can be detected from $\sqrt{P_{\rm{c}}}s_{{\rm{c}},\ell}\sqrt{\lvert{\mathcal{A}}\rvert} h_{\rm{u}}(\mathbf{r})+n_{\ell}(\mathbf{r})$ without interference. Therefore, the uplink CR is given by
\begin{align}
	{\mathcal{R} _{\mathrm{u},{\mathrm{c}}}^{\mathrm{c}}}=\log _2\left( 1+\tilde{\gamma}_{\mathrm{c}}g_{\mathrm{u}} \right). 
\end{align}

\subsection{Sensing-Centric SIC}
In the context of S-C SIC, the communication signal is detected first by treating the sensing echo signal as interference. Once the communication signal is detected, it is subtracted from the superposed signal, allowing the receiver to use the remaining part for sensing. Since we are considering the S-C design, the beamformer $w({\mathbf{t}})$ should be designed as $w_{\rm{s}}({\mathbf{t}})$ (as per \eqref{Sensing_Centric_Beamforming}) to maximize the SR.
\subsubsection{Performance of Communications}
To recover the data information from the superposed signal, the receiver exploits a detector $v_{\rm{c}}({\mathbf{r}})$ along with an ML decoder, which yields
\begin{equation}\label{Sensing_Centric}
\begin{split}
&\int_{{\mathcal{A}}_{\rm{r}}}v_{\rm{c}}^{*}({\mathbf{r}})y_{\ell}({\mathbf{r}}){\rm{d}}{\mathbf{r}}\approx\sqrt{P_{\rm{c}}}s_{{\rm{c}},\ell}\sqrt{\lvert{\mathcal{A}}\rvert} \int_{{\mathcal{A}}_{\rm{r}}}v_{\rm{c}}^{*}({\mathbf{r}})h_{\rm{u}}(\mathbf{r}){\rm{d}}{\mathbf{r}}\\
&+\sqrt{P_{\rm{s}}}s_{{\rm{s}},\ell}
\int_{{\mathcal{A}}_{\rm{r}}}v_{\rm{c}}^{*}({\mathbf{r}})a_{\rm{r}}(\mathbf{r}){\rm{d}}{\mathbf{r}}\beta_{\rm{s}}
\sqrt{\int_{\mathcal{A} _{\mathrm{t}}}\lvert a_{\mathrm{t}}(\mathbf{t}) \rvert^2\mathrm{d}\mathbf{t}}\\
&+\int_{{\mathcal{A}}_{\rm{r}}}v_{\rm{c}}^{*}({\mathbf{r}})n_{\ell}(\mathbf{r}){\rm{d}}{\mathbf{r}}.
\end{split}
\end{equation}
Without loss of generality, we focus on a single slot and assume $\lvert s_{{\rm{s}},\ell}\rvert^2=1$ for $\ell=1,\ldots,L$. The interference-plus-noise term $z_{{\rm{s}},\ell}=\sqrt{P_{\rm{s}}}s_{{\rm{s}},\ell}
\int_{{\mathcal{A}}_{\rm{r}}}v_{\rm{c}}^{*}({\mathbf{r}})a_{\rm{r}}(\mathbf{r}){\rm{d}}{\mathbf{r}}\beta_{\rm{s}}
\sqrt{\int_{\mathcal{A} _{\mathrm{t}}}\lvert a_{\mathrm{t}}(\mathbf{t}) \rvert^2\mathrm{d}\mathbf{t}}
+\int_{{\mathcal{A}}_{\rm{r}}}v_{\rm{c}}^{*}({\mathbf{r}})n_{\ell}(\mathbf{r}){\rm{d}}{\mathbf{r}}$ can be regarded as Gaussian noise, with $z_{{\rm{s}},\ell}\sim{\mathcal{CN}}(0,\sigma_{{\rm{s}},\ell}^2)$ and
\begin{equation}
    \begin{split}
      \sigma_{{\rm{s}},\ell}^2&={P_{\rm{s}}}\alpha_{\rm{s}} \left\lvert\int_{{\mathcal{A}}_{\rm{r}}}v_{\rm{c}}^{*}({\mathbf{r}})a_{\rm{r}}(\mathbf{r}){\rm{d}}{\mathbf{r}}\right\rvert^2
\int_{\mathcal{A} _{\mathrm{t}}}{\!}\left| a_{\mathrm{t}}(\mathbf{t}) \right|^2\mathrm{d}\mathbf{t}\\
&+\sigma^2\int_{{\mathcal{A}}_{\rm{r}}}\lvert v_{\rm{c}}({\mathbf{r}})\rvert^2{\rm{d}}{\mathbf{r}}.  
    \end{split}
\end{equation}
Therefore, the CR under the S-C SIC design is given by
\begin{equation}
\mathcal{R} _{\mathrm{u},\mathrm{c}}^{\mathrm{s}}(v_{\mathrm{c}}(\mathbf{r}))=\log _2\left( 1+\frac{P_{\mathrm{c}}\lvert \mathcal{A} \rvert\lvert \int_{\mathcal{A} _{\mathrm{r}}}{v_{\mathrm{c}}^{*}}(\mathbf{r})h_{\mathrm{u}}(\mathbf{r})\mathrm{d}\mathbf{r}  \rvert^2}{\sigma _{\mathrm{s},\ell}^{2}}  \right). 
\end{equation}

The problem of maximizing the CR can be formulated as follows:
\begin{align}
\max_{v_{\mathrm{c}}(\mathbf{r})}~ \mathcal{R} _{\mathrm{u},\mathrm{c}}^{\mathrm{s}}(v_{\mathrm{c}}(\mathbf{r}))\Leftrightarrow
\max_{v_{\mathrm{c}}(\mathbf{r})}~ \frac{P_{\mathrm{c}}\lvert \mathcal{A} \rvert\lvert \int_{\mathcal{A} _{\mathrm{r}}}{v_{\mathrm{c}}^{*}}(\mathbf{r})h_{\mathrm{u}}(\mathbf{r})\mathrm{d}\mathbf{r}  \rvert^2}{\sigma _{\mathrm{s},\ell}^{2}}.
\end{align}
This problem can be solved using a subspace approach similar to that used for problem \eqref{problem_sr}. 
\vspace{-5pt}
\begin{lemma}\label{SC_optimal_detector}
The optimal detector $v_{\mathrm{c}}^{\star}(\mathbf{r})$ is given by
\begin{align}
\!\!v_{\mathrm{c}}^{\star}(\mathbf{r})\!\propto\!h_{\mathrm{u}}(\mathbf{r})\!-\!\frac{\tilde{\gamma}_{\mathrm{s}}\int_{\mathcal{A} _{\mathrm{t}}}\!{\left| a_{\mathrm{t}}(\mathbf{t}) \right|^2}\mathrm{d}\mathbf{t}\!\int_{\mathcal{A} _{\mathrm{r}}}\!{a_{\mathrm{r}}^{*}(\mathbf{r})h_{\mathrm{u}}(\mathbf{r})}\mathrm{d}\mathbf{r}}{1\!+\!\tilde{\gamma}_{\mathrm{s}}\int_{\mathcal{A} _{\mathrm{t}}}\!{\left| a_{\mathrm{t}}(\mathbf{t}) \right|^2}\mathrm{d}\mathbf{t}\!\int_{\mathcal{A} _{\mathrm{r}}}\!{\left| a_{\mathrm{r}}(\mathbf{r}) \right|^2}\mathrm{d}\mathbf{r}}a_{\mathrm{r}}(\mathbf{r}).
\end{align}
\end{lemma}
Consequently, the uplink CR achieved by the S-C SIC design can be expressed as follows:
\begin{equation}
    \begin{split}
        \mathcal{R} _{\mathrm{u},\mathrm{c}}^{\mathrm{s}}&=\log _2\left( 1+\tilde{\gamma}_{\mathrm{c}}\left( \int_{\mathcal{A} _{\mathrm{r}}}\lvert h_{\mathrm{u}}(\mathbf{r}) \rvert^2\mathrm{d}\mathbf{r}\right.\right.\\
        &\left.\left.-\frac{\tilde{\gamma}_{\mathrm{s}}\int_{\mathcal{A} _{\mathrm{t}}}\lvert a_{\mathrm{t}}(\mathbf{t}) \rvert^2\mathrm{d}\mathbf{t}\lvert \int_{\mathcal{A} _{\mathrm{r}}}{a_{\mathrm{r}}(\mathbf{r})h_{\mathrm{u}}^{*}(\mathbf{r})}\mathrm{d}\mathbf{r} \rvert^2}{1+\tilde{\gamma}_{\mathrm{s}}\int_{\mathcal{A} _{\mathrm{t}}}\lvert a_{\mathrm{t}}(\mathbf{t}) \rvert^2\mathrm{d}\mathbf{t}\int_{\mathcal{A} _{\mathrm{r}}}\lvert a_{\mathrm{r}}(\mathbf{r}) \rvert^2\mathrm{d}\mathbf{r}} \right) \right). 
    \end{split}
\end{equation}
A closed-form expression for $\mathcal{R} _{\mathrm{u},\mathrm{c}}^{\mathrm{s}}$ is given as follows.
\vspace{-5pt}
\begin{theorem}
The uplink CR under the S-C design can be expressed in closed form as follows:
\begin{equation}
\mathcal{R} _{\mathrm{u},\mathrm{c}}^{\mathrm{s}}=\log _2\left( 1+\tilde{\gamma}_{\mathrm{c}}\left( g_{\mathrm{u}}-\frac{\tilde{\gamma}_{\mathrm{s}}g_{\mathrm{t}}\left| \rho _{\mathrm{u}} \right|^2}{1+\tilde{\gamma}_{\mathrm{s}}g_{\mathrm{t}}g_{\mathrm{r}}} \right) \right) .
\end{equation}
\end{theorem}
\vspace{-5pt}
\begin{IEEEproof}
Similar to the proof of Theorem~\ref{the_do_cc_sr}.
\end{IEEEproof}
\subsubsection{Performance of Sensing}
After removing the decoded communication signal from the observation, the remaining signal can be used for sensing without interference. In this case, the achievable SR is expressed as follows:
\begin{equation}
{\mathcal{R} _{\mathrm{u},{\mathrm{s}}}^{\mathrm{s}}}=\frac{1}{L}\log _2\left( 1+L\tilde{\gamma }_{\mathrm{s}}g_{\mathrm{t}}g_{\mathrm{r}} \right).    
\end{equation}

\subsection{Rate-Region Characterization}
{ The uplink SR-CR region can be characterized using a time-sharing strategy \cite{goldsmith}. Specifically, the S-C SIC is applied with probability $\varsigma$, while the C-C SIC is applied with probability $1-\varsigma$. For a given $\varsigma$, the achievable SR-CR pair $(\mathcal{R}_{\rm{u},\rm{s}}^\varsigma,\mathcal{R}_{\rm{u},\rm{c}}^\varsigma)$ satisfies
\begin{subequations}
\begin{align}
&\mathcal{R}_{\rm{u},\rm{s}}^\varsigma=\varsigma\mathcal{R}_{\rm{u},\rm{s}}^{\rm{s}}+(1-\varsigma)\mathcal{R}_{\rm{u},\rm{s}}^{\rm{c}},\\
&\mathcal{R}_{\rm{u},\rm{c}}^\varsigma=\varsigma\mathcal{R}_{\rm{u},\rm{c}}^{\rm{s}}+(1-\varsigma)\mathcal{R}_{\rm{u},\rm{c}}^{\rm{c}}.
\end{align} 
\end{subequations}
Consequently, the achievable SR-CR region for the uplink CAPA-based ISAC system can be characterized as follows:
\begin{equation}
\mathcal{C}_{\mathrm{u}}=\left\{\left({\mathcal{R}}_{\mathrm{s}},{\mathcal{R}}_{\mathrm{c}}\right)\left|\begin{matrix}{\mathcal{R}}_{\mathrm{s}}\in[0,\mathcal{R}_{\mathrm{u},\mathrm{s}}^{\varsigma}],
{\mathcal{R}}_{\mathrm{c}}\in[0,\mathcal{R}_{\mathrm{u},\mathrm{c}}^{\varsigma}],\\\varsigma\in\left[0,1\right]\end{matrix}\right.\right\}.
\end{equation}}

\vspace{-15pt}
\section{Numerical Results}\label{section_numerical}
\vspace{3pt}
Numerical results are presented to evaluate the performance of both sensing and communication in CAPA-based downlink and uplink ISAC systems. Unless specified otherwise, the simulation parameters are set as follows: $\lambda=0.125$ m, $L_x=L_z=0.5$ m ($\lvert{\mathcal{A}}_{\rm{t}}\rvert=\lvert{\mathcal{A}}_{\rm{r}}\rvert=0.25$ m$^2$), $\lvert{\mathcal{A}}\rvert=\frac{\lambda^2}{4\pi}$, $\frac{P}{\sigma_{\rm{c}}^2}=\frac{P}{\sigma_{\rm{s}}^2}=\frac{P_{\rm{c}}}{\sigma^2}=\frac{P_{\rm{s}}}{\sigma^2}=10$ dB, $\eta=120\pi~\Omega$, $L=8$, $\alpha_{\rm{s}}=1$, and $T=50$. The locations of the target and CU are given by $(r_{\rm{s}},\theta_{\rm{s}},\phi_{\rm{s}})=(10~\rm{m}, \frac{\pi}{4},\frac{\pi}{4})$ and $(r_{\rm{c}},\theta_{\rm{c}},\phi_{\rm{c}})=(20~\rm{m}, \frac{\pi}{3},\frac{\pi}{3})$, respectively.

\begin{figure}[!t]
	\centering
	\subfigbottomskip=0pt
	\subfigcapskip=-1pt
	\setlength{\abovecaptionskip}{4pt}
	\subfigure[CR versus transmit SNR.]
	{
		\includegraphics[height=0.27\textwidth]{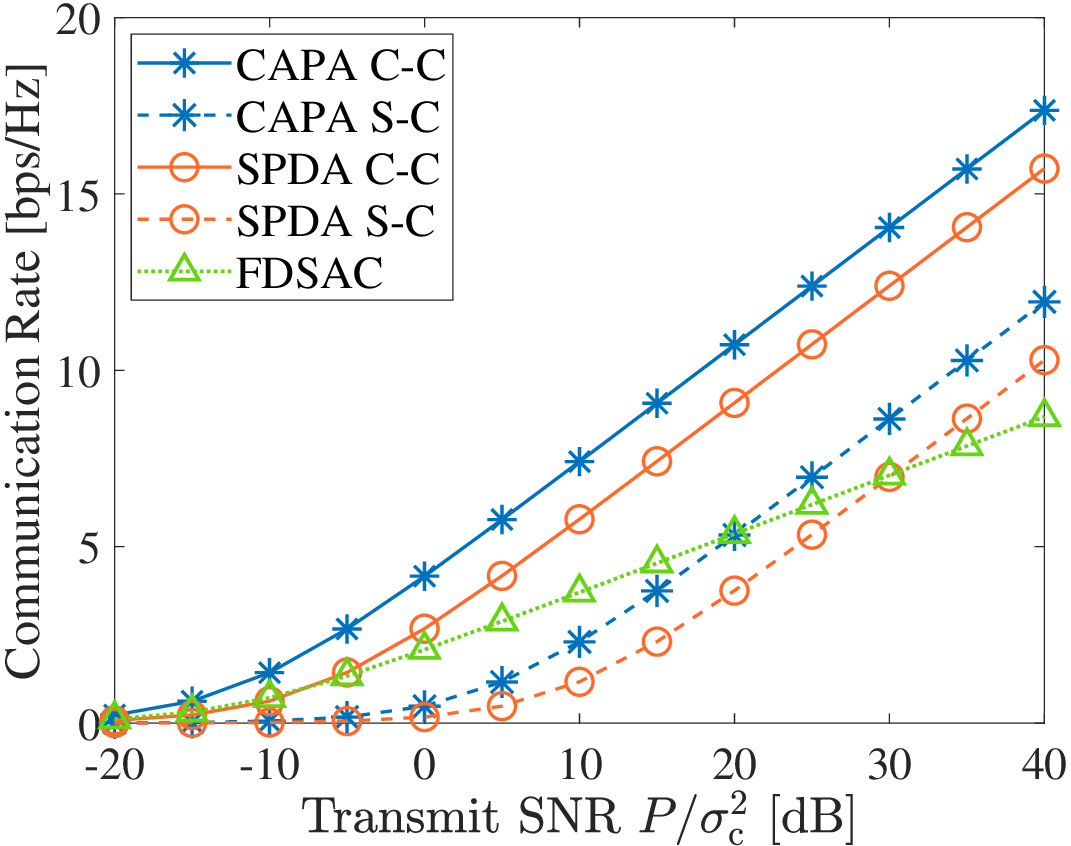}
		\label{do_cr_snr}	
	}
	\subfigure[SR versus transmit SNR.]
	{
		\includegraphics[height=0.27\textwidth]{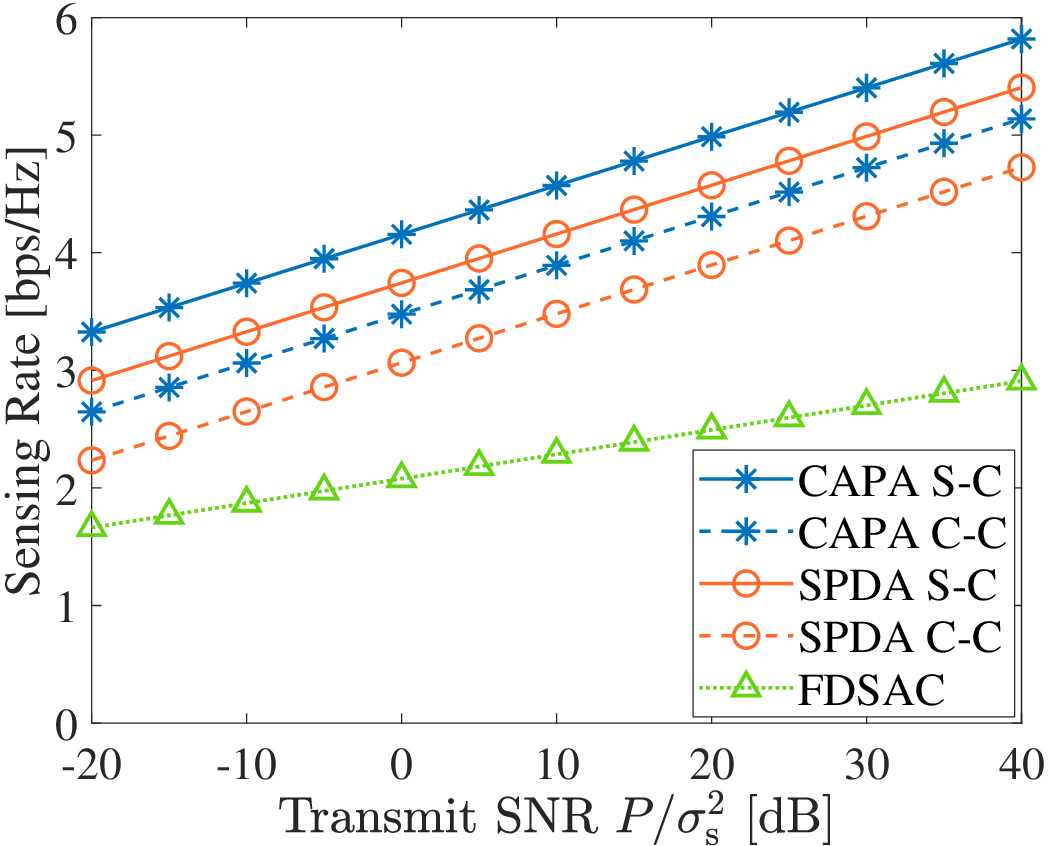}
		\label{do_sr_snr}	
	}
	\caption{Downlink performance versus transmit SNR.}
	\label{do_snr}
	\vspace{-5pt}
\end{figure}

{ For performance comparison, we consider two baseline schemes as follows.}
{ \begin{itemize}
    \item \textbf{SPDA-based ISAC:} In the SPDA scheme, both the transmit and receive continuous apertures are partitioned into $\lfloor \frac{L_x}{d} \rfloor \times\lfloor \frac{L_z}{d} \rfloor $ spatially discrete elements, each with a size of $A_{\rm{s}}=\frac{\lambda^2}{4\pi}$. Here, $d=\frac{\lambda}{2}$ is the inter-element spacing. By denoting $d_{\mathrm{s}}=\frac{\sqrt{A_{\mathrm{s}}}}{2}$, the apertures of the transmit and receive arrays are given by
    \begin{subequations}
     \begin{align}
    \mathcal{A}_{\rm{t}} =&\Big\{ \Big(\frac{2n_x-1}{2}d+\ell _x,0,\frac{2n_z-1}{2} d-\frac{L_z}{2}+\ell _z\Big)\nonumber\\
    &\left.\left| \ell _x,\ell _z\in \left[ -d_{\mathrm{s}},d_{\mathrm{s}} \right],n_x\in \mathcal{N} _x,n_z\in \mathcal{N} _z \right. \right\},\\
    \mathcal{A}_{\rm{r}} =&\Big\{ \Big(\frac{2n_x\!-\!1}{2} d-L_x+\ell _x,0,\frac{2n_z\!-\!1}{2} d-\!\frac{L_z}{2}\!+\ell _z\Big)\nonumber\\
    &\left.\left| \ell _x,\ell _z\in \left[ -d_{\mathrm{s}},d_{\mathrm{s}} \right],n_x\in \mathcal{N} _x,n_z\in \mathcal{N} _z \right. \right\},
    \end{align}   
    \end{subequations}
    where $\mathcal{N} _x\triangleq \left\{ 1,...,\lfloor \frac{L_x}{d} \rfloor \right\} $ and $\mathcal{N} _z\triangleq \left\{ 1,...,\lfloor \frac{L_z}{d} \rfloor \right\} $. We assume that the SPDA is based on a fully digital architecture, where each antenna element is connected with a dedicated RF chain. On the other hand, the CAPA considered in our work only requires a single RF chain, as the number of spatially multiplexed signals is one.
    \item \textbf{CAPA-based FDSAC:} In downlink FDSAC, both the total bandwidth and power are divided between sensing and communications. In particular, a fraction $\kappa=0.5$ of the bandwidth and $\iota =0.5$ of the power are allocated to sensing, while the remaining fractions are used for communications. The SR and CR for downlink FDSAC are given by $\mathcal{R} _{\mathrm{d},\mathrm{s}}^{\mathrm{f}}=\frac{\kappa}{L}\log _2( 1+\frac{\iota}{\kappa}L\overline{\gamma }_{\mathrm{s}}g_{\mathrm{t}}g_{\mathrm{r}} ) $ and $\mathcal{R} _{\mathrm{d},\mathrm{c}}^{\mathrm{f}}=( 1-\kappa ) \log _2( 1+\frac{1-\iota}{1-\kappa}\overline{\gamma }_{\mathrm{c}}g_{\mathrm{d}} ) $, respectively. The achievable SR-CR region is characterized by
    \begin{align}\label{do_fd_region}
    \mathcal{C} _{\mathrm{d}}^{\mathrm{f}}=\left\{ \!( \mathcal{R} _{\mathrm{s}},\mathcal{R} _{\mathrm{c}} )\! \left| \begin{matrix}
	\mathcal{R} _{\mathrm{s}}\in [ 0,\mathcal{R} _{\mathrm{d},\mathrm{s}}^{\mathrm{f}} ] ,\mathcal{R} _{\mathrm{c}}\in [ 0,\mathcal{R} _{\mathrm{d},\mathrm{c}}^{\mathrm{f}} ] ,\\
	\kappa \in [ 0,1 ] ,\iota \in [ 0,1 ]
\end{matrix} \right.\!\!\right\}.
    \end{align}
In uplink FDSAC, a fraction $\kappa=0.5$ of the total bandwidth is allocated to sensing, while the remaining portion is used for communications. The SR and CR for uplink are given by $\mathcal{R} _{\mathrm{u},\mathrm{s}}^{\mathrm{f}}=\frac{\kappa}{L}\log _2( 1+\frac{L\tilde{\gamma}_{\mathrm{s}}g_{\mathrm{t}}g_{\mathrm{r}}}{\kappa} ) $ and $\mathcal{R} _{\mathrm{u},\mathrm{c}}^{\mathrm{f}}=( 1-\kappa) \log _2( 1+\frac{\tilde{\gamma}_{\mathrm{c}}g_{\mathrm{u}}}{1-\kappa} ) $, respectively. The achievable uplink SR-CR region is characterized by
\begin{equation}
\mathcal{C} _{\mathrm{u}}^{\mathrm{f}}=\left\{ \!( \mathcal{R} _{\mathrm{s}},\mathcal{R} _{\mathrm{c}} )\! \left| \begin{matrix}
	\mathcal{R} _{\mathrm{s}}\in [ 0,\mathcal{R} _{\mathrm{u},\mathrm{s}}^{\mathrm{f}} ] ,\mathcal{R} _{\mathrm{c}}\in [ 0,\mathcal{R} _{\mathrm{u},\mathrm{c}}^{\mathrm{f}} ] ,\\
	\kappa \in [ 0,1 ]
\end{matrix} \right.\!\! \right\}.
\end{equation}
\end{itemize}}

\begin{figure}[!t]
	\centering
	\subfigbottomskip=0pt
	\subfigcapskip=-1pt
	\setlength{\abovecaptionskip}{4pt}
	\subfigure[CR versus aperture size.]
	{
		\includegraphics[height=0.27\textwidth]{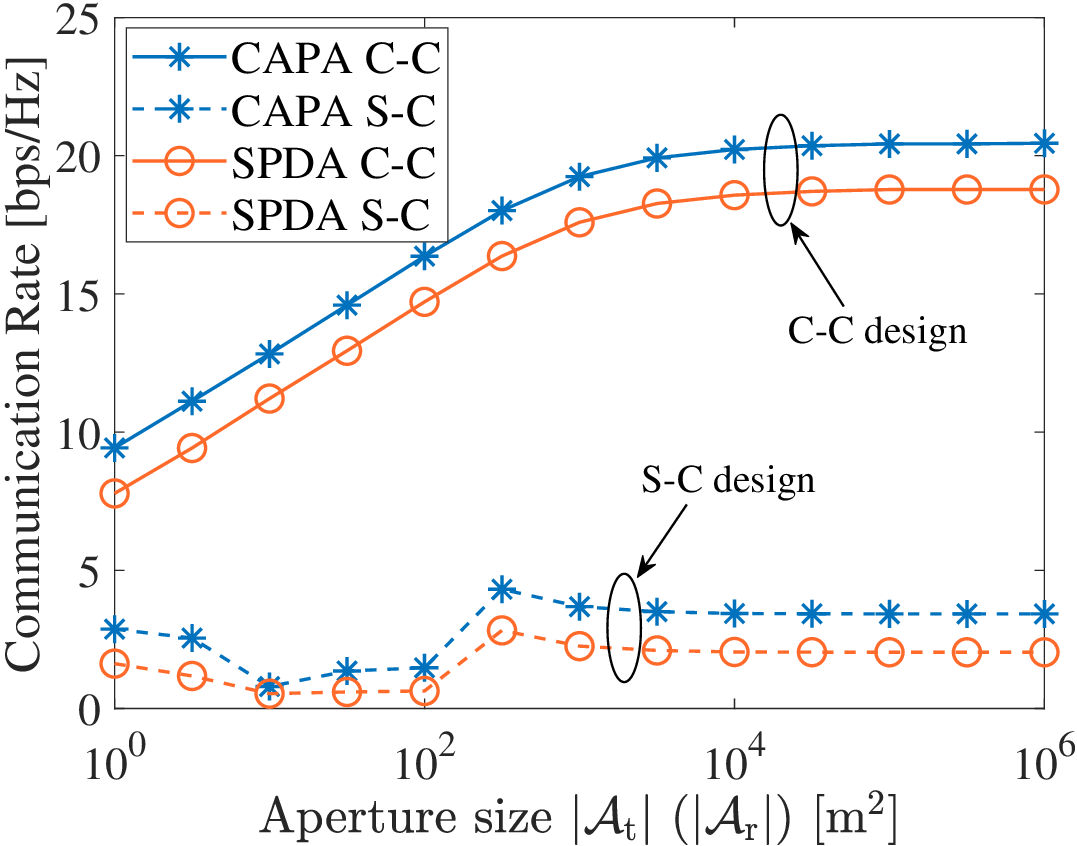}
		\label{do_cr_size}	
	}
	\subfigure[SR versus aperture size.]
	{
		\includegraphics[height=0.27\textwidth]{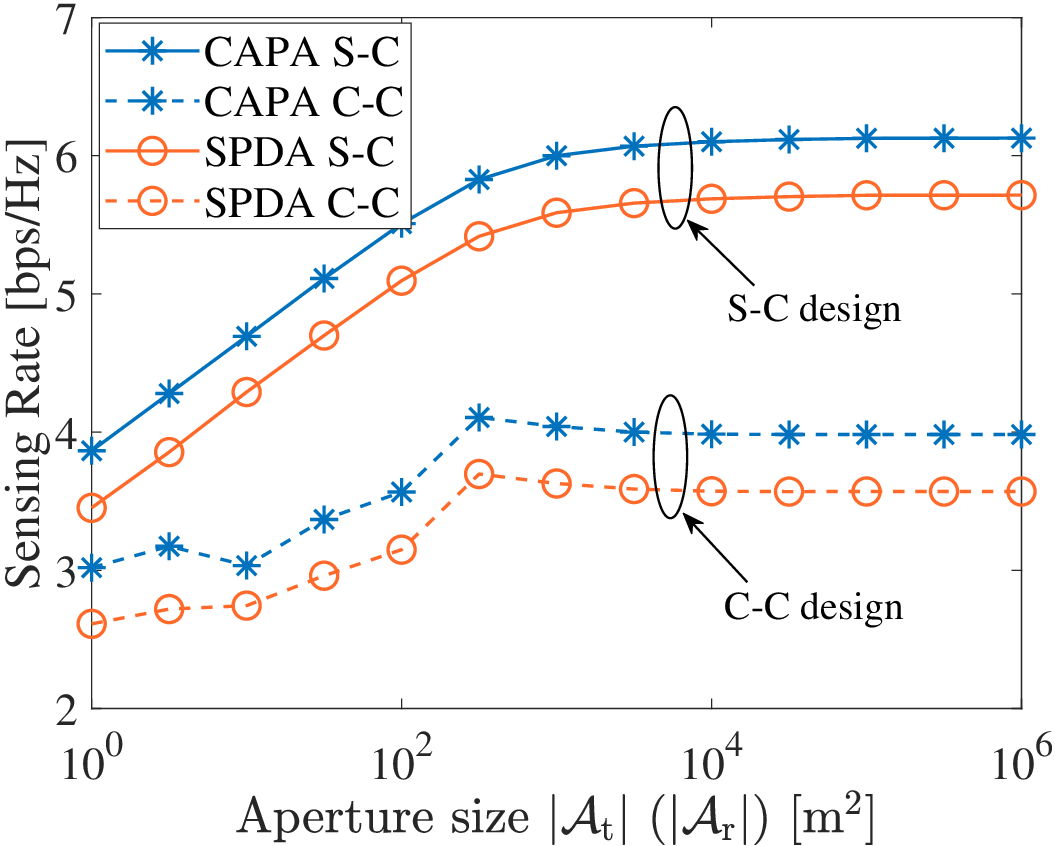}
		\label{do_sr_size}	
	}
	\caption{{ Downlink performance versus aperture size.}}
	\label{do_size}\vspace{-5pt}
\end{figure}

\subsection{Downlink Performance}
{\figurename} {\ref{do_cr_snr}} and {\figurename} {\ref{do_sr_snr}} present the downlink CR and SR as a function of the transmit SNR, respectively. The results indicate that CAPA outperforms the traditional SPDA scheme in both communication and sensing performance. This improvement is attributed to CAPA's efficient utilization of available spatial resources. Additionally, we observe that the C-C design provides superior communication performance, while the S-C design achieves better sensing performance, as expected. In the high-SNR regime, the CR and SR curves for CAPA-based C-C and S-C designs exhibit steeper slopes compared to those of FDSAC. These findings highlight the performance advantages of ISAC over FDSAC, particularly at higher SNR levels.

{\figurename} {\ref{do_cr_size}} and {\figurename} {\ref{do_sr_size}} illustrate the downlink CR and SR achieved by ISAC under both CAPA and SPDA schemes, which are plotted against the BS's aperture size. As can be seen from these two graphs, both the C-C CR and the S-C SR increase monotonically with the aperture size. This is because these designs focus energy in optimal directions for communications and sensing, respectively, which results in enhanced array gain and performance as the aperture size grows. In contrast, the C-C SR and S-C CR do not exhibit a strictly monotonic trend. Since the beamformers in these cases are not perfectly aligned with the respective sensing or communication channels, variations in the correlation term $\rho _{\mathrm{d}}=\int_{\mathcal{A}_{\mathrm{t}}}{a_{\mathrm{t}}(\mathbf{t})h_{\mathrm{d}}^{*}(\mathbf{t})}\mathrm{d}\mathbf{t}$ cause fluctuations in the array gain and performance rates. { Moreover, as the aperture size grows, both the CRs ans SRs eventually converge to finite values larger than zero, adhering to the principle of energy conservation, which is consistent with the statements in Remark~\ref{rem_asy}.}

\begin{figure} [!t]
	\centering
	\setlength{\abovecaptionskip}{5pt}
	\includegraphics[height=0.28\textwidth]{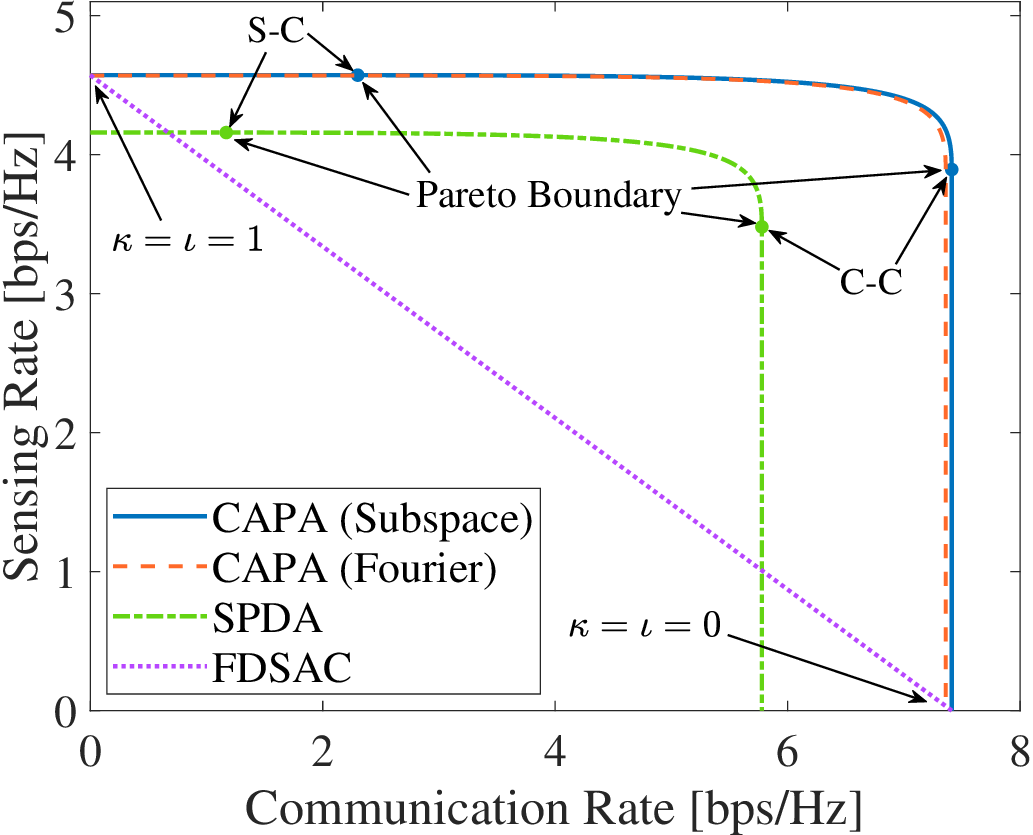}
	\caption{{ Downlink SR-CR regions.}}
	\vspace{-5pt}
	\label{do_region}
\end{figure}

{\figurename} {\ref{do_region}} presents the downlink SR-CR regions achieved by three systems: the CAPA-based ISAC system (as defined in \eqref{do_capa_region}), SPDA-based ISAC system, and CAPA-based FDSAC system (as defined in \eqref{do_fd_region}). For the ISAC system, two key corner points are highlighted, which correspond to the S-C and C-C designs, respectively. The curve connecting these points represents the Pareto boundary of the downlink ISAC rate region, which is obtained by varying the trade-off parameter $\epsilon$ from 1 to 0. An important observation from {\figurename} {\ref{do_region}} is that the rate regions of both the SPDA-based ISAC and CAPA-based FDSAC systems are fully contained within the rate region of the CAPA-based ISAC system. This underscores the superior performance of the proposed CAPA-based ISAC scheme in enhancing and balancing both communication and sensing rates.

{ While the Fourier-based approach can achieve a Pareto boundary comparable to the proposed subspace method, Table~\ref{runtime} presents the runtime for each method to obtain the Pareto boundary under varying CAPA aperture sizes and carrier frequencies. All experiments were conducted on a Windows 10 laptop equipped with an AMD Ryzen 7 5800H processor, using MATLAB R2023a. Each method was executed 100 times, and the average runtime was recorded. We observe from the table that the runtime of the Fourier-based approach increases significantly with the aperture size and frequency. This is attributed to the rapid growth in the number of integrals---computed via Gaussian quadrature---required by the Fourier method as these parameters scale. For example, when increasing the aperture size and frequency from $0.25\,\text{m}^2$ and $2.4$ GHz to $1\,\text{m}^2$ and $10$ GHz, the number of integral calculated within the Fourier-based approach rises from $162$ to $9522$. In contrast, the subspace approach consistently maintains a low runtime of approximately $0.005$ seconds, as it involves only a single integral whose complexity is independent of both aperture size and frequency. These results underscore the computational efficiency of the proposed subspace approach.

  }

\begin{table}[!t]
	\centering
	%\scalebox{0.8}{
		\caption{{ Comparison of runtime for different methods}}
		\label{runtime}
		\scalebox{0.73}{
			{ \begin{tabular}{c|cc|cc|cc}
					\toprule
					\multirow{2}{*}{Frequency}  
					& \multicolumn{2}{c|}{$L_x=L_z=0.5\,\text{m}$}
					& \multicolumn{2}{c|}{$L_x=L_z=0.7\,\text{m}$}
					& \multicolumn{2}{c}{$L_x=L_z=1\,\text{m}$}\\
					& Subspace & Fourier & Subspace & Fourier & Subspace & Fourier  \\
					\midrule
					2.4 GHz & 0.005 s & 0.163 s & 0.005 s & 0.335 s & 0.005 s & 0.642 s  \\
					5 GHz & 0.005 s & 0.721 s & 0.005 s & 1.272 s & 0.005 s & 2.495 s \\
					10 GHz  & 0.005 s & 2.539 s & 0.005 s & 4.926 s & 0.005 s & 9.885 s  \\
					\bottomrule
		\end{tabular}}}
	\end{table}

\begin{figure}[!t]
   \centering
    \subfigbottomskip=0pt
	\subfigcapskip=-2pt
\setlength{\abovecaptionskip}{3pt}
    \subfigure[CR versus transmit SNR.]
    {
        \includegraphics[height=0.27\textwidth]{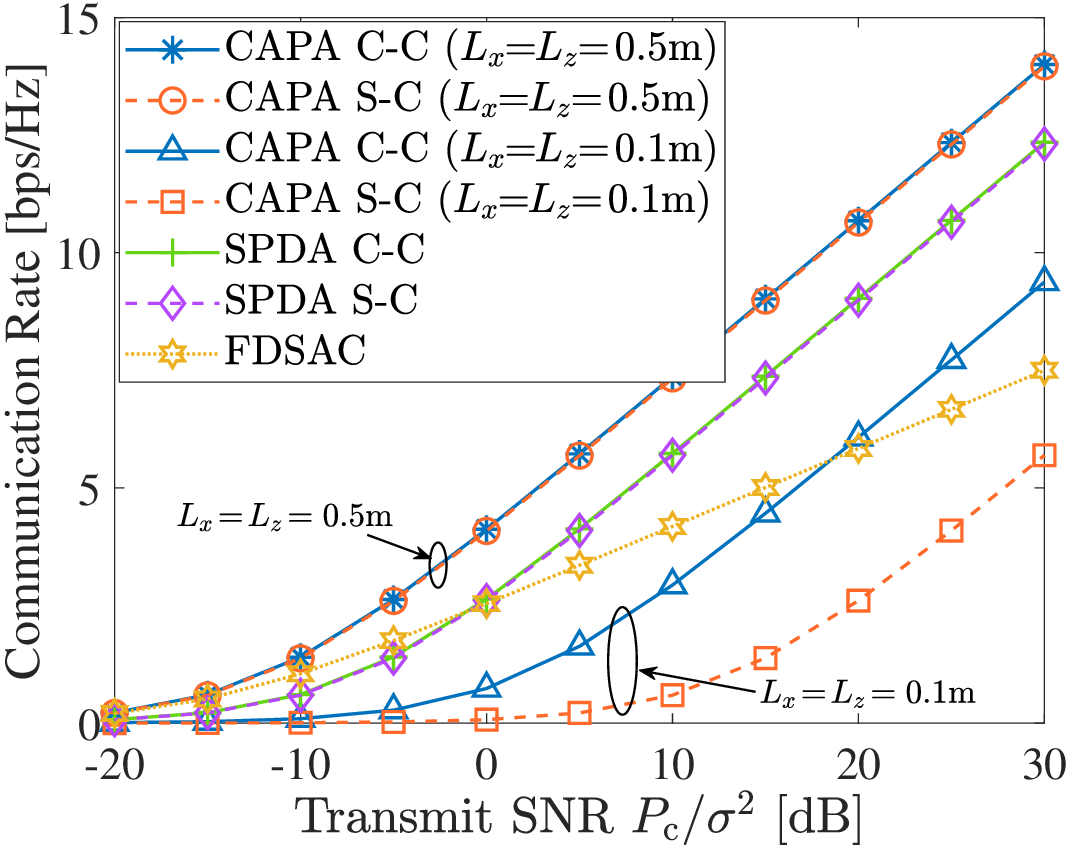}
	   \label{up_cr_snr}	
    }
    \subfigure[SR versus transmit SNR.]
    {
        \includegraphics[height=0.27\textwidth]{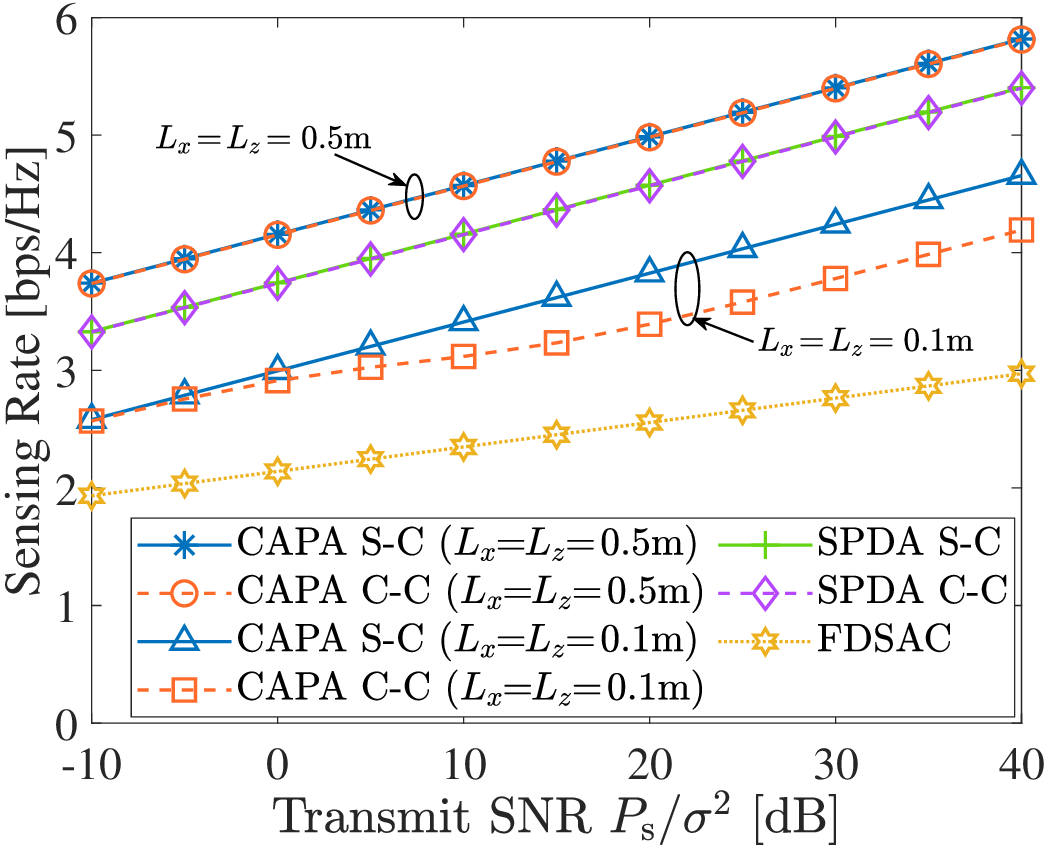}
	   \label{up_sr_snr}	
    }
\caption{Uplink performance versus transmit SNR.}
    \label{up_snr}
    \vspace{-7pt}
\end{figure}

\subsection{Uplink Performance}
We now use numerical results to demonstrate the performance of communications and sensing in CAPA-based uplink ISAC systems. Specifically, {\figurename} {\ref{up_cr_snr}} and {\figurename} {\ref{up_sr_snr}} plot the uplink CRs and SRs as functions of the communication SNR $\frac{P_{\rm{c}}}{\sigma^2}$ and sensing SNR $\frac{P_{\rm{s}}}{\sigma^2}$, respectively. As expected, the C-C SIC achieves superior communication performance, while the S-C SIC excels in sensing performance. Furthermore, both C-C ISAC and S-C ISAC exhibit identical high-SNR slopes for both CR and SR, and these slopes are larger than those achieved by the FDSAC design. This indicates the performance advantages of ISAC over FDSAC in high-SNR scenarios. Interestingly, for larger aperture sizes (more specifically, increased from $L_x=L_z=0.1$ m to $L_x=L_z=0.5$ m), the performance gap between C-C and S-C ISAC becomes negligible. This behavior is explained by the degradation of the term $\frac{\left| \rho _{\mathrm{u}} \right|^2}{g_{\mathrm{t}}g_{\mathrm{u}}}$ towards zero as the aperture size increases. This degradation reflects a weak correlation between the uplink sensing and communication channels, causing them to become nearly orthogonal. In this scenario, IFI is significantly reduced, which leads to nearly identical performance between the C-C SIC and S-C SIC designs.

\begin{figure}[!t]
    \centering
    \subfigbottomskip=0pt
	\subfigcapskip=-2pt
\setlength{\abovecaptionskip}{3pt}
    \subfigure[CR versus aperture size.]
    {
        \includegraphics[height=0.27\textwidth]{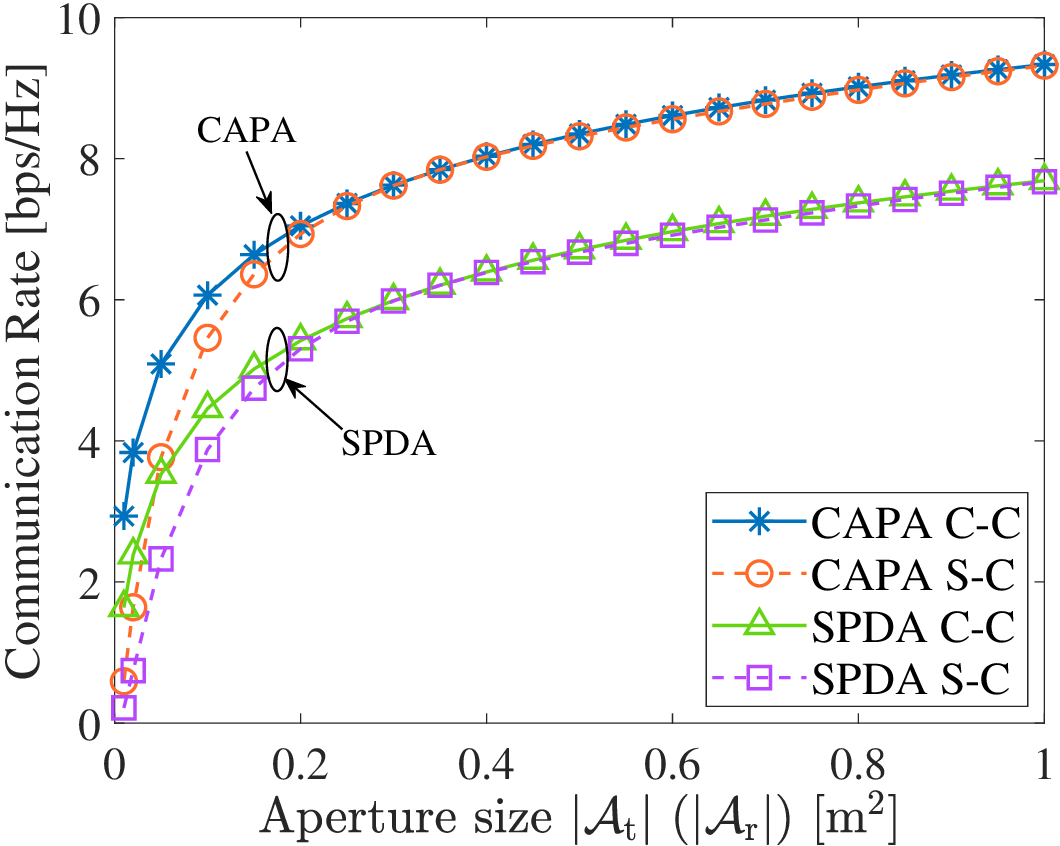}
	   \label{up_cr_size}	
    }
    \subfigure[SR versus aperture size.]
    {
        \includegraphics[height=0.27\textwidth]{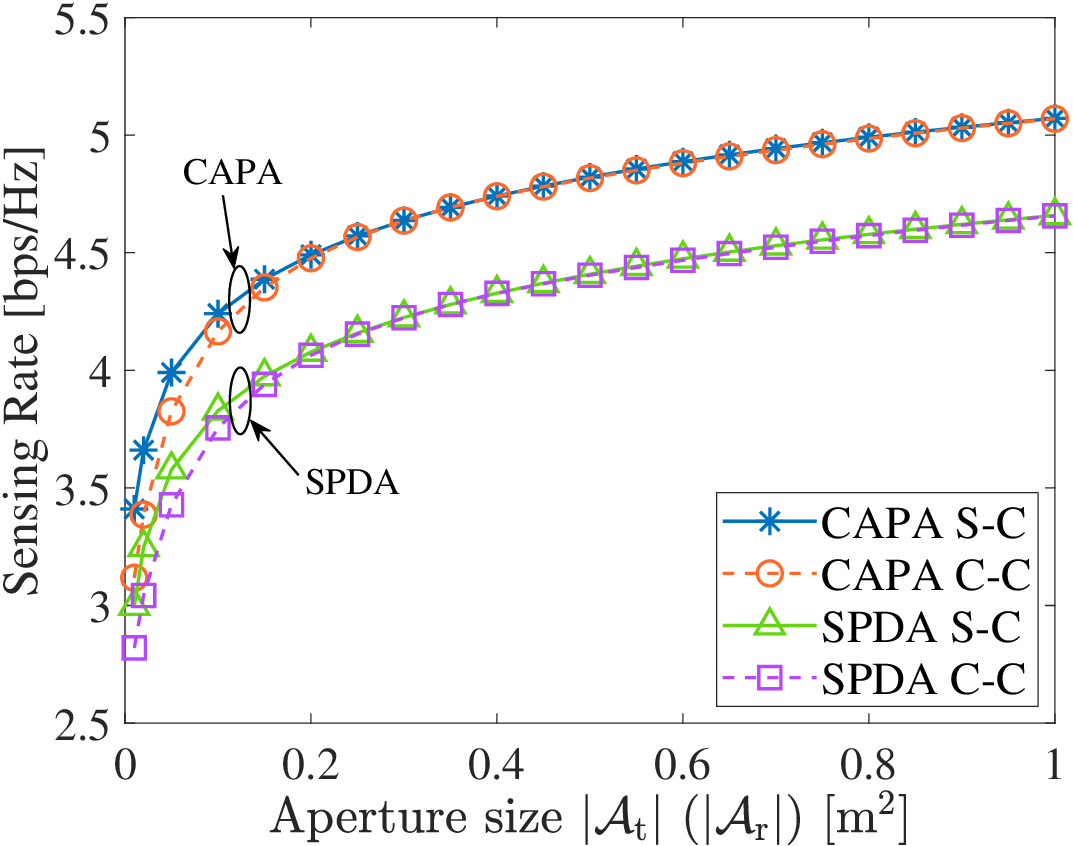}
	   \label{up_sr_size}	
    }
\caption{Uplink performance versus aperture size.}
    \label{up_size}
    \vspace{-5pt}
\end{figure}

\begin{figure} [!t]
\centering
\setlength{\abovecaptionskip}{5pt}
\includegraphics[height=0.28\textwidth]{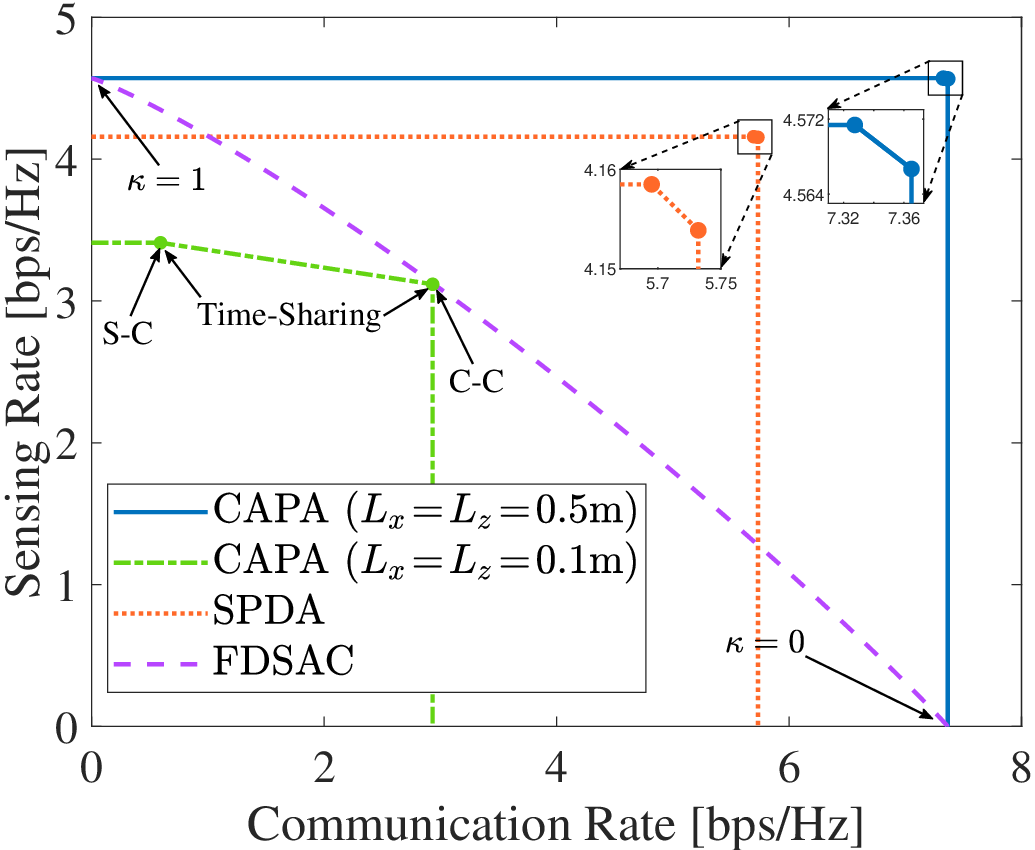}
 \caption{Uplink SR-CR regions.}
 \vspace{-7pt}
 \label{up_region}
\end{figure}

To further support the previous conclusions, {\figurename} {\ref{up_cr_size}} and {\figurename} {\ref{up_sr_size}} depict the uplink CRs and SRs as functions of the aperture size. It is observed that the performance of the S-C and C-C designs becomes nearly identical when the aperture size reaches $0.25$ m$^2$. This behavior is due to the rapid decline in the correlation between the uplink sensing and communication channels as the aperture size increases. Another observation is that, for each aperture size, CAPA outperforms SPDA in both communication and sensing performance. This trend is consistent with the results shown in {\figurename} {\ref{do_size}}.

Finally, {\figurename} {\ref{up_region}} compares the uplink SR-CR regions achieved by the CAPA-based ISAC system with those of the SPDA-based ISAC and CAPA-based FDSAC systems. For the ISAC systems, the two highlighted corner points on the plot represent the rate pairs achieved by the S-C SIC and C-C SIC designs, respectively. The line segment connecting these points corresponds to the rates obtained through a time-sharing strategy between the two designs. A significant observation is that, for the same aperture size, the rate regions of both the uplink SPDA-based ISAC and CAPA-based FDSAC systems are entirely enclosed within the rate region of the uplink CAPA-based ISAC system. This demonstrates the superior performance of CAPA-based ISAC. Furthermore, as mentioned earlier, due to minimal IFI in the uplink, the S-C and C-C designs yield nearly identical sensing and communication performance for large aperture sizes. For example, when $L_x=L_z=0.5$ m (as shown in {\figurename} {\ref{up_region}}), the corner points of the rate region are very close to each other, causing the rate region to shift from a pentagonal to an almost rectangular shape.

\section{Conclusion}\label{section_conclusion}
This article proposed an analytically tractable framework for both downlink and uplink CAPA-based ISAC by leveraging continuous operators to model the continuous nature of the EM field. Based on this framework, we analyzed the sensing and communication performance in terms of the SR and CR. For downlink ISAC, we proposed three continuous beamforming designs: the S-C design, the C-C design, and the Pareto-optimal design. For uplink ISAC, we investigated the S-C SIC and C-C SIC methods. By employing a low-complexity signal subspace-based approach, we derived closed-form expressions for optimal continuous beamformers as well as the achievable CRs, SRs, and rate regions in both downlink and uplink scenarios. Numerical results demonstrated that CAPA-based ISAC achieves better sensing and communication performance than both traditional SPDA-based ISAC and CAPA-based FDSAC systems. These findings validate the significant advantages of CAPA, making it a promising technology for next-generation ISAC systems. {As an extension, future work will consider more general CAPA-based ISAC scenarios involving multiple CUs and targets, where joint beamforming and interference management become critical challenges.}

\begin{appendix}
	\setcounter{equation}{0}
	\renewcommand\theequation{A\arabic{equation}}
{ \subsection{Proof of Theorem \ref{the_mse}}\label{proof_mse}
By defining $\hat{\mathbf{h}}_{\mathrm{s}}\triangleq \sqrt{P}\hat{v}_{\mathrm{s}}\hat{w}_{\mathrm{s}}\mathbf{s}$, \eqref{Sensing_Model_Transformed} is rewritten as 
\begin{equation}\label{a1}
\hat{\mathbf{y}}_{\mathrm{s}}=\hat{\mathbf{h}}_{\mathrm{s}}\beta _{\mathrm{s}}+\hat{\mathbf{n}}_{\mathrm{s}}.
\end{equation}	
Given the observation $\hat{\mathbf{y}}_{\mathrm{s}}$, the MSE of the estimator $f_{\mathrm{est}}\left( \hat{\mathbf{y}}_{\mathrm{s}} \right)$ for the reflection coefficient $\beta_{\mathrm{s}}$ is expressed as follows:
\begin{equation}\label{mse}
	\mathsf{MSE}=\mathbbmss{E} \left\{ \left| f_{\mathrm{est}}\left( \hat{\mathbf{y}}_{\mathrm{s}} \right) -\beta _{\mathrm{s}} \right|^2 \right\}. 
\end{equation}
It is well known that the minimum value of \eqref{mse} is attained
by the conditional mean estimator $f_{\mathrm{est}}\left( \hat{\mathbf{y}}_{\mathrm{s}} \right) =\mathbbmss{E} \left\{ \beta _{\mathrm{s}}|\hat{\mathbf{y}}_{\mathrm{s}} \right\} $. To calculate the conditional mean, we need to obtain the conditional probability density function (PDF) of $\beta _{\mathrm{s}}$ given $\hat{\mathbf{y}}_{\mathrm{s}}$, denoted as $f_{\beta _{\mathrm{s}}|\hat{\mathbf{y}}_{\mathrm{s}}}\left( x|\mathbf{y} \right) $. Based on Bayes’ theorem, we have
\begin{equation}\label{bayes}
	f_{\beta _{\mathrm{s}}|\hat{\mathbf{y}}_{\mathrm{s}}}\left( x|\mathbf{y} \right) =\frac{f_{\beta _{\mathrm{s}}}\left( x \right) f_{\hat{\mathbf{y}}_{\mathrm{s}}|\beta _{\mathrm{s}}}\left( \mathbf{y}|x \right)}{f_{\hat{\mathbf{y}}_{\mathrm{s}}}\left( \mathbf{y} \right)},
\end{equation}
where $f_{\beta _{\mathrm{s}}}\left( \cdot \right) $, $f_{\hat{\mathbf{y}}_{\mathrm{s}}|\beta _{\mathrm{s}}}\left( \cdot|\cdot \right)$, and $f_{\hat{\mathbf{y}}_{\mathrm{s}}}\left( \cdot \right)$ denote the PDFs of $\beta _{\mathrm{s}}$, $\hat{\mathbf{y}}_{\mathrm{s}}$ conditioned on $\beta _{\mathrm{s}}$, and $\hat{\mathbf{y}}_{\mathrm{s}}$, respectively. Given that $\beta _{\mathrm{s}}\sim \mathcal{C} \mathcal{N} (0,\alpha _{\mathrm{s}})$ and $\hat{\mathbf{n}}_{\mathrm{s}}\sim \mathcal{C} \mathcal{N} (0,\hat{\sigma}_{\mathrm{s}}^{2})$ with $\hat{\sigma}_{\mathrm{s}}^{2}\triangleq \sigma _{\mathrm{s}}^{2}\int_{\mathcal{A} _{\mathrm{r}}}{\left| v_{\mathrm{s}}(\mathbf{r}) \right|^2}\mathrm{d}\mathbf{r}$, we have
\begin{subequations}\label{pdfs}
	\begin{align}
		&f_{\beta _{\mathrm{s}}}\left( x \right) =\frac{1}{\pi \alpha _{\mathrm{s}}}\mathrm{e}^{-\frac{\left| x \right|^2}{\alpha _{\mathrm{s}}}},\\
		&f_{\hat{\mathbf{y}}_{\mathrm{s}}|\beta _{\mathrm{s}}}( \mathbf{y}|x) =\frac{1}{\left( \pi \hat{\sigma}_{\mathrm{s}}^{2} \right) ^L}\exp \Big( \!-{\hat{\sigma}_{\mathrm{s}}^{-2}}{\big( \mathbf{y}-\hat{\mathbf{h}}_{\mathrm{s}}x \big) ^{\mathsf{H}}\big( \mathbf{y}-\hat{\mathbf{h}}_{\mathrm{s}}x \big)} \Big), \\
		&f_{\hat{\mathbf{y}}_{\mathrm{s}}}\left( \mathbf{y} \right) =\frac{\exp \left( -\mathbf{y}^{\mathsf{H}}\big( \alpha _{\mathrm{s}}\hat{\mathbf{h}}_{\mathrm{s}}\hat{\mathbf{h}}_{\mathrm{s}}^{\mathsf{H}}+\hat{\sigma}_{\mathrm{s}}^{2}\mathbf{I}_L \big) ^{-1}\mathbf{y} \right)}{\pi ^L\det \big( \alpha _{\mathrm{s}}\hat{\mathbf{h}}_{\mathrm{s}}\hat{\mathbf{h}}_{\mathrm{s}}^{\mathsf{H}}+\hat{\sigma}_{\mathrm{s}}^{2}\mathbf{I}_L \big)}  \notag\\
		&~~~~~\overset{( a )}{=}\frac{\exp \left( -\hat{\sigma}_{\mathrm{s}}^{-2}\mathbf{y}^{\mathsf{H}}\left( \mathbf{I}_L-\frac{\alpha _{\mathrm{s}}}{\hat{\sigma}_{\mathrm{s}}^{2}+\alpha _{\mathrm{s}}\lVert \hat{\mathbf{h}}_{\mathrm{s}} \rVert ^2}\hat{\mathbf{h}}_{\mathrm{s}}\hat{\mathbf{h}}_{\mathrm{s}}^{\mathsf{H}} \right) \mathbf{y} \right)}{\left( \pi \hat{\sigma}_{\mathrm{s}}^{2} \right) ^L\big( 1+{\hat{\sigma}_{\mathrm{s}}^{-2}}{\alpha _{\mathrm{s}}}\lVert \hat{\mathbf{h}}_{\mathrm{s}} \rVert ^2 \big)},
	\end{align}
\end{subequations}
where equality $(a)$ follows from the matrix determinant lemma, i.e., $\det \big( \alpha _{\mathrm{s}}\hat{\mathbf{h}}_{\mathrm{s}}\hat{\mathbf{h}}_{\mathrm{s}}^{\mathsf{H}}+\hat{\sigma}_{\mathrm{s}}^{2}\mathbf{I}_L \big) =\left( \hat{\sigma}_{\mathrm{s}}^{2} \right) ^L\big( 1+{\hat{\sigma}_{\mathrm{s}}^{-2}}{\alpha _{\mathrm{s}}}\lVert \hat{\mathbf{h}}_{\mathrm{s}} \rVert ^2 \big)$ and Woodbury matrix identity, i.e., $\big( \alpha _{\mathrm{s}}\hat{\mathbf{h}}_{\mathrm{s}}\hat{\mathbf{h}}_{\mathrm{s}}^{\mathsf{H}}+\hat{\sigma}_{\mathrm{s}}^{2}\mathbf{I}_L \big) ^{-1}=\hat{\sigma}_{\mathrm{s}}^{-2}\big( \mathbf{I}_L-\frac{\alpha _{\mathrm{s}}}{\hat{\sigma}_{\mathrm{s}}^{2}+\alpha _{\mathrm{s}}\lVert \hat{\mathbf{h}}_{\mathrm{s}} \rVert ^2}\hat{\mathbf{h}}_{\mathrm{s}}\hat{\mathbf{h}}_{\mathrm{s}}^{\mathsf{H}} \big)$. Substituting \eqref{pdfs} into \eqref{bayes} gives
\begin{equation}
	\begin{split}
		&f_{\beta _{\mathrm{s}}|\hat{\mathbf{y}}_{\mathrm{s}}}\left( x|\mathbf{y} \right) =\frac{\alpha _{\mathrm{s}}\lVert \hat{\mathbf{h}}_{\mathrm{s}} \rVert ^2+\hat{\sigma}_{\mathrm{s}}^{2}}{\pi \alpha _{\mathrm{s}}\hat{\sigma}_{\mathrm{s}}^{2}}\times\\
		&~~\exp \!\left( \!-\frac{\alpha _{\mathrm{s}}\lVert \hat{\mathbf{h}}_{\mathrm{s}} \rVert ^2+\hat{\sigma}_{\mathrm{s}}^{2}}{\alpha _{\mathrm{s}}\hat{\sigma}_{\mathrm{s}}^{2}}\left| x-\frac{\alpha _{\mathrm{s}}}{\alpha _{\mathrm{s}}\lVert \hat{\mathbf{h}}_{\mathrm{s}} \rVert ^2+\hat{\sigma}_{\mathrm{s}}^{2}}\hat{\mathbf{h}}_{\mathrm{s}}^{\mathsf{H}}\mathbf{y} \right|^2 \right) .
	\end{split}
\end{equation}
As a result, we can obtain the conditional mean estimator as
\begin{align}\label{estimator}
	f_{\mathrm{est}}\left( \hat{\mathbf{y}}_{\mathrm{s}} \right) =\mathbbmss{E} \left\{ \beta _{\mathrm{s}}|\hat{\mathbf{y}}_{\mathrm{s}} \right\} &=\int_0^{\infty}{xf_{\beta _{\mathrm{s}}|\hat{\mathbf{y}}_{\mathrm{s}}}\left( x|\mathbf{y} \right) \mathrm{d}}x\notag\\
	&=\frac{\alpha _{\mathrm{s}}}{\alpha _{\mathrm{s}}\lVert \hat{\mathbf{h}}_{\mathrm{s}} \rVert ^2+\hat{\sigma}_{\mathrm{s}}^{2}}\hat{\mathbf{h}}_{\mathrm{s}}^{\mathsf{H}}\mathbf{y}.
\end{align}
Inserting \eqref{a1} and \eqref{estimator} into \eqref{mse} yields
\begin{align}\label{mse_2}
	\mathsf{MSE}&=\mathbbmss{E} \left\{ \left| \frac{\alpha _{\mathrm{s}}}{\alpha _{\mathrm{s}}\lVert \hat{\mathbf{h}}_{\mathrm{s}} \rVert ^2+\hat{\sigma}_{\mathrm{s}}^{2}}\hat{\mathbf{h}}_{\mathrm{s}}^{\mathsf{H}}\left( \hat{\mathbf{h}}_{\mathrm{s}}\beta _{\mathrm{s}}+\hat{\mathbf{n}}_{\mathrm{s}} \right) -\beta _{\mathrm{s}} \right|^2 \right\} \notag\\
	&=\mathbbmss{E} \left\{ \left| \frac{-\sigma _{\mathrm{s}}^{2}\beta _{\mathrm{s}}}{\alpha _{\mathrm{s}}\lVert \hat{\mathbf{h}}_{\mathrm{s}} \rVert ^2+\hat{\sigma}_{\mathrm{s}}^{2}}+\frac{\alpha _{\mathrm{s}}\hat{\mathbf{h}}_{\mathrm{s}}^{\mathsf{H}}\hat{\mathbf{n}}_{\mathrm{s}}}{\alpha _{\mathrm{s}}\lVert \hat{\mathbf{h}}_{\mathrm{s}} \rVert ^2+\hat{\sigma}_{\mathrm{s}}^{2}} \right|^2 \right\} .
\end{align}
Since $\beta _{\mathrm{s}}$ and $\hat{\mathbf{n}}_{\mathrm{s}}$ are mutually independent, \eqref{mse_2} can be further simplified as follows:
\begin{align}\label{mse_3}
	\mathsf{MSE}&=\mathbbmss{E} \left\{ \left| \frac{-\sigma _{\mathrm{s}}^{2}\beta _{\mathrm{s}}}{\alpha _{\mathrm{s}}\lVert \hat{\mathbf{h}}_{\mathrm{s}} \rVert ^2+\hat{\sigma}_{\mathrm{s}}^{2}} \right|^2 \right\} +\mathbbmss{E} \left\{ \left| \frac{\alpha _{\mathrm{s}}\hat{\mathbf{h}}_{\mathrm{s}}^{\mathsf{H}}\hat{\mathbf{n}}_{\mathrm{s}}}{\alpha _{\mathrm{s}}\lVert \hat{\mathbf{h}}_{\mathrm{s}} \rVert ^2+\hat{\sigma}_{\mathrm{s}}^{2}} \right|^2 \right\} \notag\\
	&=\frac{\alpha _{\mathrm{s}}}{1+\alpha _{\mathrm{s}}\lVert \hat{\mathbf{h}}_{\mathrm{s}} \rVert ^2\hat{\sigma}_{\mathrm{s}}^{-2}}=\frac{\alpha _{\mathrm{s}}}{1+L\overline{\gamma }_{\mathrm{s}}\frac{\left| \hat{v}_{\mathrm{s}} \right|^2\left| \hat{w}_{\mathrm{s}} \right|^2}{\int_{\mathcal{A} _{\mathrm{r}}}{\left| v_{\mathrm{s}}(\mathbf{r}) \right|^2}\mathrm{d}\mathbf{r}}}.
\end{align}
By observing \eqref{mse_3} and \eqref{Downlink_SR_Pre_Opt1}, the optimal sensing detector that minimizes the MSE follows ${{v}}_{\rm{s}}({\mathbf{r}})\propto a_{\rm{r}}(\mathbf{r})$, which corresponds to the detector that maximizes the SR. Consequently, the minimum MSE (MMSE) can be expressed as follows:
\begin{align}\label{mmse}
	\mathsf{MMSE}({{w}}({\mathbf{t}}))=\frac{\alpha _{\mathrm{s}}}{1+L\overline{\gamma }_{\mathrm{s}}\left| \hat{w}_{\mathrm{s}} \right|^2\int_{\mathcal{A} _{\mathrm{r}}}{\left| a_{\mathrm{r}}(\mathbf{r}) \right|^2}\mathrm{d}\mathbf{r}}.
\end{align}

The BCRB provides a fundamental lower bound on the MSE achievable by any unbiased Bayesian estimator. Unlike the conventional CRB which assumes deterministic unknowns, the BCRB incorporates prior information about the parameter being estimated. Based on \cite{bcrb_1}, the Bayesian Fisher information for estimating $\beta _{\mathrm{s}}$ is given by
\begin{equation}
{\mathsf{BFI}}={\mathbbmss{E}}\left\{\left|\frac{\partial\ln{f_{\beta_{\mathrm{s}}}\left( \beta_{\mathrm{s}} \right)}}{\partial\beta_{\mathrm{s}}}\right|^2\right\}
+{\mathbbmss{E}}\left\{\left|\frac{\partial\ln{f_{\hat{\mathbf{y}}_{\mathrm{s}}|\beta _{\mathrm{s}}}(\hat{\mathbf{y}}_{\mathrm{s}}|\beta _{\mathrm{s}})}}{\partial\beta_{\mathrm{s}}}\right|^2\right\}.
\end{equation}
Referring to \eqref{pdfs} and performing some basic mathematical manipulations, we obtain
\begin{subequations}
	\begin{align}
		&{\mathbbmss{E}}\left\{\left|\frac{\partial\ln{f_{\beta_{\mathrm{s}}}\left( \beta_{\mathrm{s}} \right)}}{\partial\beta_{\mathrm{s}}}\right|^2\right\}=\frac{1}{\alpha_{\mathrm{s}}},\\ &{\mathbbmss{E}}\left\{\left|\frac{\partial\ln{f_{\hat{\mathbf{y}}_{\mathrm{s}}|\beta _{\mathrm{s}}}(\hat{\mathbf{y}}_{\mathrm{s}}|\beta _{\mathrm{s}})}}{\partial\beta_{\mathrm{s}}}\right|^2\right\}=\frac{1}{\hat{\sigma}_{\mathrm{s}}^{2}}\lVert \hat{\mathbf{h}}_{\mathrm{s}} \rVert ^2.
	\end{align}
\end{subequations} 
It follows that the BCRB can be written as follows \cite{bcrb_1}:
\begin{align}\label{bcrb}
	\mathsf{BCRB}=\frac{1}{{\mathsf{BFI}}}=\frac{1}{\frac{1}{\alpha_{\mathrm{s}}}+\frac{1}{\hat{\sigma}_{\mathrm{s}}^{2}}\lVert \hat{\mathbf{h}}_{\mathrm{s}} \rVert ^2}=\mathsf{MSE},
\end{align}
where the last equality is due to \eqref{mse_3}. By further referring to \eqref{mmse}, we find that the minimum BCRB is given by $\mathsf{BCRB}({{w}}({\mathbf{t}}))=\frac{\alpha _{\mathrm{s}}}{1+L\overline{\gamma }_{\mathrm{s}}\left| \hat{w}_{\mathrm{s}} \right|^2\int_{\mathcal{A} _{\mathrm{r}}}{\left| a_{\mathrm{r}}(\mathbf{r}) \right|^2}\mathrm{d}\mathbf{r}}$. This implies that, under the considered model, the MMSE estimator attains the BCRB, making it an efficient estimator.

By comparing the results of \eqref{mmse} and \eqref{bcrb} with the definition of the SR given in \eqref{SR_define_2_step1}, it is clearly shown that both the MSE and BCRB are negatively correlated to the SR, which means that maximizing SR is equivalent to minimizing the MMSE and BCRB. This completes the proof.
}
	
\subsection{Proof of Theorem \ref{the_do_cc_cr}}\label{proof_do_cc_cr}
The distance between ${\mathbf{p}}_{\mathrm{c}}=[r_\mathrm{c}\Phi_\mathrm{c}, r_\mathrm{c}\Psi_\mathrm{c}, r_\mathrm{c}\Theta_\mathrm{c}]^{\mathsf{T}}$ and $\mathbf{t}=[x,0,z]\in\mathcal{A}_\mathrm{t}$ is calculated as follows:
\begin{align}\label{distance}
\lVert \mathbf{p}_{\mathrm{c}}-\mathbf{t} \rVert =(x^2+z^2-2r_{\mathrm{c}}( \Phi _{\mathrm{c}}x+\Theta _{\mathrm{c}}z ) +r_{\mathrm{c}}^{2})^{\frac{1}{2}}.	
\end{align}
By inserting $\mathbf{u}=\mathbf{p}_{\mathrm{c}}$ and \eqref{distance} into \eqref{Communication_Channel_LoS_Model}, the channel response $h_{\rm{d}}({\mathbf{t}})$ can be expressed as follows:
\begin{equation}
\begin{split}
	h_{\rm{d}}({\mathbf{t}})&=\frac{\mathrm{j}\eta k_0\sqrt{\frac{r_{\mathrm{c}}\Psi _{\mathrm{c}}}{4\pi}}\mathrm{e}^{-\mathrm{j}k_0(x^2+z^2-2r_{\mathrm{c}}\left( \Phi _{\mathrm{c}}x+\Theta _{\mathrm{c}}z \right) +r_{\mathrm{c}}^{2})^{\frac{1}{2}}}}{(x^2+z^2-2r_{\mathrm{c}}\left( \Phi _{\mathrm{c}}x+\Theta _{\mathrm{c}}z \right) +r_{\mathrm{c}}^{2})^{\frac{3}{4}}}\\
	&=\tilde{h}_{\mathrm{c}}(x,z),\ x\in [0,L_x],z\in [-{L_z}/{2},{L_z}/{2}].
\end{split}
\end{equation}
As a result, the channel gain $\int_{\mathcal{A} _{\mathrm{t}}}{\left| h_{\rm{d}}({\mathbf{t}}) \right|}^2\mathrm{d}\mathbf{t}$ is given by
\begin{equation}
\begin{split}
	&\int_{\mathcal{A} _{\mathrm{t}}}{\left| h_{\rm{d}}({\mathbf{t}}) \right|^2}\mathrm{d}\mathbf{t}=\int_{-\frac{L_z}{2}}^{\frac{L_z}{2}}{\int_0^{L_x}{\tilde{h}_{\mathrm{c}}(x,z)\tilde{h}_{\mathrm{c}}^{*}(x,z)\mathrm{d}x\mathrm{d}z}}\\
	&=\int_{-\frac{L_z}{2}}^{\frac{L_z}{2}}\int_0^{L_x}\frac{\frac{1}{4\pi}\eta^2k_0^2r_{\mathrm{c}}\Psi _{\mathrm{c}}\mathrm{d}x\mathrm{d}z}{\left(x^2\!+\!z^2\!-\!2r_{\mathrm{c}}\!\left( \Phi _{\mathrm{c}}x\!+\!\Theta _{\mathrm{c}}z \right) \!+\!r_{\mathrm{c}}^{2}\right)^{\frac{3}{2}}}.
\end{split}
\end{equation}
The inner integral can be evaluated using \cite[Eq. (2.264.5)]{integral}, and the outer integral using\cite[Eq. (2.284)]{integral}, which yields $\int_{\mathcal{A} _{\mathrm{t}}}{\left| h_{\rm{d}}({\mathbf{t}}) \right|^2}\mathrm{d}\mathbf{t}=g_{\mathrm{d}}$. Thus, the results of \eqref{do_cc_cr} follow directly from this calculation.
\subsection{Proof of Lemma \ref{Lemma_Subspace}}\label{proof_Lemma_Subspace}
We prove Lemma \ref{Lemma_Subspace} by contradiction. Suppose $w({\mathbf{t}})$ does not lie in the subspace spanned by $\{h_{\rm{d}}^{*}({\mathbf{t}}),a_{\rm{t}}^{*}(\mathbf{t})\}$. In that case, $w({\mathbf{t}})$ would have a component orthogonal to both $h_{\rm{d}}^{*}({\mathbf{t}})$ and $a_{\rm{t}}^{*}(\mathbf{t})$, which can be expressed as follows:
\begin{align}
w({\mathbf{t}})=\alpha_1h_{\rm{d}}^{*}({\mathbf{t}})+\alpha_2a_{\rm{t}}^{*}(\mathbf{t})+\check{w}(\mathbf{t}),
\end{align}
where $\alpha_1,\alpha_2\in{\mathbbmss{C}}$, and $\alpha_1h_{\rm{d}}^{*}({\mathbf{t}})+\alpha_2a_{\rm{t}}^{*}(\mathbf{t})$ is an arbitrary function that lies in the subspace spanned by $\{h_{\rm{d}}^{*}({\mathbf{t}}),a_{\rm{t}}^{*}(\mathbf{t})\}$. The term $\check{w}(\mathbf{t})$ denotes the orthogonal component that does not belong to this subspace. By orthogonality, we have
\begin{align}\label{Subspace_Approach_Step1}
\int_{{\mathcal{A}}_{\rm{t}}}h_{\rm{d}}^{*}({\mathbf{t}})
	\check{w}^{*}(\mathbf{t}){\rm{d}}{\mathbf{t}}=\int_{{\mathcal{A}}_{\rm{t}}}a_{\rm{t}}^{*}(\mathbf{t})
	\check{w}^{*}(\mathbf{t}){\rm{d}}{\mathbf{t}}=0.
\end{align} 
Using these orthogonality conditions along with \eqref{CC_Beamforming_Design} and \eqref{SC_Beamforming_Design}, we obtain
\begin{subequations}\label{Subspace_Approach_Step2}
\begin{align}
\hat{\gamma}_{\rm{c}}({{w}}({\mathbf{t}}))&=\left\lvert\int_{{\mathcal{A}}_{\rm{t}}}h_{\rm{d}}({\mathbf{t}})\check{w}_{\bot}(\mathbf{t}){\rm{d}}{\mathbf{t}}\right\rvert^2
=\hat{\gamma}_{\rm{c}}(\check{w}_{\bot}(\mathbf{t})),\\
\hat{\gamma}_{\rm{s}}({{w}}({\mathbf{t}}))&=\left\lvert\int_{{\mathcal{A}}_{\rm{t}}}a_{\rm{t}}(\mathbf{t})\check{w}_{\bot}(\mathbf{t}){\rm{d}}{\mathbf{t}}\right\rvert^2
=\hat{\gamma}_{\rm{s}}(\check{w}_{\bot}(\mathbf{t})),
\end{align}
\end{subequations}
where $\check{w}_{\bot}(\mathbf{t})\triangleq\alpha_1h_{\rm{d}}^{*}({\mathbf{t}})+\alpha_2a_{\rm{t}}^{*}(\mathbf{t})$. Additionally, the power constraint is given by
\begin{align}\label{Subspace_Approach_Step3}
\int_{{\mathcal{A}}_{\rm{t}}}\lvert w({\mathbf{t}})\rvert^2{\rm{d}}{\mathbf{t}}=
\int_{{\mathcal{A}}_{\rm{t}}}\lvert \check{w}_{\bot}(\mathbf{t})\rvert^2{\rm{d}}{\mathbf{t}}+
\int_{{\mathcal{A}}_{\rm{t}}}\lvert w({\mathbf{t}})\rvert^2{\rm{d}}{\mathbf{t}}.
\end{align}
The results in \eqref{Subspace_Approach_Step2} and \eqref{Subspace_Approach_Step3} imply that the orthogonal component $\check{w}_{\bot}(\mathbf{t})$ contributes nothing to either the CR or the SR but still consumes additional power. 

Since $\check{w}_{\bot}(\mathbf{t})$ has no positive impact on system performance and only increases power consumption, it should be set to zero. Therefore, the optimal beamformer satisfies $w({\mathbf{t}})=\check{w}_{\bot}(\mathbf{t})=\alpha_1h_{\rm{d}}^{*}({\mathbf{t}})+\alpha_2a_{\rm{t}}^{*}(\mathbf{t})$, which lies in the subspace spanned by $\{h_{\rm{d}}^{*}({\mathbf{t}}),a_{\rm{t}}^{*}(\mathbf{t})\}$. This concludes the proof of Lemma \ref{Lemma_Subspace}.
\subsection{Proof of Lemma \ref{lem_w_star}}\label{proof_lem_w_star}
The optimal solution of problem \eqref{pareto_problme_2} can be derived from the KKT conditions as follows:
\begin{align}
	&\nabla(-\gamma)+\nu \nabla(\lVert{\mathbf{w}}\rVert^2-1)+\mu_1\nabla (\epsilon\gamma-\lvert{{\mathbf{h}}}_{\mathrm{d}}^{\mathsf{T}}\mathbf{w}\rvert^2)\notag\\
	&+\mu_2\nabla ((1-\epsilon)\gamma-\lvert{\mathbf{a}}_{\mathrm{t}}^{\mathsf{T}}\mathbf{w}\rvert^2)={\mathbf{0}}, \label{KKT_1} \\ 
	&\mu_1(\epsilon\gamma-\lvert{{\mathbf{h}}}_{\mathrm{d}}^{\mathsf{T}}\mathbf{w}\rvert^2)=0,\label{KKT_2} \ \mu_2((1-\epsilon)\gamma-\lvert{\mathbf{a}}_{\mathrm{t}}^{\mathsf{T}}\mathbf{w}\rvert^2)=0,\\
	&\mu_1\geq0,\ \mu_2\geq0,\ \nu\in{\mathbbmss{R}},
\end{align}
where $\nu$, $\mu_1$, and $\mu_2$ are real-valued Lagrangian multipliers. From \eqref{KKT_1}, we obtain
\begin{align}
		&\epsilon\mu_1+(1-\epsilon)\mu_2=1,\label{KKT_1_Dev2}\\		&(\mu_1{{\mathbf{h}}}_{\mathrm{d}}^*{{\mathbf{h}}}_{\mathrm{d}}^{\mathsf{T}}+\mu_2{\mathbf{a}}_{\mathrm{t}}^*{\mathbf{a}}_{\mathrm{t}}^{\mathsf{T}}){\mathbf{w}}=\nu{\mathbf{w}}.\label{KKT_1_Dev1}   
\end{align}
It follows from \eqref{KKT_1_Dev2} that $\mu_1$ and $\mu_2$ cannot both be zero simultaneously. We now discuss three cases as follows.
\subsubsection{$\mu_1>0$ and $\mu_2=0$} From \eqref{KKT_2}, this gives $\gamma=\frac{\lvert{\mathbf{h}}_{\mathrm{d}}^{\mathsf{T}}\mathbf{w}\rvert^2}{\epsilon}$. The optimal solution in this case is ${\mathbf{w}_\epsilon}=\frac{{\mathbf{h}}_{\mathrm{d}}^*}{\sqrt{g_{\mathrm{d}}}}$ and $\gamma_\star=\frac{g_{\mathrm{d}}}{\epsilon}$. Substituting this result into the constraint $\lvert{\mathbf{a}}_{\mathrm{t}}^{\mathsf{T}}{\mathbf{w}}\rvert^2\geq(1-\epsilon)\gamma$ yields $\epsilon\in \left[ \frac{g_{\mathrm{d}}^2}{g_{\mathrm{d}}^2+\left| \rho _{\mathrm{d}} \right|^2},1 \right] $.
\subsubsection{$\mu_1=0$ and $\mu_2>0$} Following similar steps to Case 1, we obtain ${\mathbf{w}_\epsilon}=\frac{{\mathbf{a}}_{\mathrm{t}}^*}{\sqrt{g_{\rm{t}}}}$ with $\epsilon\in \left[ 0,\frac{\left| \rho _{\mathrm{d}} \right|^2}{\left| \rho _{\mathrm{d}} \right|^2+g_{\mathrm{t}}^2} \right] $.
\subsubsection{$\mu_1>0$ and $\mu_2>0$}
In this case, from \eqref{KKT_2}, we have
\begin{equation}\label{KKT_1_Dev3}	\mathcal{G}=\frac{1}{\epsilon}\lvert{\mathbf{h}}_{\mathrm{d}}^{\mathsf{T}}\mathbf{w}\rvert^2=\frac{1}{1-\epsilon}\lvert{\mathbf{a}}_{\mathrm{t}}^{\mathsf{T}}{\mathbf{w}}\rvert^2.
\end{equation}
From \eqref{KKT_1_Dev1} and \eqref{KKT_1_Dev3}, $\mathbf{w}$ can be expressed as a linear combination of $\mathbf{h}_{\mathrm{d}}$ and $\mathbf{a}_{\mathrm{t}}$:
\begin{align}\label{wab}
	\mathbf{w}=\frac{\mu _1\mathbf{h}_{\mathrm{d}}^{\mathsf{T}}\mathbf{w}}{\nu}\mathbf{h}_{\mathrm{d}}^*+\frac{\mu _2\mathbf{a}_{\mathrm{t}}^{\mathsf{T}}\mathbf{w}}{\nu}\mathbf{a}_{\mathrm{t}}^*\propto \hat{a}\mathbf{h}_{\mathrm{d}}^*+\hat{b}\mathbf{a}_{\mathrm{t}}^*,
\end{align}
where $\frac{\hat{a}}{\hat{b}}=\frac{\mu _1\sqrt{\epsilon}}{\mu _2\sqrt{1-\epsilon}\mathrm{e}^{-\mathrm{j}\angle \rho _{\mathrm{d}}}}$. Inserting \eqref{wab} into \eqref{KKT_1_Dev1} gives
\begin{align}\label{mu_eta}
	{\mu_1}(g_{\mathrm{d}}+\rho_{\mathrm{d}}{\hat{b}}/{\hat{a}})={\mu_2}(g_{\mathrm{t}}+\rho _{\mathrm{d}}^*{\hat{a}}/{\hat{b}})=\nu.    
\end{align}
It follows from \eqref{mu_eta} and \eqref{KKT_1_Dev2} that 
\begin{align}
&\mu _1=\frac{g_{\mathrm{t}}-\sqrt{\left( 1-\epsilon \right) /\epsilon}\left| \rho _{\mathrm{d}} \right|}{\left( 1-\epsilon \right) g_{\mathrm{d}}+\epsilon g_{\mathrm{t}}-2\sqrt{\epsilon \left( 1-\epsilon \right)}\left| \rho _{\mathrm{d}} \right|},\\
&\mu _2=\frac{g_{\mathrm{d}}-\sqrt{\epsilon /\left( 1-\epsilon \right)}\left| \rho _{\mathrm{d}} \right|}{\left( 1-\epsilon \right) g_{\mathrm{d}}+\epsilon g_{\mathrm{t}}-2\sqrt{\epsilon \left( 1-\epsilon \right)}\left| \rho _{\mathrm{d}} \right|}.
\end{align}
Given the prerequisites $\mu_1>0$ and $\mu_2>0$, we have $\epsilon \in \left( \frac{\left| \rho _{\mathrm{d}} \right|^2}{\left| \rho _{\mathrm{d}} \right|^2+g_{\mathrm{t}}^2},\frac{g_{\mathrm{d}}^2}{g_{\mathrm{d}}^2+\left| \rho _{\mathrm{d}} \right|^2} \right) $. Consequently, the optimal $\mathbf{w}$ in this case can be expressed as follows: 
\begin{equation}
	\mathbf{w}_\epsilon=\frac{\upsilon _1}{\tau}\mathbf{h}_{\mathrm{d}}^*+\frac{\upsilon _2\mathrm{e}^{-\mathrm{j}\angle \rho _{\mathrm{d}}}}{\tau}\mathbf{a}_{\mathrm{t}}^*,
\end{equation}
where $\upsilon _1$ and $\upsilon _2$ are given by \eqref{Appendix_Used_Results}, and $\tau$ normalizes the solution. Taken together, the proof of Lemma \ref{lem_w_star} is concluded.
\end{appendix}

\bibliographystyle{IEEEtran}
\bibliography{IEEEabrv}
\end{document}